\makeatletter
\def\cl@chapter{}
\makeatother
\documentclass[twocolumn]{arxiv}
\usepackage{iftex}
\ifPDFTeX
  \usepackage{cmap}
\fi
\usepackage{fix-cm}
\usepackage{soul}

\usepackage{iftex}
\ifPDFTeX
  \usepackage[utf8]{inputenc}
\fi

\usepackage[T1]{fontenc}

\usepackage{natbib}
\usepackage{doi}

\usepackage{graphicx}

\usepackage[section]{placeins} 

\usepackage[switch]{lineno}
\modulolinenumbers[1]

\usepackage{nameref}

\usepackage{amssymb}

\usepackage[inline]{enumitem}
\usepackage[normalem]{ulem} 

\usepackage[table,xcdraw,dvipsnames]{xcolor}

\usepackage{multirow}
\usepackage{threeparttable, tablefootnote}
\usepackage{booktabs, makecell}
\usepackage{siunitx}
\usepackage{arydshln}
\usepackage{pifont}
\usepackage{marvosym}

\usepackage{url}
\makeatletter
\g@addto@macro{\UrlBreaks}{\UrlOrds}
\makeatother

\RequirePackage{hyperref}
\PassOptionsToPackage{hyphens}{url}\usepackage{hyperref}

\newcommand\myshade{85}
\colorlet{mylinkcolor}{violet}
\colorlet{mycitecolor}{YellowOrange}
\colorlet{myurlcolor}{Aquamarine}

\hypersetup{
    linkcolor  = mylinkcolor!\myshade!black,
    citecolor  = mycitecolor!\myshade!black,
    urlcolor   = myurlcolor!\myshade!black,
    colorlinks = true,
    pdftitle={FADE-CTP: A Framework for the Analysis and Design of Educational Computational Thinking Problems},
    pdfpagemode=FullScreen,
}
\usepackage{xurl}

\usepackage[capitalize,nameinlink]{cleveref} 
\crefname{figure}{Fig.}{Figs.}%
\crefname{section}{Section}{Sections}

\usepackage{orcidlink}

\usepackage[skipabove=2mm, skipbelow=1mm]{mdframed} 
\mdfdefinestyle{ndef-frame}{
	linewidth=1pt, %
	linecolor= gray, %
	backgroundcolor= gray!10,
	topline=false, %
	bottomline=false, %
	rightline=false,%
	leftmargin=0pt, %
	innerleftmargin=.6em, %
	innerrightmargin=.6em, 
	rightmargin=0pt, %
	innertopmargin=.5em,%
	innerbottommargin=.5em, %
}%
\surroundwithmdframed[style=ndef-frame]{ndef}
\newtheorem{ndef}{Definition}

\begin{document}
\sloppy
\title{FADE-CTP: A Framework for the Analysis and Design of Educational Computational Thinking Problems}



\author{Giorgia Adorni         \and
        Alberto Piatti \and
        Engin Bumbacher \and 
        Lucio Negrini \and 
        Francesco Mondada \and 
        Dorit Assaf \and 
        Francesca Mangili \and 
        Luca Maria Gambardella
}
\institute{
G. Adorni \orcidlink{0000-0002-2613-4467}{ 0000-0002-2613-4467} \at Università della Svizzera Italiana (USI), Dalle Molle Institute for Artificial Intelligence (IDSIA), Lugano, Switzerland \\\email{giorgia.adorni@usi.ch} 
\and
A. Piatti \orcidlink{0000-0002-5196-4630}{ 0000-0002-5196-4630} \at University of Applied Sciences and Arts of Southern Switzerland (SUPSI), Department of Education and Learning (DFA), Locarno, Switzerland \\\email{alberto.piatti@supsi.ch} 
\and
E. Bumbacher \orcidlink{0000-0002-4322-7059}{ 0000-0002-4322-7059} \at Haute école pédagogique du canton de Vaud (HEP-VD), Lausanne, Switzerland, \\\email{engin.bumbacher@hepl.ch} 
\and 
L. Negrini \orcidlink{0000-0001-6793-6258}{ 0000-0001-6793-6258} \at University of Applied Sciences and Arts of Southern Switzerland (SUPSI), Department of Education and Learning (DFA), Locarno, Switzerland \\\email{lucio.negrini@supsi.ch} 
\and 
F. Mondada \orcidlink{0000-0001-8641-8704}{ 0000-0001-8641-8704} \at Ecole Polytechnique Fédérale de Lausanne (EPFL), Mobile Robotic Systems Group (MOBOTS), Lausanne, Switzerland \\\email{francesco.mondada@epfl.ch} 
\and
D. Assaf \orcidlink{0000-0002-7877-3041}{ 0000-0002-7877-3041} \at University of Applied Sciences and Arts Northwestern Switzerland (FHNW), School of Education, Windisch, Switzerland \\\email{dorit.assaf@fhnw.ch} 
\and 
F. Mangili \orcidlink{0000-0002-3215-1028}{ 0000-0002-3215-1028} \at University of Applied Sciences and Arts of Southern Switzerland (SUPSI), Dalle Molle Institute for Artificial Intelligence (IDSIA), Lugano, Switzerland \\\email{francesca.mangili@supsi.ch} 
\and 
L. M. Gambardella \at Università della Svizzera Italiana (USI), Dalle Molle Institute for Artificial Intelligence (IDSIA), Lugano, Switzerland \\\email{luca.gambardella@usi.ch}    
}


\maketitle
\bigskip
{~}
\\ \bigskip

\begin{abstract}
In recent years, the emphasis on computational thinking (CT) has intensified as an effect of accelerated digitalisation.
While most researchers are concentrating on defining CT and developing tools for its instruction and assessment, we focus on the characteristics of computational thinking problems (CTPs) -- activities requiring CT to be solved -- and how they influence the skills students can develop.
{In this paper, we present a comprehensive framework for systematically profiling CTPs by identifying specific components and characteristics, while establishing a link between these attributes and a structured catalogue of CT competencies.
} 
{The purposes of this framework are (i) facilitating the analysis of existing CTPs to identify which abilities can be developed or measured based on their inherent characteristics, and (ii) guiding the design of new CTPs targeted at specific skills by outlining the necessary characteristics required for CT activation. 
}
{
To illustrate the framework functionalities, we begin by analysing prototypical activities in the literature, a process that leads to the definition of a taxonomy of CTPs across various domains, and we conclude with a case study on the design of a different version of one of these activities, the Cross Array Task (CAT), set in different cognitive environments. 
This approach allows an understanding of how CTPs in different contexts display unique and recurring characteristics that promote the development of distinct skills.
}
%
{
In conclusion, this framework can inform the development of assessment tools, improve teacher training, and facilitate the analysis and comparison of existing CT activities, contributing to a deeper understanding of competency activation and guiding curriculum design in CT education.}
\end{abstract}


\section{Introduction and Background}
\label{sec:intro}

{In an era defined by rapid technological advancement and increasing digitalisation, computational thinking (CT) has gained significant attention across educational sectors as a foundational skill that equips students with the ability to navigate and engage with complex systems in an increasingly data-driven world} \citep{weintrop2021assessing,wing2014computational,wing2006computational,Kafai2020}. 

{Computational thinking problems (CTPs) are activities that necessitate the application of CT to arrive at a solution. These problems can vary widely in their design and complexity, influencing the type of skills students can cultivate during the learning process.}

{Despite the growing emphasis on defining CT and developing pedagogical tools for its instruction and assessment, there remains a critical gap in understanding the specific characteristics of CTPs and their implications for skill development} \citep{brennan2012new,lafuente2022assessing,weintrop2016defining}.

Prevailing approaches, like those from \cite{brennan2012new} and \cite{grover2017} have sought to decompose CT into sub-dimensions (e.g., decomposition, generalisation, pattern recognition) to categorise existing tasks based on the underlying skills they entail.
This breakdown aims to facilitate the development of activities that specifically target each skill, making the teaching and assessment of CT more systematic \citep{lafuente2022assessing}.
However, recent studies have underscored the challenges of isolating these abilities in applied models.
For instance, while \cite{lafuente2022assessing} successfully developed a reliable and validated CT assessment for adults by integrating existing items, the statistical analyses revealed a preference for a one-dimensional model rather than a multi-dimensional approach.
{Although experts identified various CT sub-dimensions that the assessment items should address, the close interrelationship between them complicates their separation} \citep{lafuente2022assessing}. This intertwining of skills presents a significant challenge, as it suggests that the development and assessment of CT skills cannot be effectively achieved by merely focusing on distinct sub-dimensions.
This phenomenon is also evident in other complex competencies, such as scientific inquiry and practices, where the interdependencies among sub-dimensions make clear distinctions increasingly difficult \citep{osborne2014,ford2015}.

{This limitation suggests that CT is not simply a collection of isolated skills, but a set of intertwined competencies that emerge more effectively in authentic, meaningful contexts. In other words, CTPs require a more integrated approach, where skills are not merely isolated but activated and connected through problems that reflect real-world scenarios.}

{From this perspective, the theory of situated cognition becomes crucial: learning CT is more effective when it takes place in relevant contexts, as competencies develop through interaction with the environment and the learning community} \citep{roth2013situated,heersmink2013taxonomy}.
{This perspective suggests that CT is not merely a collection of generic skills but a process that adapts and evolves based on the situations in which it is applied. Consequently, the design of CTPs should consider not only the target skills but also the learning context to promote a more nuanced, contextualised understanding of CT complexity.}

{However, as noted in the works of} \cite{shute2017demystifying} and \cite{tikva_mapping_2021}, {current frameworks tend to focus on defining skills or creating tools for instruction and assessment, often neglecting the specific characteristics of CTPs that can impact student engagement and learning outcomes. 
This oversight can limit our understanding of how the structure and design of CTPs can effectively foster or impede specific competencies. 
A situated cognition approach, therefore, can clarify how CTP design influences the learning experience and student's ability to apply CT in authentic, meaningful ways} \citep{piatti_2022}.
 

{To address these gaps, this study proposes a framework that provides (i) a methodology for systematically analysing existing CTPs to identify which CT skills can be developed or assessed in real educational contexts based on their structural and contextual characteristics and (ii) a design approach for developing new CTPs that deliberately target specific skills by incorporating essential attributes required for activating CT within varied learning contexts.}

{
To achieve this, our framework first establishes a general set of core components and characteristics that define CTPs, including aspects such as artefactual environments and problem domains.
We then develop a hierarchical catalogue of CT skills, organised to reflect interrelated competencies.
By systematically linking these characteristics with the competencies they can foster, the framework creates a structured profile for CTPs that allows educators to target specific CT skills more precisely within their instructional design.
This profiling approach is intended to facilitate a clearer alignment between CTP design and competency development, addressing a significant need in CT education for structured, skill-focused tools.}

{To showcase the capabilities of our framework, first, we analyse existing prototypical CTPs in the literature, categorising these across unplugged, robotic and virtual domains.
This process leads us to develop a taxonomy of CTPs that reveals patterns in how activities in different domains have different characteristics and thus can target specific CT competencies, ultimately providing educators with a structured guide for selecting or designing CT activities based on the desired skill outcomes.
A second application is a design-focus case study centred on the Cross Array Task (CAT), an activity for algorithmic thinking assessment. We started from the unplugged version of the CAT, which we had analysed during the previous phase, and we showed how to design a virtual variant of the activity.
This comparative case study illustrates how design elements, such as environment and interaction type, influence the development of specific CT skills. By demonstrating how these variations activate different competencies, our framework guides educators in designing contextually relevant CTPs.
}

{In conclusion, this work introduces a novel framework that can empower educators to systematically profile CTPs by identifying their core components, characteristics and competencies.
This structured approach can offer practical guidance for the analysis and design of activities targeted at specific CT skills.}

Although CT has become a key educational objective, especially following Jeannette Wing's foundational definition as ``the thought processes involved in formulating problems and their solutions so that the solutions are represented in a form that an information-processing agent can effectively carry out'' \citep{wing2006computational}, the lack of a precise, universally accepted definition has limited its broader application in educational contexts \citep{shute2017demystifying}. 
Currently, there is a proliferation of overlapping definitions and multiple interpretations of similar aspects of CT, while many focus on specific facets with distinct objectives, such as promoting problem-solving skills or understanding algorithmic processes, yet none provides a comprehensive view that encapsulates CT's multifaceted nature \citep{lafuente2022assessing}. 
This lack of consensus has constrained the field's advancement, with progress remaining largely exploratory rather than standardised \citep{weintrop2021assessing}.
{By defining a structured set of CT competencies and linking them to the CTPs characteristics, our framework aims to contribute to a more comprehensive and widely applicable definition of CT, in turn supporting the broader goal of integrating CT into curricula in contextually relevant ways.
}

\section{Method}
\label{sec:method}

{This section presents our methodological approach to developing a framework for profiling computational thinking problems (CTPs). 
We begin with the definition of CTPs, followed by a presentation of their components and distinctive characteristics. 
Next, we will create a catalogue of CT competencies, before examining the link between the characteristics of CTPs and specific CT skills. 
This structure will illustrate how each element contributes to a deeper understanding and practical application of CT in education.
}



\subsection{{Definition of CTP}}\label{sec:ctp}

{Our definition of CTPs is grounded in the theoretical framework proposed by} \cite{piatti_2022}, {which integrates foundational concepts of CT from} \cite{wing2006computational} {with situated theories of learning by} \cite{roth2013situated} {and} \cite{heersmink2013taxonomy}.
Specifically, they emphasise that CT should be considered a situated activity, contextualised and embedded in real-world scenarios, and understood as a dynamic and adaptive process rather than a fixed set of competencies.

{Building on this integrated perspective, we define CTPs as tasks designed to engage learners in applying CT skills to derive solutions, within realistic environments that reflect problem-solving complexities.
In alignment with} \cite{piatti_2022}, {our approach underscores the influence of physical and social contexts on CT activities and recognises the role of cognitive artefacts in supporting problem resolution.
}

\begin{figure}[b]
	\centering
	\includegraphics[width=.95\columnwidth]{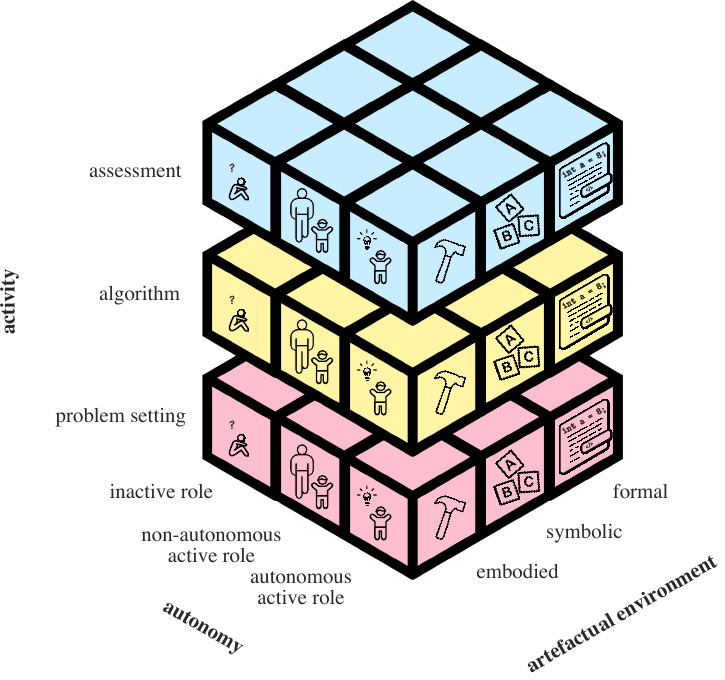}
	\caption{\textbf{Visualisation of the CT-cube, from \cite{piatti_2022}}.
		This model considers the type of activity (problem setting, algorithm, assessment), the artefactual environment (embodied, symbolic, formal), and the autonomy (inactive role, non-autonomous active role, or autonomous active role).}
	\label{fig:ctcube}
\end{figure}
The computational thinking cube (CT-cube), illustrated in \Cref{fig:ctcube}, serves as a foundational framework for understanding the design and assessment of CTPs, emphasising the interconnectedness of various elements that influence CT activities \citep{piatti_2022}. 
{Specifically, the CT-cube encompasses three critical dimensions, which will later inform our presentation of the components and characteristics of CTPs.
The first dimension addresses the \textit{type of activity} being performed or required, which may involve problem setting, algorithm development, or assessment. 
%
The second dimension highlights the \textit{artefactual environment} in which the activities occur. This encompasses the resources utilised, ranging from embodied tools (such as physical manipulatives) to symbolic or formal representations. 
%
 The third dimension considers the concept of \textit{autonomy}, comprising social interactions and individual's level of independence, underscoring the varying levels of learner engagement a priori and during the CTPs. Depending on whether learners assume an inactive, non-autonomous active, or autonomous role, their interaction with the task and the application of their CT skills may differ significantly.} 

{
The interplay of the type of activity, the artefactual environment, and the learner's autonomy collectively determines the characteristics of the CTP, shaping the competencies developed during engagement and influencing both the learning experience and the effectiveness of problem-solving approaches.
}


\subsubsection{{Components}}
\begin{figure}[htb]
	\centering
	\includegraphics[width=.95\columnwidth]{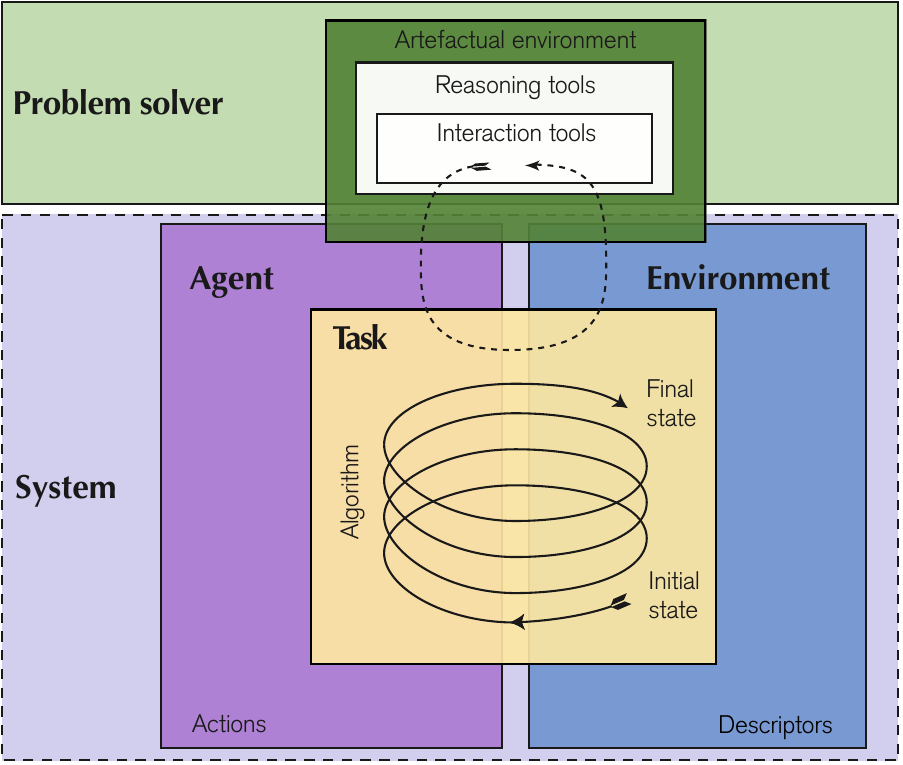}
	\caption{\textbf{Visualisation of the components of a CTP.}
        CTPs include (1) the problem solver (in green) characterised by the artefactual environment, i.e., the set of reasoning and interaction tools, (2) the system, which consists of an environment with its descriptors (in blue) and an agent with its actions (in violet), and (3) the task (in yellow) characterised by the set of initial states, algorithms and final states.}
	\label{fig:framework}
\end{figure}

{We identified several components that constitute CTPs, illustrate in } \Cref{fig:framework}: the system, comprising the environment and the agent, the problem solver, and the task. 

The \textit{environment} is a physical and/or a virtual external space, characterised by one or more variables, called ``descriptors'', which may change over time according to the dynamics of this space.

The \textit{agent} is a human, robotic or virtual being {that interacts with the environment by} performing ``actions'' to change the value of its descriptors and, therefore, alter the state of the environment.
An ``algorithm'' is a finite set of instructions an agent should follow to perform actions in the environment to solve the task.
Algorithms for different agents can take various forms: code for a virtual agent, behaviour for a robot, or a verbal or written set of instructions for a human.


The \textit{problem solver} is a human or group of people who can solve tasks that require the use of algorithms, such as designing, implementing, or communicating them to an agent to change the state of an environment. 
They have access to \textit{reasoning tools}, which are cognitive artefacts {that assist in thinking} about the task, such as whiteboards used to organise ideas and understand the logic of a problem or solution.
Some of these tools, known as \textit{interaction tools}, also allow the problem solver to interface with the system.
For example, a programming platform may serve as both a reasoning tool, enabling the problem solver to plan and design code, and an interaction tool, facilitating the execution of the algorithm and allowing the observation of its effect on the system.
{Collectively, these tools form the \textit{artefactual environment}, which according to the definition of} \cite{piatti_2022} {and the model of the three worlds of mathematics by} \cite{tall2006theory,tall2013humans,tall2020three}, {can also be categorised in: ``embodied'', iconic representational or ecological tools, based on sensory perception and embodiment; ``symbolic'' tools, used to conceive and apply procedures and rules; and ``formal'' tools,  used to create, generalise and represent structures.} 

The \textit{task} is the activity that the problem solver performs to find one or more solutions to a CTP.
A solution is a combination of ``initial states'', ``algorithms'', and ``final states'' that meets the system's requirements for a particular environment, with its set of states, and a given agent, with its set of algorithms. 
The initial state is the starting configuration of the environment, while the final state is the state of the environment after the algorithm is performed.
For a solution to be valid, the algorithm must be executed on the initial state and then produce the final state.
{Each element that composes a task (initial state, algorithm, final states) can be ``given'' or is ``to be found''. 
Based on the number and the epistemic nature of elements to be found, it is possible to divide tasks into six types.
Those with a single objective are: 
(1) \emph{find the initial state}: given the final state and the algorithm that produced it, the problem solver must infer the initial state on which the algorithm was applied;
(2) \emph{find the algorithm}: given the initial and the final states, the problem solver must devise and describe an algorithm, or a part of it, that the agent can execute to transform the system from the initial to the final state;
(3) \emph{find the final state}: given the initial state and an algorithm, the problem solver must derive the final state.
Pairs of single-objective tasks form those with multiple objectives:
(4) \emph{creation act}: a combination of find the algorithm and find the final state; 
(5) \emph{application act}: a combination find the initial state and find the final state; 
(6) \emph{project act}: a combination find the initial state and find the algorithm. 
}

\subsubsection{{Characteristics}}

\begin{figure*}[ht]
	\centering
	\includegraphics[width=\textwidth]{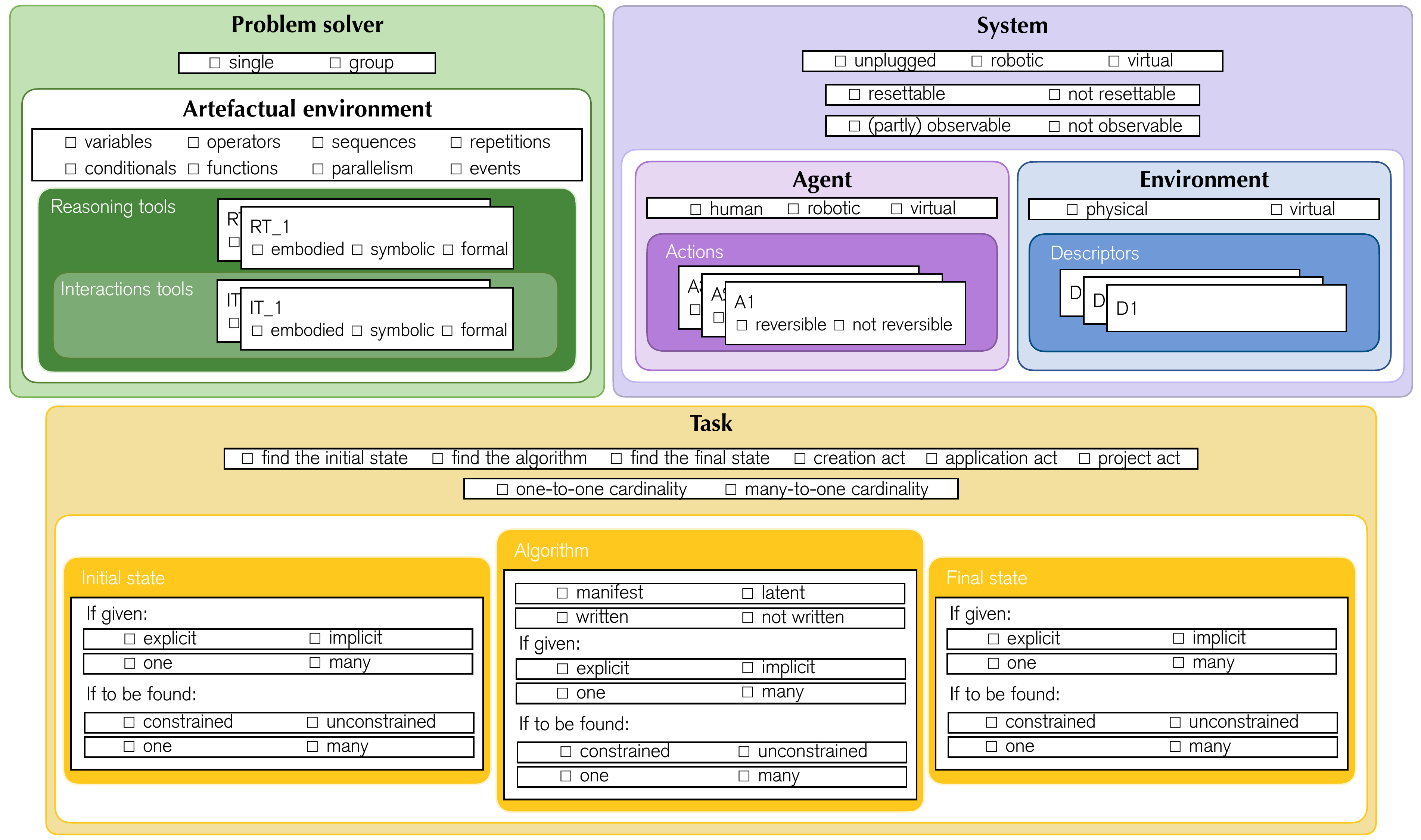}
	\caption{\textbf{Graphical template for the analysis of CTPs components and characteristics.} 
        The same colour scheme as in \Cref{fig:framework} is applied.}
\label{fig:framework-components}
\end{figure*}
{After defining the components of CTPs, we identify key characteristics that further clarify their nature. These attributes, along with their role, are illustrated in} \Cref{fig:framework-components}, {which serves as a template for CTP analysis.  
}

\begin{ndef}[Problem domain] 
	\normalfont 
	~ \newline  
	The category of an activity, determined by the nature of the agent and of the environment. 
\end{ndef}
Three main categories of domains are commonly recognised in cognitive tasks, including: ``unplugged'' activities, which involve a human agent and a physical environment; ``robotic'' activities, in which the agent is a robot and the environment is physical; and ``virtual'' activities, where both agent and environment are virtual, such as in a simulated scenario. 


\begin{ndef}[Tool functionalities] 
	\normalfont 
	~ \newline  
	The {artefactual environment's capabilities} enabling the problem solver to {construct the algorithm}.
\end{ndef}
{
The functionalities we included in this categorisation are tailored for beginner-level CT education to introduce foundational algorithmic concepts, such as} ``variables'', ``operators'', ``sequences'', ``repetitions'', ``conditionals'', ``functions'', ``parallelism'' and ``events''. 
For example, a symbolic artefact, such as a block-based programming platform, may have many functionalities, such as sequences, repetitions, conditionals, etc. 
In contrast, the programming interface may have limited functionalities during a robotic activity, for example, it could only consent using operators (like moving forward) or events. 

\begin{ndef}[System resettability] 
	\normalfont  	~ \newline  
        The property of a system to be restored to its initial state, either through the direct intervention of the problem solver on the system or indirectly via the reversibility of actions within the system.
\end{ndef}
{Resettability can be ``direct'' when the problem solver can directly intervene on the system by manually returning the robot to its starting position and restoring the environment; or ``indirect'' when the problem solver can use a system-provided reset mechanism.
If neither option is available, the system is ``non-resettable'', for example, when the problem solver can move the robot back to the starting position only through an algorithm, but any alterations to the environment remain irreversible.}



\begin{ndef}[System observability] 
	\normalfont  	~ \newline  
 The property of a system that {allow} the problem solver to observe the effects of {the agent actions in the environment and their impact on its state}. 
\end{ndef}
Systems can be classified as ``totally observable'' if every action and their effects are visible, {e.g., if the problem solver and the robot are in the same room, and all changes to the system state are visible in real-time;} ``partially observable'' when only the aggregate effects of a set of actions are visible, {e.g., if the problem solver can enter the room only at the end of the task and observe the final state of the system, without seeing the actions that led to it}; or ``not observable'', if none of the agent's actions or their results are visible, {e.g., if the problem solver cannot enter the room and must infer the system state from other information, such as sensor data.}
{It is worth noting that, in the unplugged domain, problem solver and agent can be the same entity. When they overlap, the system is totally observable.}


\begin{ndef}[Task cardinality] 
	\normalfont  	~ \newline  
	The {relationship} between the number of given elements {and those} to be found to solve a task. 
\end{ndef}
{CTPs can present three types of cardinality:} ``one-to-one'', ``many-to-one'' or ``many-to-many''. 
{In a one-to-one task, each provided element corresponds directly to one element to be found, e.g., if a single initial and a final states are given, a single algorithm has to be found. 
In a many-to-one task, multiple given elements are intended to be resolved by a single solution element, e.g., if several initial states are provided, and the goal is to find a single algorithm that can transform each of these initial states into the same final state.
In a many-to-many task, both the provided and target elements are multiple, requiring the problem solver to find various solutions. 
For example, a task might provide multiple initial states and a single final state, and the solver would need to identify several algorithms, each capable of transforming one or more of the initial states into the specified final state.
This type of task can be traced back to multiple many-to-one tasks.}

\begin{ndef}[Task explicitness] 
\normalfont  	~ \newline 
{The level of detail in the presentation of the task's elements.} 
\end{ndef}
{In a CTP, the given elements can be ``explicit'' if they are directly provided and immediately usable in the problem-solving process, or ``implicit'' if they are expressed with constraints that require further interpretation to be understood.
For example, in a task where the problem solver must find the algorithm for a robot to turn on its lights after finding a ball, the ball's position can be given explicitly (e.g., coordinates) or implicitly (e.g., in the playground).}

\begin{ndef}[Task constraints] 
\normalfont  	~ \newline 
The limitations or specific requirements that {the task elements to be found must meet to consider the solution valid.}
\end{ndef}
{In a CTP, the elements to be found can be} ``unconstrained'' if they can be freely selected among all possible states and algorithms, with no limitations or specific requirements that need to be met to consider the solution valid; or ``constrained'' if they must belong to a restricted subset of states or algorithms.
{Referring to the same example presented to explain the task explicitness characteristic,} the algorithm to be found can be unconstrained if the robot can perform any action to find the ball ({e.g., moving randomly, using sensors, etc.}) or constrained if the programming platform limits the {robot's actions (e.g., restrict movement to specific directions, using only specific sensors).}

\begin{ndef}[Algorithm representation] 
\normalfont  	~ \newline 
The mean by which an algorithm is given.
\end{ndef}
An algorithm is considered ``manifest'' if directly expressed, while ``latent'' if not stated but should be inferred by the problem solver.
Manifest algorithms can be ``written'' if represented by an external {and persistently, like the code in a programming language}, or ``not written'' if communicated verbally or through other non-permanent means.

\subsection{{Catalogue of CT Competencies}}\label{sec:skills}

\begin{figure*}[!ht]
\centering
\includegraphics[width=.95\textwidth]{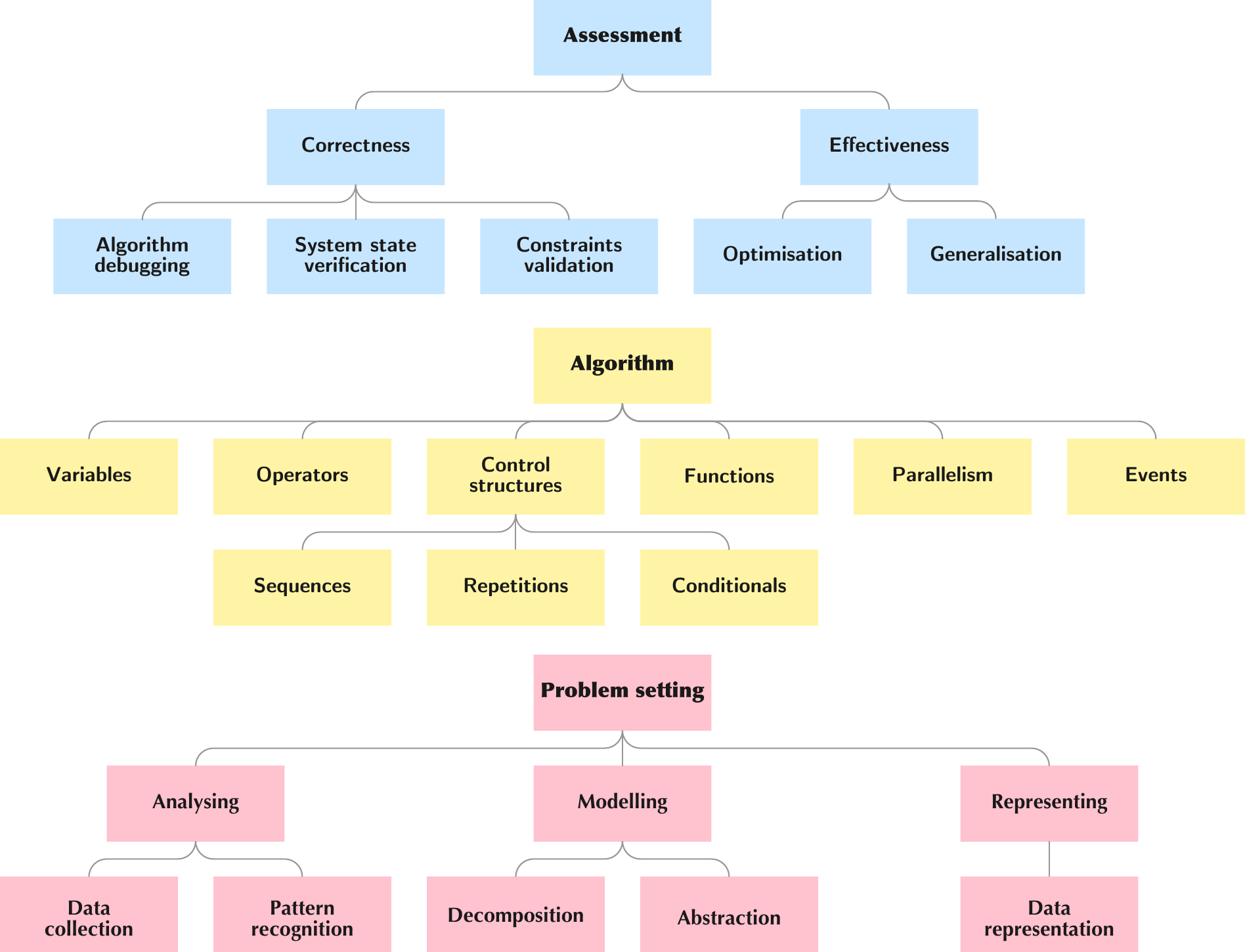}
\caption{\textbf{Visualisation of our taxonomy of CT competencies.}
	The overall structure is based on the CT-cube \citep{piatti_2022}. The sub-skills are derived from validated CT models \citep{brennan2012new,weintrop2016defining,shute2017demystifying}.
	The same colour scheme as in \Cref{fig:ctcube} is applied.
}
\label{fig:taxonomy}
\end{figure*}

{Alongside the definition of CTPs, their components, and characteristics, we have developed a catalogue of CT competencies, also referred to as skills, that are fundamental abilities students need to solve CTPs effectively.
We aimed to create a practical resource synthesising insights from existing frameworks and contribute to a more extensive and universal definition of CT.
}

{To ensure a thorough approach, we drew from multiple state-of-the-art competency models and frameworks.} 
Our selection was inspired by the literature reviews of \cite{tikva_mapping_2021} and \cite{bocconi2016developing,bocconi2022reviewing}, which provide a comprehensive overview of CT skills in compulsory education.
One significant framework that guided the development of our catalogue is that of \cite{brennan2012new}, which categorises CT skills into {computational concepts}, {practices}, and {perspectives}. 
{While this model is commonly referenced in literature, it primarily focuses on activities conducted in digital environments, such as programming and software development.
Although the CT aspects covered are essential, they do not encompass the broader range of CTPs we explore, including hands-on robotics and unplugged activities. To bridge this gap, we extended the framework to incorporate competencies applicable across these varied contexts,} such as elements from the STEM taxonomy proposed by \cite{weintrop2016defining}, which includes data practices, modelling and simulation practices, computational problem-solving, and systems thinking practices.
We also referenced the work of \cite{shute2017demystifying}, which {expands on the previous frameworks by offering a broader, more adaptable set of CT competencies with a focus on cognitive processes and applicability across diverse contexts.}

To {provide a comprehensive and organised framework for CT competencies,} we arranged our catalogue into a hierarchy of skills and sub-skills, illustrated in \Cref{fig:taxonomy}.
{This structure clarifies the relationships among competencies, making it easier to identify specific skills within broader categories, and supporting a more precise and targeted approach for educators and researches working with CT skill development and assessment.
The first layer of competencies is based on the activity dimension of the CT-cube} \citep{piatti_2022}, {introduced in} \Cref{sec:ctp}, {while the additional layers are based on the frameworks of} \cite{brennan2012new}, \cite{weintrop2016defining}, {and} \cite{shute2017demystifying}.

\begin{table}[!ht]
\renewcommand{\arraystretch}{1.6}

\centering
\begin{threeparttable}
\scriptsize
	\caption{\label{tab:h1} \textbf{Core skills definition.}
		} 
	\begin{tabular}{m{2.3cm}m{5.2cm}}
		\toprule
        \textbf{Competence}\tnote{*}& \textbf{Definition}\\ \midrule
		\textit{Problem setting}\tnote{a} & 
		Recognise, understand, reformulate or model a CTP and its components so that its solution can be computed.
		\\
		\textit{Algorithm}\tnote{b} & 
		Conceive and represent a set of agent's actions that should be executed by a human, artificial or virtual agent to solve the task.\\
		{\textit{Assessment}}\tnote{c} & 
		Evaluate the quality and validity of the solution in relation to the original task.\\
		\bottomrule
	\end{tabular}
	\begin{tablenotes}\scriptsize
            \item[*] The skills listed are based on the values of the activity dimension of the CT-cube framework \citep{piatti_2022}.
		\item[a] See \Cref{tab:h21} for ``problem setting'' sub-competencies. 
		\item[b] See \Cref{tab:h22} for ``algorithm'' sub-competencies. 
		\item[c] See \Cref{tab:h23} for ``assessment'' sub-competencies. 
	\end{tablenotes}
\end{threeparttable}
\end{table}

\Cref{tab:h1} summarises the competencies in the first level of the hierarchy while \Cref{tab:h21,tab:h22,tab:h23} provide a detailed breakdown of competencies in the lower levels of the hierarchy.
Each row in the tables represents a specific skill, with the parent skill separated from the lower-level skills by a dashed line and its lower-level competencies by dotted lines. 

\begin{table}[!htb]
\renewcommand{\arraystretch}{1.6}
\scriptsize
\centering
\begin{threeparttable}  
	\caption{\label{tab:h21} \textbf{Problem setting sub-skills definition.}
		}
\begin{tabular}{m{1.7cm}m{5.8cm}}
	\toprule
	\textbf{Competence}\tnote{*}& \textbf{Definition} \\ \midrule
	
	{\textit{Analysing}} 
& Collect, examine and interpret data about the system: environment descriptors and agent actions. 
\\ \hdashline[4pt/2pt]

{\textit{Data \newline collection}} 
& Gather details about the system. 
\\  \hdashline[1pt/2pt]

{\textit{Pattern \newline recognition}} 
& Identify similarities, trends, ideas and structures within the system. 
\\\hline
{\textit{Modelling}} 
& 
Restructure, clean and update knowledge about the system.
\\ \hdashline[4pt/2pt]

{\textit{Decomposition}} 
& Divide the original task into sub-tasks that are easier to be solved.
\\ \hdashline[1pt/2pt]

{\textit{Abstraction}} 
&
Simplify the original task, focus on key concepts and omit unimportant ones.
\\\hline
{\textit{Representing}} 
& Illustrate or communicate information about the system and the task. 
\\ 

\bottomrule
\end{tabular}
	\begin{tablenotes}\scriptsize
            \item[*] The skills listed are based on leading-edge competence models \citep{brennan2012new,shute2017demystifying,thalheim2000database,weintrop2016defining,wing2011research,bocconi2016developing,selby2013computational,angeli2016k,csizmadia2015computational,selby2014can,barr}.
	\end{tablenotes}
 
\end{threeparttable}
\end{table}
~
\begin{table}[!htb]
\renewcommand{\arraystretch}{1.6}
\scriptsize
\centering
\begin{threeparttable}
\caption{\label{tab:h22} \textbf{Algorithm sub-skills definition.}
}
\begin{tabular}{m{1.7cm}m{5.8cm}}

\toprule
\textbf{Competence}\tnote{*}& \textbf{Definition} \\ \midrule

{\textit{Variables}} 
& Entity that stores values about the system or intermediate data.
\\\hline
{\textit{Operators}} 
& Mathematical operators (e.g., addition ($+$), subtraction ($-$)), logical symbols (e.g., and (\&), or (|), not (!)) or for comparison (e.g., equal to (==), greater than (>), less than (<)), or even specific commands or actions (e.g., ``turn left'', ``go straight'').
\\ \hline
{\textit{Control \newline structures}} 
& Statements that define the agent actions flow's direction, such as sequential, repetitive, or conditional.
\\ \hdashline[4pt/2pt]

{\textit{Sequences}} 
& Linear succession of agent actions. 
\\ \hdashline[1pt/2pt]

{\textit{Repetitions}} 
& Iterative agent actions.
\\ \hdashline[1pt/2pt]

{\textit{Conditionals}} 
& Agent actions dependent on conditions. 
\\\hline
{\textit{Functions}} 
& Set of reusable agent actions which produce a result for a specific sub-task.
\\\hline
{\textit{Parallelism}} 
& Simultaneous agent actions.
\\\hline
{\textit{Events}} 
& Variations in the environment descriptors that trigger the execution of agent actions.\\ 


\bottomrule
\end{tabular}
	\begin{tablenotes}\scriptsize
            \item[*] The skills listed are based on leading-edge competence models \citep{brennan2012new,bocconi2016developing,bocconi2022reviewing,shute2017demystifying,rodriguez2020computational,cui2021interplay}.
	\end{tablenotes}

\end{threeparttable}
\end{table}

\begin{table}[!htb]
\renewcommand{\arraystretch}{1.6}
\scriptsize
\centering
\begin{threeparttable}
\caption{\label{tab:h23} \textbf{Assessment sub-skills definition.}
}
\begin{tabular}{m{1.7cm}m{5.8cm}}
\toprule
\textbf{Competence}\tnote{*}& \textbf{Definition}\\ \midrule

{\textit{Correctness}} & Assess whether the task solution is correct. \\ \hdashline[4pt/2pt]

{\textit{Algorithm \newline debugging}} 
&  
Evaluate whether the algorithm is correct, identifying errors and fixing bugs that prevent it from functioning correctly.  
\\ \hdashline[1pt/2pt]

{\textit{System states \newline verification}} 
& Evaluate whether the system is in the expected state, detecting and solving potential issues.  \\\hdashline[1pt/2pt]

{\textit{Constraints \newline validation}} 
& Evaluate whether the solution satisfies the constraints established for the system and the algorithm, looking for and correcting eventual problems. \\\hline

{\textit{Effectiveness}} & Assess how effective is the task solution. \\ \hdashline[4pt/2pt]

{\textit{Optimisations}} 
& Evaluate whether the solution meets the standards in a timely and resource-efficient manner, and eventually identify ways to optimise the performance.\\\hline

{\textit{Generalisation}} 
& 
Formulate the task solution in such a way that can be reused or applied to different situations.
\\
\bottomrule
\end{tabular}
	\begin{tablenotes}\scriptsize
            \item[*] The skills listed are based on leading-edge competence models \citep{brennan2012new,shute2017demystifying,weintrop2016defining,bocconi2016developing}.
	\end{tablenotes}
\end{threeparttable}
\end{table}

\subsection{{Framework for CTP Profiling}}\label{sec:link}

\begin{table*}[!ht]
\centering
\caption{\textbf{Comprehensive overview of the relationship between different CTP characteristics and CT competencies.}
The table shows the relationship between the characteristics of CTPs (columns) and CT competencies (rows). 
The CTP features considered include the tools' functionalities, the system's property, and the task trait.
{The same colour scheme as in} \Cref{fig:ctcube,fig:framework} {is applied}.
}
\label{tab:static-framework}
\includegraphics[width=\textwidth]{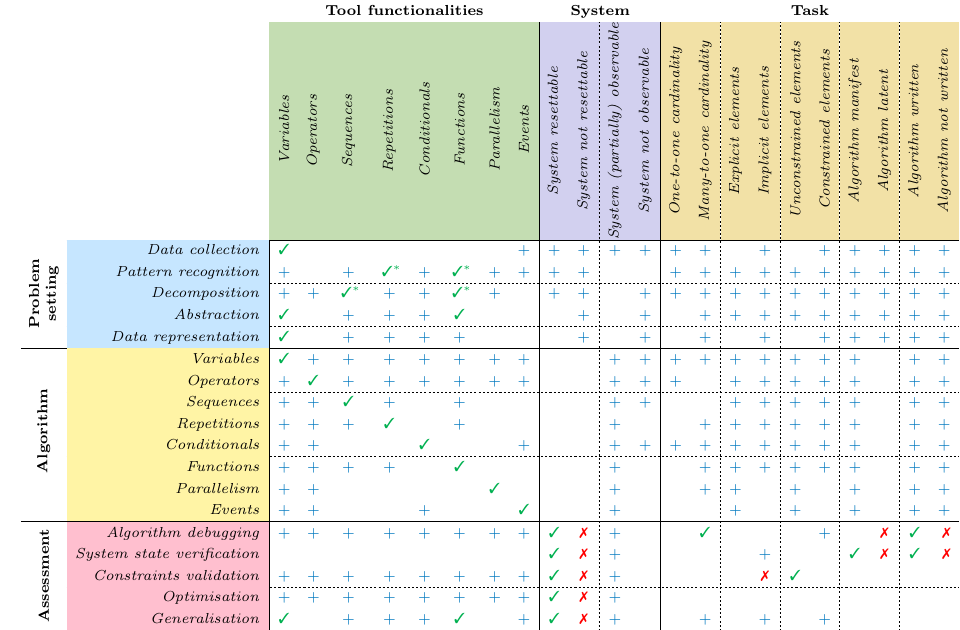}
\\\vspace{1em}
\begin{minipage}{\linewidth}
{{\color[HTML]{00B050}\ding{51}} indicates that the characteristics is required for the development of the competence.}

{{\color[HTML]{00B050}\ding{51}$^*$} indicates that at least one of several characteristics in a group is required for the development of the competence.
}

{{\color[HTML]{FF0000}\ding{55}} indicates that the characteristic prevent the development of the competence.}

{{\color[HTML]{0078BF}$+$} indicates that the characteristic can support the development of the competence.}

{Blank cells indicate that the characteristic is irrelevant for the development of the competence.}
\end{minipage}
\end{table*}

{Building on our earlier discussions of CTP components and characteristics, in this section, we introduce our framework for profiling CTPs.
We defined specific relationships between CTP characteristics and CT competencies, outlining, for each skill, the set of characteristics essential for their development and those that inhibit it. To develop a specific competence, all required characteristics must be present, and none may be inhibitory. 
}

{While identifying only the required characteristics and those to be absent reveals which competencies can technically be developed, it provides a limited perspective. 
For this reason, we decided to include also characteristics that can enhance and support skill development beyond basic requirements, strengthening the overall framework for competency development.
For example, a manifest written algorithm can significantly facilitate the development of algorithmic skills at different levels of abstraction, such as repetitions, by helping learners understand how loops work, recognise them and practice their application, ultimately leading to assimilation} \citep{bloom1956taxonomy,CTF}.

{All relationships between CTP characteristics and CT competencies are examined in detail in \textit{Supplementary information},} where we outline (i) how various CTP characteristics influence the development of CT competencies and (ii) which CT skills are more frequently developed and/or employed when solving CTPs with specific traits. 

\Cref{tab:static-framework} {illustrates our framework and serves as a template for analysing and designing CTPs by creating a profile of each specific CTP.
For analysis, this profile is built by identifying all the CTP characteristics. This approach allows us to pinpoint which competencies can be developed and assessed based on the presence of required characteristics and the absence of inhibitory ones.
For design, the process begins by determining the specific competencies to be targeted for development and/or assessment. This requires identifying the necessary characteristics to include while excluding any that may inhibit skill development. 
Additionally, supportive characteristics that are not essential but may enhance skill development can be selectively integrated, leading to a richer CTP. 
}

\section{Results and Experimental Findings}\label{sec:results}

{This section is organised into two main parts. 
The first demonstrates the framework's functionality in analysing CTPs, through which a taxonomy of CTPs across various domains has been developed, including unplugged activities, robotics, and virtual environments. 
The second part showcases the framework's application in the design of CTPs, highlighting how specific competencies can be effectively targeted and developed.
}

\subsection{A Taxonomy of CTPs}\label{sec:taxonomy_result1}

\begin{table*}[!ht]
    \renewcommand{\arraystretch}{1.6}
    \scriptsize
    \centering
    \caption{\textbf{Characteristics taxonomy across CTPs domains.}}
    \label{tab:charcateristics_taxonomy}
    \begin{tabular}{l|m{3.5cm}m{3cm}m{3cm}}
         & {\textbf{Unplugged}} & {\textbf{Robotic}} & {\textbf{Virtual}} \\
        \midrule
        
        \cellcolor[HTML]{C4DDB2}\textit{Tool functionalities} & Mostly variables, operators, sequences, repetitions, and functions; \newline Rarely conditionals and parallelism \newline Never events  & All & Mostly variables, operators, sequences, repetitions, conditionals, and functions; \newline Rarely parallelism and events \\\hline    
        \cellcolor[HTML]{D2D0EF}\textit{System resettability} & Resettable (43\%) & Resettable (40\%) & Resettable (100\%) \\

        \cellcolor[HTML]{D2D0EF}\textit{System observability} & Observable (57\%) & Observable (100\%) & Observable (100\%) \\\hline       
   
        \cellcolor[HTML]{F2E1A6}\textit{Task cardinality} & One-to-one (100\%) & One-to-one (100\%) & One-to-one (75\%); \newline Many-to-one (25\%) \\
     
        \cellcolor[HTML]{F2E1A6}\textit{Task explicitness} & Explicit (71\%); \newline Implicit (29\%) & Explicit (80\%); \newline Implicit (20\%) & Explicit (100\%) \\
    
        \cellcolor[HTML]{F2E1A6}\textit{Task constraints} &  Unconstrained (57\%); \newline Constrained (43\%) & Unconstrained (100\%) & Unconstrained (100\%) \\

        \cellcolor[HTML]{F2E1A6}\textit{Algorithm representation} & Manifest written (72\%); \newline Manifest non-written (14\%); \newline Latent (14\%) & Manifest written (100\%) & Manifest written (67\%); \newline Latent (33\%) \\
    \end{tabular}
\end{table*}

\begin{table}[!ht]
\renewcommand{\arraystretch}{1.6}
\addtolength{\tabcolsep}{-0.4em}
\scriptsize
\centering
\caption{\textbf{Competencies taxonomy across CTPs domains.}}
\label{tab:competencies_taxonomy}
\begin{tabular}{cl|ccc}
& & \textbf{Unplugged} & \textbf{Robotic} & \textbf{Virtual} \\
\midrule
\multirow{5}{*}{\rotatebox[origin=c]{90}{\parbox[c]{1.6cm}{\centering\textbf{Problem setting}}}} &\cellcolor[HTML]{C6E6FF}\textit{Data collection}           &     100\% &     100\% &     100\% \\
&\cellcolor[HTML]{C6E6FF}\textit{Pattern recognition       }&     100\% &     100\% &     100\% \\ \cdashline{2-5}[1pt/1pt]
&\cellcolor[HTML]{C6E6FF}\textit{Decomposition             }&     100\% &     100\% &     100\% \\
&\cellcolor[HTML]{C6E6FF}\textit{Abstraction               }&     100\% &     100\% &     100\% \\ \cdashline{2-5}[1pt/1pt]
&\cellcolor[HTML]{C6E6FF}\textit{Data representation       }&     100\% &     100\% &     100\% \\\hline
\multirow{8}{*}{\rotatebox[origin=c]{90}{\parbox[c]{1.6cm}{\centering\textbf{Algorithm}}}}&\cellcolor[HTML]{FFF4A5}\textit{Variables}                 &     100\% &     100\% &     100\% \\
&\cellcolor[HTML]{FFF4A5}\textit{Operators                 }&     100\% &     100\% &     100\% \\\cdashline{2-5}[1pt/1pt]
&\cellcolor[HTML]{FFF4A5}\textit{Sequences                 }&     100\% &     100\% &     100\% \\
&\cellcolor[HTML]{FFF4A5}\textit{Repetitions               }&     100\% &   \; 60\% &   \; 67\% \\
&\cellcolor[HTML]{FFF4A5}\textit{Conditionals              }&   \; 50\% &   \; 60\% &   \; 33\% \\\cdashline{2-5}[1pt/1pt]
&\cellcolor[HTML]{FFF4A5}\textit{Functions                 }&     100\% &     100\% &     100\% \\
&\cellcolor[HTML]{FFF4A5}\textit{Parallelism               }&   \; 17\% &   \; 40\% & \; \; 0\% \\
&\cellcolor[HTML]{FFF4A5}\textit{Events                    }& \; \; 0\% &   \; 80\% &   \; 33\% \\\hline
\multirow{5}{*}{\rotatebox[origin=c]{90}{\parbox[c]{1.6cm}{\centering\textbf{Assessment}}}}&\cellcolor[HTML]{FFBFCF}\textit{Algorithm debugging}       &   \; 50\% &   \; 40\% &   \; 67\% \\
&\cellcolor[HTML]{FFBFCF}\textit{System state verification }&   \; 33\% &   \; 40\% &   \; 67\% \\
&\cellcolor[HTML]{FFBFCF}\textit{Constraints validation    }&   \; 50\% & \; \; 0\% & \; \; 0\% \\\cdashline{2-5}[1pt/1pt]
&\cellcolor[HTML]{FFBFCF}\textit{Optimisation              }&   \; 50\% &   \; 40\% &     100\%  \\
&\cellcolor[HTML]{FFBFCF}\textit{Generalisation            }&   \; 50\% &   \; 40\% &     100\%  \\
\end{tabular}
\end{table}

To validate our method {and demonstrate how it can be applied to analyse CTPs}, we focused on three key domains: unplugged, robotic, and virtual activities.
{This choice was motivated by the desire to explore how different domains inherently shape the other characteristics of CTPs and their associated CT competencies.
While CTPs can be classified according to many characteristics, such as task type or algorithm representation, categorising by domain can provide a more holistic perspective on how the context and medium of activity influence the development of CT skills.
}

{To this end, we applied our framework to a selection of prototypical CTPs, recognised as representative in educational settings, by outlining their components, identifying their characteristics, and determining which competencies could be developed and which could not.
This selection is only illustrative of the framework's use, and it is not exhaustive.}

{This process led to the creation of a ``taxonomy'' that maps the relation between domain-specific characteristics and the CT competencies that can be developed within each domain. This systematic classification highlights how each domain offers unique opportunities and challenges that influence how learners engage with CTPs and develop CT skills.
}


\Cref{tab:charcateristics_taxonomy,tab:competencies_taxonomy} {give a broad picture of the key characteristics and CT competencies developed in each domain. 
In the following sections, we will explore each domain in detail, analysing the characteristics and competencies taxonomy.}
\nameref{appendix:appB} { provide a detailed analysis of each CTP used to define this taxonomy, including (i) a description of the CTP, its components, and characteristics, illustrated through a graphical template, (ii) an overview of the competencies that can be developed, as well as those that cannot, and (iii) the resulting profile of the CTP, presented in a table that illustrates the relationship between the CTP's characteristics and the CT competencies that are activated.
}

\subsubsection{{Unplugged CTPs}}
\label{sec:unplugged}
Unplugged CTPs refer to activities that do not require the use of computers or technology \citep{brackmann2017development, del2020computational}.
As per our framework, {what denotes this domain is the presence of a human agent} rather than a virtual or robotic one, {and a physical environment. Typically, they involve hands-on activities using physical objects like blocks, puzzles, or cards, or even non-digital tasks such as traditional closed-ended questions and paper-and-pencil tests.}

{To construct a taxonomy of characteristics and competencies for unplugged CTPs, we analysed four representative unplugged activities, focusing on core CT skills in accessible, non-digital formats, making them ideal for introductory learners.
The activities include:}
\begin{enumerate}[noitemsep,nolistsep]
    \item {\textit{Cross Array Task (CAT)}} (see \cref{sec:cat}){: designed to assess algorithmic thinking, this task requires students to communicate a reference pattern through verbal instructions or gestures} \citep{piatti_2022}.

    \item {\textit{Graph Paper Programming}} (see \cref{sec:GPP}){: designed to introduce core programming concepts, this offline programming task has two variants: one requires students to write instructions using symbols to reproduce a given pattern; the other requires students to recreate the pattern based on a given a set of instructions} \citep{org2015web,gpp}.

    \item {\textit{Triangular Peg Solitaire}} (see \cref{sec:TPS}){: designed to develop and test algorithm thinking, this strategy-based task has two variants: one requires students to eliminate pegs on a board by moving them under specific rules and constraints} \citep{pegsolitaire,berlekamp2004winning}; {the other requires students to solve the game by documenting the strategy with paper and pencil} \citep{barbero2018analysing,bell2008solving,bell_unlimited}.

    \item {\textit{Computational Thinking test (CTt)}} (see \cref{sec:CTt}){: designed to assess algorithmic CT skills, this task requires students to answer multiple-choice questions within familiar programming contexts, prompting to sequence commands, complete partial code, and debug errors} \citep{roman2015computational,roman2017complementary,roman2017cognitive,roman2018detected}.
\end{enumerate}


\paragraph{Taxonomy of characteristics for unplugged CTPs}

{Unplugged CTPs typically involve simple \textit{tool functionalities} such as variables, operators, sequences, repetitions, and functions. Although conditionals are present in some activities, parallelism is infrequent, and events are not featured in this domain.
These characteristics primarily support the development of problem setting competencies, such as pattern recognition, decomposition, and abstraction, while also introducing core algorithmic concepts, allowing students to grasp these fundamental skills through hands-on and tangible experiences before advancing to more abstract, technology-based tasks} \citep{bell_csu}.

{The \textit{system resettability} of unplugged CTPs is generally limited.
While the lack of this characteristic} can encourage students to plan and think more carefully about their actions, fostering a more thoughtful approach to problem setting, developing competencies like abstraction and decomposition, it may also make it more difficult for students to experiment with different approaches, hindering the development of competencies related to assessment and evaluation, such as debugging.


{The level of \textit{system observability} in unplugged CTPs varies significantly.}
Activities with higher observability allow learners to easily identify and apply a broad range of CT competencies, particularly in algorithmic and assessment skills.
Conversely, activities with lower observability tend to be more abstract or open-ended, making it harder to formulate direct solutions and encouraging learners to engage in exploratory, creative thinking, which can support the development of problem setting competencies.
{Importantly, while this characteristic does not activate any competencies directly, it shapes how learners engage with and apply different CT skills.}


{The \textit{task cardinality} is typically one-to-one, meaning each given element (e.g., initial state, final state) corresponds to exactly one element to be found (e.g., algorithm).
This direct correspondence ensures clarity and focus in the problem-solving process, supporting the targeted development of specific skills. In contrast, many-to-one tasks promote advanced skills such as generalisation, as they require applying the same principle across multiple scenarios.
It's important to note that one-to-one tasks are not confined to the unplugged domain; they are common across all beginner-level CTPs in education, as they are simpler and can be broken down into manageable steps, allowing for a more structured approach to problem solving. }

{\textit{Task explicitness} is common in unplugged CTPs, where clearly defined task elements make the task easier to understand, particularly for novice learners, and help develop essential skills such as problem setting and algorithmic thinking.
For this reason, this characteristic is not confined to the unplugged domain but is prevalent in all CTPs for beginner education.}

{Unplugged CTPs feature both \textit{constrained} and \textit{unconstrained tasks}. 
Constrained tasks, which impose strict rules or limits, are more challenging as they require students to follow precise instructions and validate constraints. 
Instead, unconstrained tasks allow for greater freedom, encouraging students to explore diverse solutions and fostering creativity.
Both types of tasks are valuable for developing various CT skills at various levels, supporting foundational skill-building in problem setting and algorithmic concepts.}

{Finally, the \textit{algorithm representation} in unplugged activities is predominantly manifest, with learners actively representing algorithms, often through visual or even written instructions. 
This approach supports the development of assessment skills, such as algorithm debugging, by allowing learners to recognise and understand errors, though effective correction may require a system reset. Manifest representations also strengthen problem setting skills and reinforce algorithmic thinking.
Conversely, latent algorithm representations are less common but valuable for cultivating problem setting skills. These tasks require learners to process the algorithm's flow and structure mentally, fostering a deeper understanding and encouraging them to identify patterns and make abstract connections.}

\paragraph{Taxonomy of competencies for unplugged CTPs}

{Unplugged CTPs are highly effective for developing foundational CT competencies.
They offer an engaging, hands-on approach that grounds CT learning in accessible, non-digital experiences, making them particularly well-suited for introductory learners. 
These activities prioritise the development of problem setting and foundational algorithmic skills while selectively introducing more advanced assessment competencies.
 
Specifically, \textit{problem setting} skills are universally cultivated across unplugged CTPs.
In terms of \textit{algorithmic} competencies, variables, operators, simple control structures like sequences and repetitions, and functions are consistently developed. 
However, more complex algorithmic concepts, like conditionals and parallelism, are less frequently tackled, and events are generally absent in unplugged tasks.
Lastly, \textit{assessment} skills, including competencies related to task correctness, such as debugging, verifying system states, and validating constraints, as well as those focused on the effectiveness of solutions, like optimisation and generalisation, can be only moderately acquired in unplugged CTPs.
}


\subsubsection{{Robotics CTPs}}
\label{sec:robotics}

{Robotics CTPs, distinguished in} educational robotics and physical computing activities, {are denoted but the presence of a robotic agent interacting with a physical environment.}
The physical robotic hardware is equipped with controllers, sensors, and actuators. {It can be programmed through a dedicated programming platform that allows users to define behaviours in response to environmental inputs} \citep{bravo2017review}.
{Various commercially available platforms support programming robots using formal textual programming languages, like Python} \citep{noone2018visual}, or symbolic visual programming languages like Blockly and Scratch \citep{shin2014visual}. 
Some platforms also allow for embodied physical interactions, {allowing users to program the robot} through touch buttons or tangible symbols that are scanned and executed \citep{bers2010tangible, mussati2019tangible}.

{To construct a taxonomy of characteristics and competencies for robotic CTPs, we analysed three educational robotics activities and physical computing activity,} each distinguished by different types of agents, such as the Thymio II \citep{riedo2013thymio,shin2014visual}, the Ozobot \citep{Bryndova}, and the Micro:bit \citep{ball2016microsoft,microbit2016web}. 
{The activities include:}
\begin{enumerate}[noitemsep,nolistsep]
    \item {\textit{Thymio Lawnmower Mission}} (see \cref{sec:tlm}){: designed to promote the development of CT skills in general, this task requires students to program the Thymio II robot to drive around a lawn area, covering as much of the area as possible, avoiding fences} \citep{Chevalier2020}.

    \item {\textit{Remote Rescue with Thymio II ((R2T2)}} (see \cref{sec:r2t2}){: designed to introduce robotic programming through a collaborative approach, this task requires teams of students to program 16 Thymio II robots to move around a simulated damaged power Mars station to restart its main generator} \citep{mondada2016r2t2}.

    \item {\textit{Ozobot Maze}} (see \cref{sec:ozobot}){: designed to introduce coding concepts, this screen-less robotic task requires students to instruct the Ozobot to cross a maze, avoiding obstacles, to reach a specific room} \citep{Bryndova}.

    \item {\textit{Mini-golf challenge}} (see \cref{sec:minigolf}){: designed to practice with CT skills, this creative task requires a group of students to define the behaviour of mini-golf lane movable obstacles, sounds, and lights by programming the BBC micro:bit} \citep{assaf2011embedit,assaf2021imake}.
\end{enumerate}

\paragraph{Taxonomy of characteristics for robotic CTPs}
Robotics CTPs typically involve a wide range of \textit{tool functionalities} that students can access, from those related to basic programming concepts, such as variables and sequences, to more advanced ones like loops, parallelism, and events.
{This abundance of functionalities enables the development of problem setting and algorithmic competencies.} 

They \textit{system resettability} in robotics activities {is very common as the} problem solver often can directly intervene on the system to restart the task, typically through a reset function included in the programming platforms, allowing students to test and refine their designs through iterative experimentation. 
{While this characteristic can be beneficial, enabling students to correct errors and develop assessment skills like debugging, it also has downsides.
In some cases, students may rely on repeated trials without fully understanding the task, using a trial-and-error approach rather than deliberate problem-solving.
To counter this, some activities are beginning to restrict system resets purposefully to minimise reliance on trial and error, encouraging students instead to engage in planning and developing problem setting skills from the start.}

{In the robotic CTPs we analysed, \textit{system observability} is a prominent characteristic.} It allows students to monitor the robot's behaviour and analyse the outcomes of their algorithms, which is crucial for the development of core algorithmic skills as well as assessment skills, like debugging. 
When paired with system resettability, observability can sometimes lead students into an endless cycle of trial and error.
{To address this, some robotics activities introduce} a physical separation between the problem solver and the system,{for instance, when the robot is programmed remotely and provides only limited, delayed, or asynchronous visual feedback. In these cases, observability becomes partial,} encouraging students to go beyond trial-and-error tactics. This setup fosters problem setting skills alongside algorithmic ones, {leading to a more intentional and strategic approach to problem-solving.}

{Similarly to the discourse made for unplugged CTPs,} the \textit{task cardinality} in robotic activities is one-to-one. 
{This structure simplifies the problem-solving process, helping students focus on foundational learning and develop core competencies before advancing to more complex tasks.}

{Regarding \textit{task explicitness}, robotic CTPs, particularly educational robotics activities, typically involve tasks with} explicit instructions and well-defined elements, {which make them more accessible, especially for novice learners.} In these cases, students are often more inclined to dive directly into programming the agent's behaviour, {fostering the development of algorithmic thinking.}
On the other hand, physical computing activities usually feature fewer explicit elements, requiring students to engage in more critical thinking and problem-solving to navigate the task, {ultimately enhancing their problem setting skills.}


Robotics CTPs are typically \textit{unconstrained tasks}, providing students with more freedom and opportunities to experiment with various approaches to algorithm design without necessarily undergoing a rigorous problem setting phase. Novice learners may benefit from this feature, as it can encourage them to be more creative and exploratory in their problem-solving process. However, it may also limit the development of more advanced competencies. 

Finally, the \textit{algorithm representation} in robotics activities is typically manifest and written, {because these tasks often involve programming languages, either textual (like Python) or visual (like Scratch or Blockly).} 
{This form of representation helps develop assessment skills} and also fosters more logical thinking and problem-solving skills, {strengthening problem setting competencies and algorithmic thinking.}

\paragraph{Taxonomy of competencies for robotic CTPs}

{Robotics CTPs offer an excellent compromise, balancing the development of a wide range of CT competencies.
By engaging students in hands-on tasks that encompass \textit{problem setting}, \textit{algorithmic} thinking, and \textit{assessment}, robotics activities promote a well-rounded skill set that supports both foundational and advanced CT skills, especially in the algorithmic competencies such as parallelism and events.
Moreover, to prevent an over-reliance on trial-and-error, strategies are often implemented to limit debugging opportunities, encouraging students to focus more on problem setting and algorithmic thinking. 
This balanced approach ensures that students not only develop technical abilities but also enhance their critical thinking and problem-solving capabilities, making robotics an effective tool for fostering comprehensive CT.
}
\subsubsection{{Virtual CTPs}}
\label{sec:virtual}
{Virtual CTPs are distinguished from the other domains for the presence of a virtual agent and environment.
Virtual CTPs are distinct from other domains due to the presence of a virtual agent and environment. 
These activities typically involve a virtual interface, often paired with a comprehensive programming platform that enables users to program the virtual agent's behaviour using various programming languages, including both textual and visual programming} \citep{Rijo2022,noone2018visual,shin2014visual}.
{Alternatively, non-programming methods, like puzzles and virtual games, allow to interact with the agent directly, for example, by clicking and using drag-and-drop functionalities} \citep{Wang2022,tsarava2017training}.
{Moreover, virtual activities often come equipped with debugging tools, allowing users to identify and correct errors in their code, further enhancing their problem-solving skills}. 
{This combination of features offers a dynamic and flexible learning environment for developing CT skills.
}



{To construct a taxonomy of characteristics and competencies for virtual CTPs, we analysed three activities, including:}
\begin{enumerate}[noitemsep,nolistsep]
    \item {\textit{Classic maze}} (see \cref{sec:classicmaze}){: designed to learn the foundational algorithmic concepts and simple debugging techniques, this task requires students to program some characters to move through a maze to accomplish some tasks} \citep{classicmaze,roman2018detected,AB,PZ}.

    \item {\textit{Store the Marbles}} (see \cref{sec:RLB}){: designed to introduce the basics of programming, this task requires students to program a virtual robot to drop marbles in some holes} \citep{RLB,FIOI}.

    \item {\textit{Zoombinis Allergic Cliffs Puzzle}} (see \cref{sec:zoombinis}){: designed to develop problem setting and algorithm thinking, this game-based task requires students to guide little blue creatures through different puzzles to escape imprisonment} \citep{zoombinis2021,zoombinis2021web}.
\end{enumerate}

\paragraph{Taxonomy of characteristics for virtual CTPs}
{Virtual CTPs, often equipped with programming platforms, typically provide many \textit{tool functionalities} that facilitate the development of a wide range of algorithmic skills.
However, concepts such as parallelism and events are often overlooked, possibly due to the simplicity or linear nature of many virtual tasks, which prioritise sequential problem-solving.
That said, given the availability of programming interfaces, more advanced functionalities can be easily incorporated into these tasks. The absence of the concepts above reflects our specific choice of tasks rather than a limitation of the domain itself.
}


Virtual activities commonly feature \textit{resettable systems}, providing the convenience of a reset function that allows learners to restart the activity easily. This characteristic makes these CTPs particularly useful for developing algorithmic thinking and performing assessment tasks, {as students can quickly test different solutions and refine their approaches.}
{However, this feature also presents a potential downside: it may encourage a trial-and-error approach. While the ability to reset and test solutions repeatedly is beneficial for learning through experimentation, over-reliance on this method can hinder the development of problem setting skills. 
Learners may focus more on fixing errors rather than engaging in thoughtful problem analysis and strategic planning, which can limit their overall learning experience. Therefore, while the reset function is advantageous, it is important to use it in moderation to encourage more deliberate and systematic problem-solving.
}

{
\textit{System observability} is a key feature of many virtual activities, allowing learners to monitor the behaviour of the virtual agent and the outcomes of their programming or procedures. 
This characteristic is crucial for developing algorithmic thinking, as students can analyse how their code affects the agent's actions in real-time, fostering a deeper understanding of cause and effect and also supporting assessment skills.
However, when the feedback is often instantaneous and visible, students may become accustomed to quick fixes and immediate corrections, leading once again to an overemphasis on debugging rather than cultivating problem solving skills. 
To counterbalance this, it is important to encourage learners to focus not just on debugging but also on understanding the underlying problem and developing a structured approach to solving it.
}


{\textit{Task cardinality} in virtual activities often differs from the typical one-to-one correspondence seen in other CTP domains, with some tasks being of the many-to-one type.
This is likely due to the inherent flexibility and complexity of virtual environments, which allow learners to apply the same programming principles in different contexts, tackling problems from various angles and experimenting with different approaches.
This characteristic promotes the development of advanced assessment skills, such as generalisation, as well as problem setting competencies, like pattern recognition, abstraction, and decomposition. }


Virtual CTPs are typically \textit{explicit tasks}.
{While this characteristic supports the prompt acquisition of algorithmic thinking,} it can limit their critical thinking and hinder the development of some problem setting skills. 

{As for robotic activities, virtual CTPs are \textit{unconstrained tasks}, where learners have significant freedom in how they approach and solve the problem.
While this flexibility encourages creativity and can help develop a range of algorithmic concepts, it may also limit the development of certain problem setting skills, as the lack of constraints makes it harder for learners to define problems clearly and systematically.
Additionally, the freedom in virtual tasks may reduce the need for structured assessment, potentially hindering the development of precise evaluation and debugging skills.
}



{The \textit{algorithm representation} in virtual activities depends on the artefactual environment. For instance, coding tasks present manifest and written algorithms,} providing opportunities to develop algorithmic thinking skills and practice debugging. 
In contrast, in virtual games, algorithms are latent, as the problem solver directly interacts with the environment without using external tools, potentially limiting the development of algorithmic skills and debugging abilities while promoting advanced problem setting skills, such as pattern recognition and abstraction,

\paragraph{Taxonomy of competencies for virtual CTPs}

{Virtual CTPs support the development of a broad range of competencies across different levels.}
They foster key \textit{problem setting} skills, such as problem decomposition, pattern recognition, and abstraction, especially through virtual games.  
Coding tasks within these activities emphasise \textit{algorithmic} skills, equipping learners with tools and functionalities to experiment with programming concepts and explore their applications.
{\textit{Assessment} skills are also well supported in virtual CTPs, where learners engage in debugging to ensure task correctness and practice optimisation and generalisation to enhance task effectiveness.}

\subsubsection{{Summary of findings}}

{This taxonomy of CTPs helps highlight how different domains are suited to developing specific CT skills.}

{Regarding problem setting skills, unplugged CTPs are particularly effective as their hands-on, tangible nature allows learners to engage with problems physically, promoting critical thinking and problem-solving.
This aligns with the literature, which emphasizes the effectiveness of unplugged activities in fostering critical thinking by providing a tangible, accessible way to interact with complex problems, something abstract or digital tasks may not offer} \citep{Relkin2020,Relkin2021,brackmann2017development,Wohl2015}.
{Robotics and physical computing activities are also suitable for developing problem setting skills, especially when constraints are applied to the task, such as blocking the programming interface.
This approach limits trial-and-error methods, encouraging deeper problem-solving strategies and promoting the development of CT skills in general} \cite{Chevalier2020}.
{Virtual games can also trigger problem setting abilities thanks to the immersive environments where learners must set and solve problems within a given context.
This stands in contrast to coding activities, which tend to be more effective in developing algorithmic and assessment skills. 
This aligns with studies suggesting that virtual games, particularly puzzle-based ones, promote pattern recognition, abstraction, and decomposition in dynamic context-rich scenarios} \citep{shute2017demystifying,Varghese2023,Lee2014}.

{When considering algorithmic skills, robotics and virtual CTPs emerge as highly effective in building both foundational and more advanced algorithmic capabilities.}
{These domains promote deeper algorithmic thinking} \citep{atmatzidou2016advancing,Saxena2019,saritepeci2017analyzing}, {while unplugged activities, though effective for introducing basic concepts and fostering problem setting and reasoning, are less effective in developing higher-order competencies like algorithm design} \citep{Relkin2021,Bell2018,Cortina2015,Lu2009,Rodriguez2016}.

{For assessment skills, virtual CTPs prove to be the most effective, providing immediate feedback that enables learners to test and refine their solutions} \citep{Ko2004,Liu2017,McCauley2008}.
{Robotics activities can also support assessment skills, but their effectiveness depends on imposed constraints, such as blocking commands or limiting system resets} \citep{Chevalier2020}.
{In contrast, unplugged activities lack real-time feedback and often don't allow for quick adjustments, limiting their ability to support iterative improvements} \citep{Ahn2021}.

\subsection{{Designing the CAT}}\label{sec:design_result2}

\begin{figure*}[ht]
  \centering
  \begin{minipage}{0.45\textwidth}
    \includegraphics[width=\textwidth]{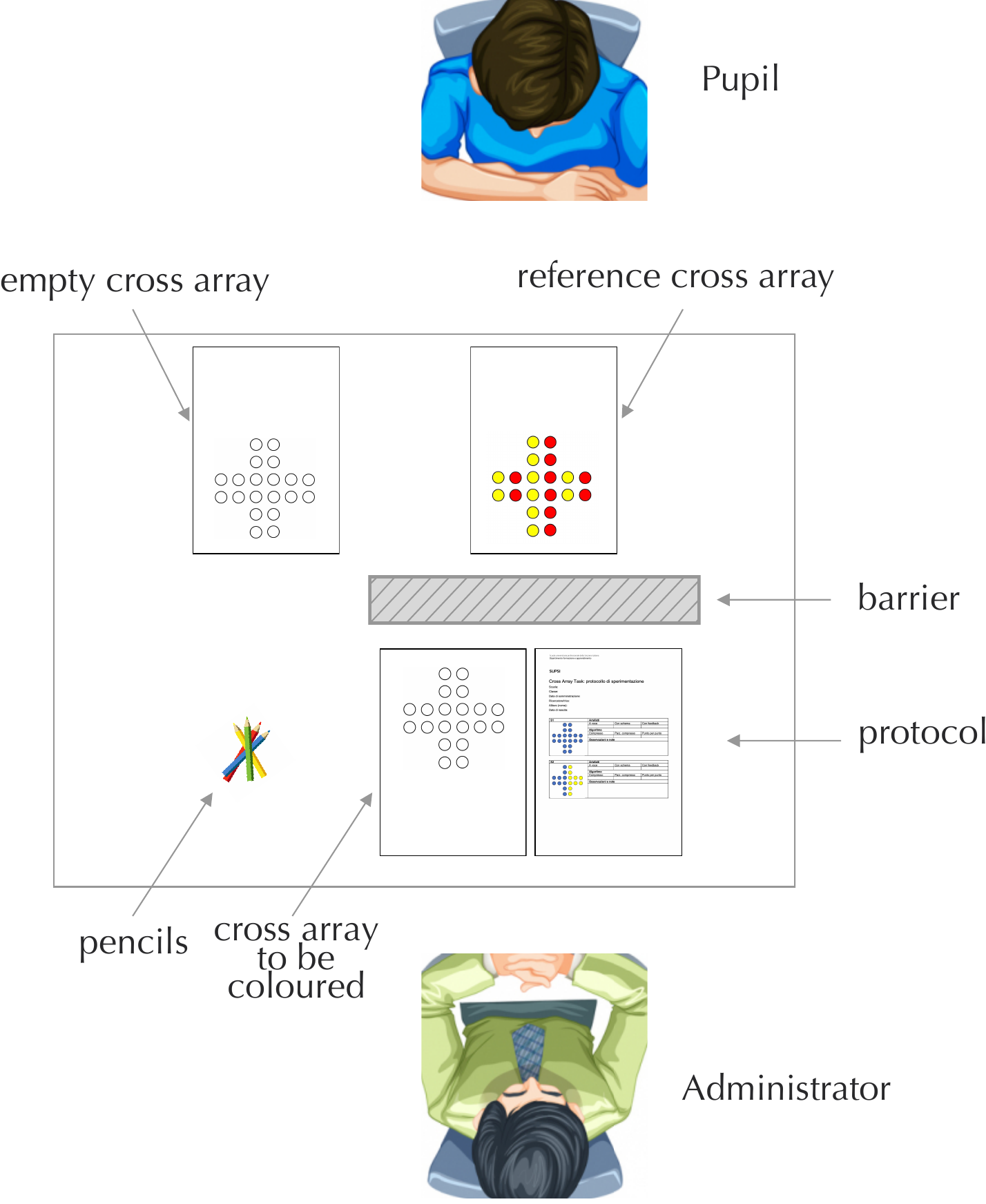} \\[\abovecaptionskip]
    \small \textbf{The unplugged {CAT}.}
         The problem solver, a student, is tasked with instructing the agent, typically a researcher, to reproduce a reference schema.
         The instructions can be communicated orally or through gestures on a support schema.
         A physical, removable barrier between the problem solver and the agent prevents visual cues between them and heightens the challenge.
         The agent interprets and records the student's instructions on a protocol and replicates the colouring pattern on a blank cross array.
  \end{minipage}
  \hfill
  \begin{minipage}{0.45\textwidth}
    \includegraphics[width=\textwidth]{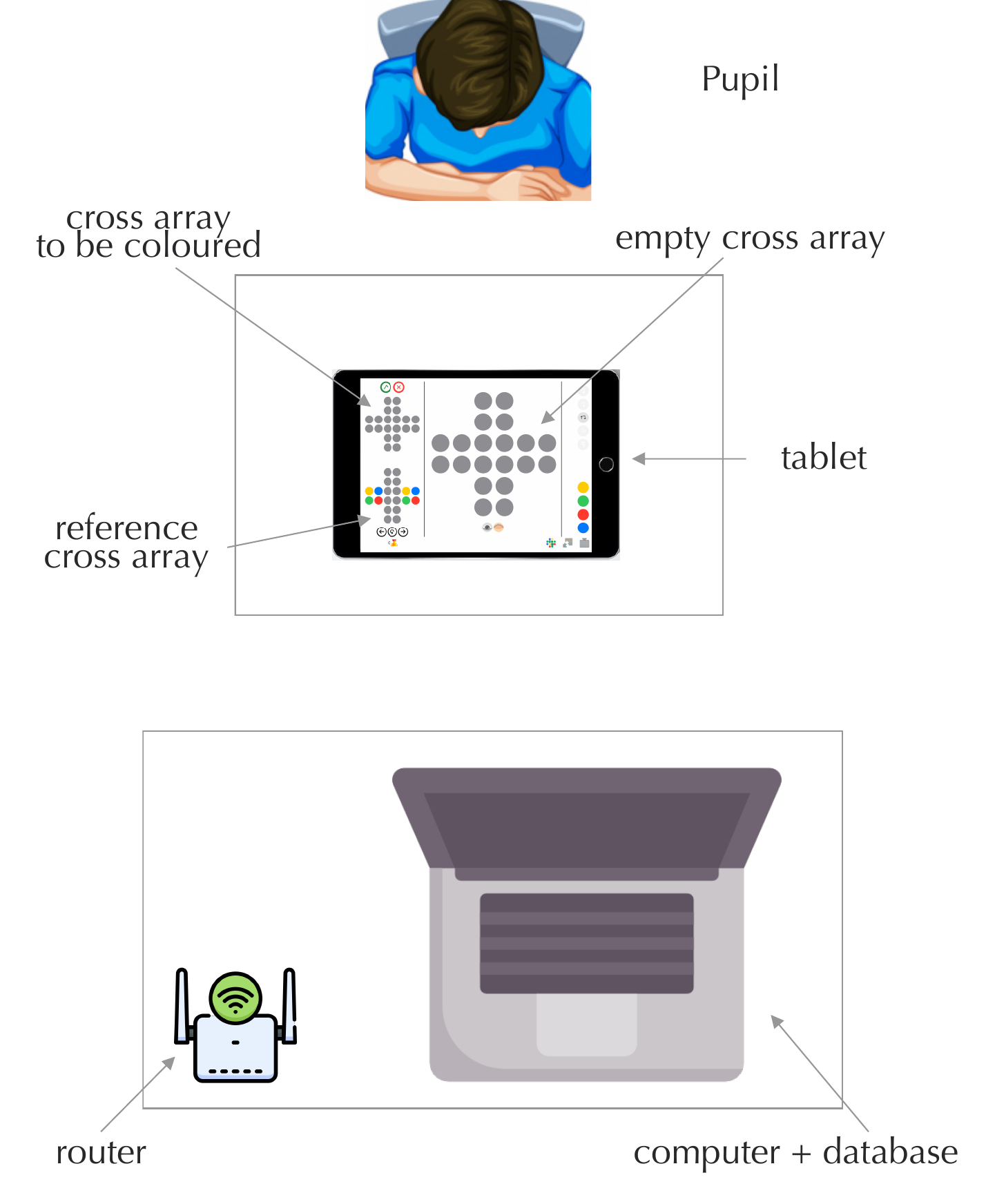}   \\[\abovecaptionskip]
    \small \textbf{The virtual {CAT}.}
         The problem solver, a student, is tasked to reproduce a reference schema using a gesture-based interface or a visual programming interface, that involves combining visual blocks to compose a set of instructions.
         The task's default setting limits the problem solver's ability to see the outcome of their actions due to a visual cue prevention feature that can be easily turned on or off.
         The system automatically logs all actions and algorithms.
  \end{minipage}
  \caption{\textbf{Comparison of the CAT setting between the unplugged and virtual CAT.} Adapted from \cite{piatti_2022}.}
    \label{fig:CATsettings}
\end{figure*}

{To validate our framework and illustrate its application in designing CTPs, we focus on the Cross Array Task (CAT), one of the activities analysed in} \cref{sec:taxonomy_result1} {to define and refine our taxonomy of CTPs. 
Originally developed to assess the development of algorithmic thinking in K-12 pupils, the CAT has served as an exemplary case for demonstrating the framework's analysis capabilities.
The CAT is a structured problem-solving activity where young learners act as problem solvers, guiding an agent (typically a researcher or educator) to replicate a coloured reference pattern on a cross-shaped grid, referred to as the cross array. 
To complete the task, the learner provides a sequence of instructions, either verbally or through gestures, explaining how the agent should recreate the pattern using coloured elements on a blank grid. 
To heighten the challenge, a removable barrier between the learner and the agent blocks visual cues, requiring clear and structured communication. 
The agent records each step of the learner's instructions on a protocol, allowing for a systematic assessment of algorithmic thinking skills.
For a detailed analysis of this activity's components, characteristics and competencies refer to} \cref{sec:cat}. 

{This CTP has proven effective in fostering essential problem setting skills, such as pattern recognition and decomposition, as well as foundational algorithmic competencies.
However, despite its effectiveness, the unplugged CAT also has inherent limitations. 
Firstly, its assessment relies on manual human observation, which is both time-intensive and susceptible to observer bias, making it difficult to scale as each session requires a facilitator to monitor and evaluate each learner's progress. 
Secondly, the unplugged format lacks immediate feedback capabilities, which limits learners' ability to learn from their mistakes in real time. 
Finally, it does not offer features such as reset functionality, which is crucial for debugging tasks; without this option, learners cannot reattempt tasks systematically, which hinders their ability to practice error correction.
}

{Automating the CAT by shifting its domain from unplugged to virtual offers a solution to these limitations. 
A digital format can streamline administration and assessment processes, enabling real-time feedback and integrating new features like reset and debugging options. 
The settings of the activities in the two domains are illustrated in} \Cref{fig:CATsettings}.

\begin{figure*}[!ht]
    \centering
    \includegraphics[width=\textwidth]{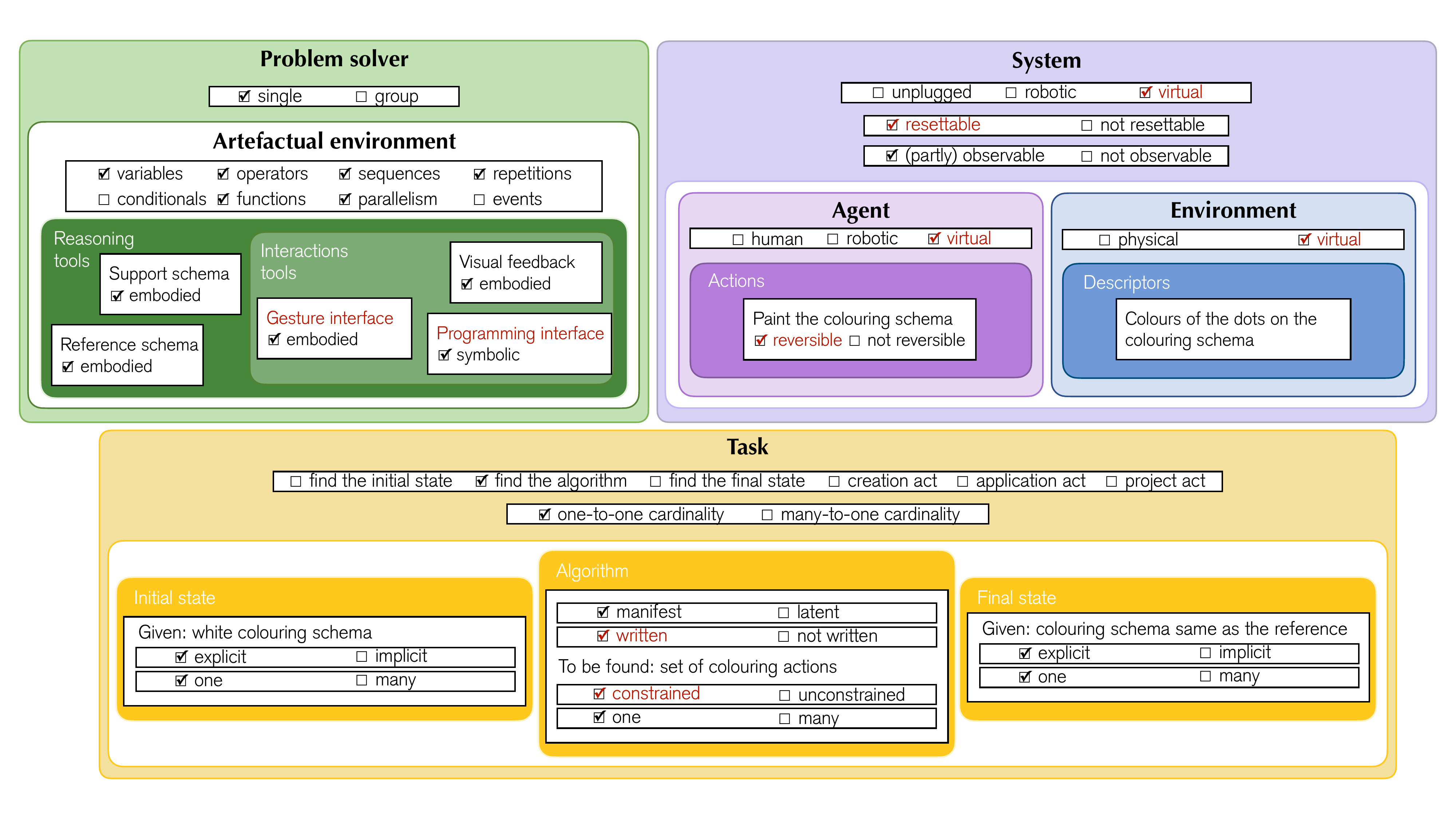}
    \caption{\textbf{Virtual CAT components and characteristics.}}
    \label{fig:virtualCAT-features}
\end{figure*}

{In the redesigned virtual CAT, our primary goal is to retain the original learning objectives of the unplugged CAT while enhancing the task's potential to evaluate additional skills, particularly assessment-related competencies like debugging. 
Debugging is a crucial skill in CT education, as it teaches learners to systematically verify the correctness of their solutions and make improvements, reinforcing a critical problem-solving mindset.
Thus, the virtual CAT design should maintain the original learning objectives while also introducing structured opportunities for debugging.}

{To achieve these expanded learning objectives, we revisited the core characteristics of the unplugged CAT and identified necessary changes to support skill development in a digital setting. 
While the functionalities of the task remain largely unchanged, certain system characteristics required modification. 
}

{First, in the unplugged version, for example, the CAT lacks resettable components, which limits opportunities for debugging practice. In the virtual format, we introduced a resettable interface that enables learners to test, revise, and correct their algorithm steps, activating the debugging process in real-time. 
Additionally, to support assessment skills, we transitioned from the implicit, hands-on interactions of the unplugged format, where learners communicate through gestures and verbal instructions, to an external persistent algorithm representation, i.e., a visual block-based programming interface. 
This digital adaptation allows learners to construct structured algorithms that remain visible for review, fostering a deeper understanding of algorithm design and validation. With this persistent visual representation, learners can systematically revisit their thought processes, identify errors, and refine their instructions, strengthening both problem setting and algorithmic skills. 
}

\begin{figure*}[!ht]
    \centering
    \includegraphics[width=\textwidth]{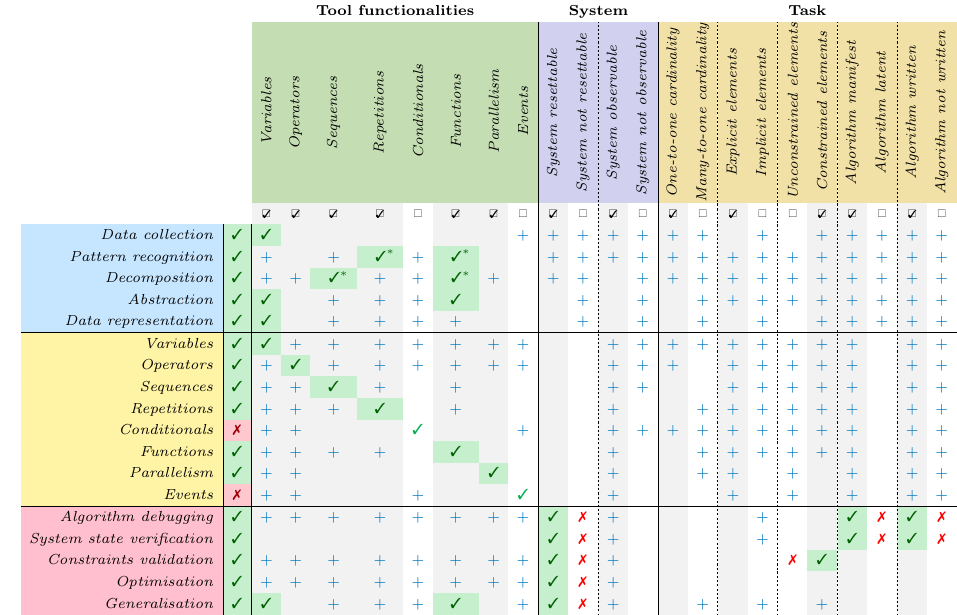}
    \caption{\textbf{Virtual CAT profile.}}
    \label{fig:virtualCAT-mapping}
\end{figure*}
{As a result of implementing new interaction methods, the virtual CAT introduces an artefactual environment with structured, predefined tools that inherently constrain the task. In the unplugged CAT, the problem solver had full freedom to use available tools creatively, employing any gestures or verbal instructions to generate an algorithm. 
However, when designing the virtual CAT as an on-tablet CTP, we introduced two specific interfaces: a gesture interface and a block-based programming platform. The gesture interface replicates common commands observed in the unplugged CAT but requires an interpreter to process commands, thus limiting the range of possible gestures. 
Similarly, the block-based programming interface allows learners to build algorithms with a predefined set of blocks, constraining the range of actions available.
}
{While this change may reduce the freedom and creativity of the problem solver, potentially limiting the variety of algorithmic concepts they might explore, it strengthens the task's problem setting phase. 
By operating within a structured set of tools, learners are encouraged to engage deeply with constraints, enhancing their skills in problem definition and constraint validation. 
}

{Components and characteristics of the virtual CAT are illustrated through the graphical template in} \Cref{fig:virtualCAT-features}, {while the profile, outlining the relationship between its characteristics and competencies, is illustrated in} \cref{fig:virtualCAT-mapping}.

{Finally,} \Cref{tab:charcateristics_comparison} {summarises the key differences between the characteristics of the unplugged and virtual CAT, including aspects such as system resettability, algorithm representation, and task constraints.}
\begin{table}[!ht]
    \renewcommand{\arraystretch}{1.6}
    \scriptsize
    \centering
    \caption{\textbf{Comparison of CAT variants characteristics.}}
    \label{tab:charcateristics_comparison}
    \begin{tabular}{m{2cm}|cc}
         & {\textbf{Unplugged CAT}} & {\textbf{Virtual CAT}} \\
        \midrule
        
        \cellcolor[HTML]{C4DDB2}\textit{Tool \newline functionalities} & \multicolumn{2}{>{\centering\arraybackslash}m{5cm}}{Variables, operators, sequences, repetitions, functions and parallelism}\\\hline        \cellcolor[HTML]{D2D0EF}\textit{System \newline resettability} & Not resettable & Resettable \\
        \cellcolor[HTML]{D2D0EF}\textit{System \newline observability} & \multicolumn{2}{c}{Partially observable}\\\hline
        \cellcolor[HTML]{F2E1A6}\textit{Task \newline cardinality} & \multicolumn{2}{c}{One-to-one} \\     
        \cellcolor[HTML]{F2E1A6}\textit{Task \newline explicitness} & \multicolumn{2}{c}{Explicit}\\    
        \cellcolor[HTML]{F2E1A6}\textit{Task \newline constraints} &  Unconstrained & Constrained \\
        \cellcolor[HTML]{F2E1A6}\textit{Algorithm \newline representation} & Manifest non-written & Manifest written \\
    \end{tabular}
\end{table}

\section{Discussion and Conclusions}\label{sec:discussion_conclusion}

{In the field of CT education, much of the research has concentrated on creating models to define CT skills and developing assessment tools} \citep{brennan2012new,grover2017,weintrop2016defining,lafuente2022assessing}. 
{While these approaches have contributed significantly to the field, we have identified several gaps that need to be addressed.
For instance, many theoretical models tend to be overly complex, incomplete or overlap with one another} \citep{tikva_mapping_2021,shute2017demystifying}. 
{Furthermore, there is limited guidance on how to effectively design activities that not only foster these competencies but also assess them accurately within specific educational contexts} \citep{Saxena2019,Relkin2019}. {
Currently, there is a lack of comprehensive frameworks that integrate these various aspects.} 

{Our research aims to address these critical gaps in the literature by offering a more integrated approach. 
In this paper, we present a comprehensive framework that focuses on computational thinking problems (CTPs), activities that require CT to be solved. 
Rather than focusing solely on defining CT competencies, we propose a shift toward identifying and analysing the components and characteristics of CTPs. 
We argue that these characteristics, particularly contextual factors, directly influence the development of CT skills} \citep{heersmink2013taxonomy,piatti_2022,roth2013situated}. 
{By linking these characteristics to a structured catalogue of CT competencies, we aim to align activity design with the specific competencies they are intended to develop.}

{The contribution of this framework is twofold: (i) it facilitates the analysis of existing CTPs by identifying which competencies they can develop or assess based on their inherent characteristics, and (ii) it guides the design of new CTPs targeted at specific CT skills by outlining the necessary components and characteristics required to activate them.}

\subsection{{Contributions}}

\paragraph{{Analysis of CTPs: Taxonomy of CTPs}}

{
The first contribution of the framework is its capability to analyse existing CTPs.
To demonstrate this, we applied the framework to a representative set of activities, leading to the development of a taxonomy of CTPs that categorises activities across three primarily domains: unplugged, robotics, and virtual. 
This taxonomy serves two key purposes: it identifies key characteristics commonly present within each domain of CTPs, and (ii) it outlines the key competencies that can be developed and fostered in each domain.
}

{The taxonomy demonstrates how different CTP domains are particularly suited to fostering specific competencies. 
It emphasises the importance of selecting the right type of activity to target these competencies, ensuring that the CTP aligns with educational goals. 
By tailoring activities to the skills being developed, educators can promote deeper and more meaningful learning outcomes.
}

\paragraph{Design of CTPs: Application to the virtual CAT}

{The second contribution of the framework is its ability to design new CTPs. 
To demonstrate this, we began with an existing unplugged activity, the CAT, used in the development of the taxonomy of CTPs, and explored how to design a new version of it within the virtual domain.
By applying the framework, we were able to adapt the unplugged CAT to a virtual setting systematically, preserving essential characteristics to activate core CT competencies while introducing new elements, such as system resettability, to enhance additional competencies, like debugging.} 

{While this study stops at the presentation of the theoretical design of the activity,} \cite{adorni_pilot,adorni_softwarex} {document its practical development and implementation.
The virtual CAT was subsequently tested in a large-scale study} \citep{adorni_main}, {which demonstrated that specific digital environment features, such as predefined visual programming blocks, more effectively support the development of algorithmic thinking compared to the unplugged version. These findings provide further validation for the framework.}

{This case study demonstrates the framework's ability to systematically design and adapt CTPs across different formats, enhancing their effectiveness in developing targeted competencies and validating its broader utility in aligning educational activities with specific learning goals.}


\subsection{{Limitations}}
Notwithstanding the contributions of this study, there are several limitations to be considered. 

{First, the scope of the CTPs taxonomy is currently limited by the specific activities analysed in this paper, necessitating} an expansion to include more varied and diverse CTPs across educational levels and subjects.


{Additionally, although the framework outlines which competencies can be developed given a CTP with specific characteristics, it does not yet specify the levels of abstraction at which they can be cultivated. Future work could aim to clarify whether competencies are fostered at foundational levels, e.g., recognising or understanding an algorithmic concept, or at more advanced stages, e.g., applying and assimilating an algorithmic concept} \citep{bloom1956taxonomy,CTF}. 

{Finally, expanding the framework to include broader competencies related to CT, like creativity, would be beneficial.}







%


\section*{Fundings}
 This research was funded by the Swiss National Science Foundation (SNSF) under the National Research Program 77 (NRP-77) Digital Transformation (project number 407740\_187246).
 
\section*{Competing Interests}
 The authors declare that they have no conflict of interest.

\newpage
\section*{Author Contributions}

\textbf{Giorgia Adorni}: Conceptualization, Methodology, Validation, Formal analysis, Investigation, Resources, Data curation, Writing - original draft \& review \& editing, Visualization, Supervision. \\
\textbf{Alberto Piatti}: Conceptualization, Methodology, Validation, Formal analysis, Writing - original draft \& review \& editing, Supervision, Project administration, Funding acquisition. \\
\textbf{Engin Bumbacher}: Conceptualization, Methodology, Validation, Formal analysis, Writing - original draft \& review \& editing. \\
\textbf{Lucio Negrini}: Validation, Formal analysis, Investigation, Writing - original draft \& review \& editing.\\
\textbf{Francesco Mondada}: Validation, Visualization, Writing - review \& editing, Supervision, Project administration, Funding
acquisition.\\
\textbf{Dorit Assaf}: Writing - review \& editing, Funding acquisition.\\
\textbf{Francesca Mangili}: Writing - review \& editing.\\
\textbf{Luca Maria Gambardella}: Funding acquisition.

\bibliographystyle{spbasic}      
\bibliography{bibliography.bib}   


%




\clearpage

\section*{Appendix: CTPs profiles}\label{appendix:appB}
\appendix

\renewcommand{\thefigure}{\thesection.\arabic{figure}}
\setcounter{figure}{0}
\renewcommand{\thetable}{\thesection.\arabic{table}}
\setcounter{table}{0}

{In this appendix section, we present the detailed analysis of selected CTPs used to construct our taxonomy. As discussed in} \Cref{sec:taxonomy_result1}: \nameref{sec:taxonomy_result1}, {these CTPs were chosen as prototypical examples commonly used in educational contexts. 
We applied our framework to each, outlining their components and characteristics and identifying which competencies they can support. This selection is an illustrative application of the framework rather than an exhaustive analysis. 
Each CTP is presented with (i) a description of its components and characteristics, accompanied by a graphical template, (ii) an overview of the competencies it supports or lacks, and (iii) a profile table showing how its characteristics align with specific CT competencies.
}
{The graphical representations of CTPs specific components and characteristics are adapted from the general template in} \Cref{fig:framework-components}, {while the profile table structure is based on} \Cref{tab:static-framework} {and has been further refined to indicate the presence of each characteristic with a tick mark, highlighting these columns in light grey for clarity. 
For each competence, an additional column uses a green checkmark to indicate if it can be developed (given the present characteristics) or a red cross if it cannot.
}



\section{{Cross Array Task (CAT)}}\label{sec:cat}
\begin{figure}[htb]
\centering
\includegraphics[width=.95\columnwidth]{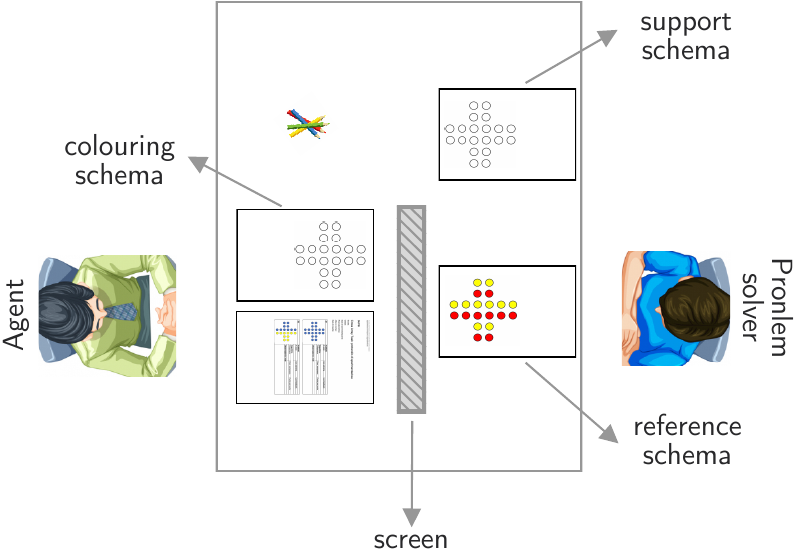}
\caption[CAT]{\textbf{The CAT activity, adapted from \cite{piatti_2022}.}
The task requires the problem solver to instruct the agent to reproduce a reference schema solely through verbal communication, with the option of supplementing instructions via gesturing on a support schema if deemed necessary. 
A removable screen separates the participant and researcher to regulate potential visual cues.}
\label{fig:CAT}
\end{figure}

\begin{figure*}[!t]
		\centering
		\includegraphics[width=\textwidth]{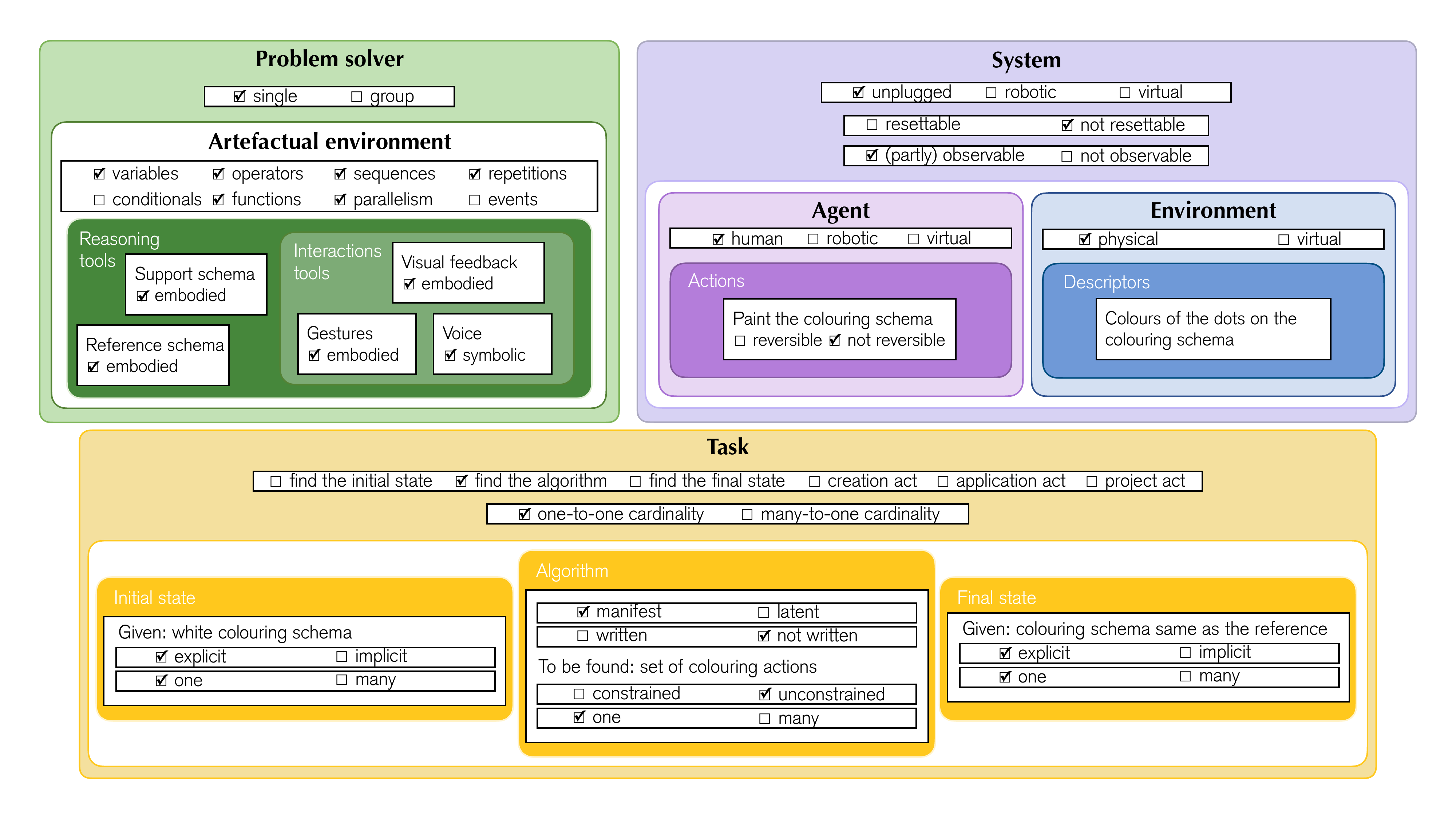}
		\caption{\textbf{CAT components and characteristics}.}
		\label{fig:CAT-features}
	\end{figure*}

\begin{figure*}[!h]
		\centering
		\includegraphics[width=\textwidth]{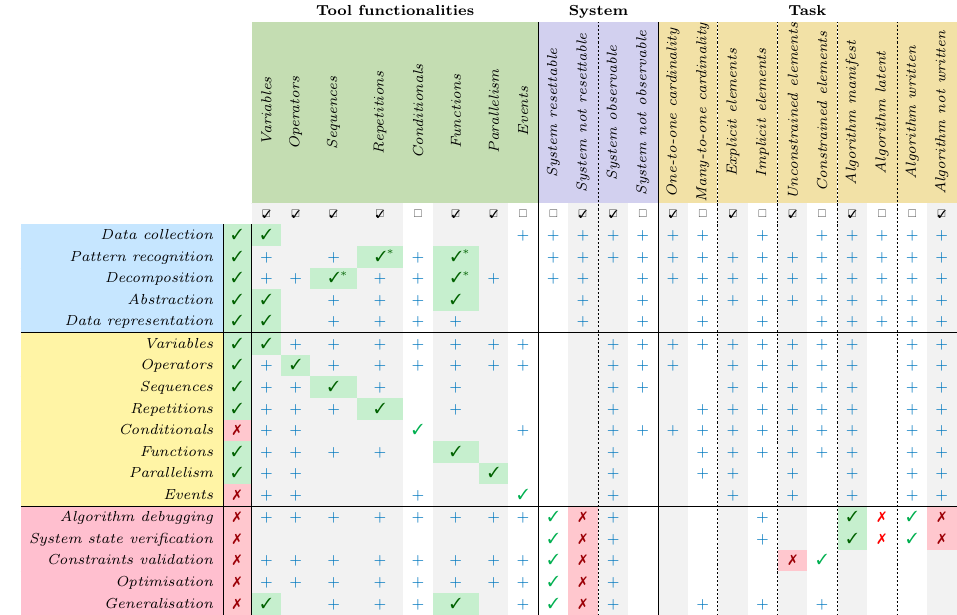}
		\caption{\textbf{CAT profile}.}
		\label{tab:CAT-mapping}
	\end{figure*}

The Cross Array Task (CAT), {illustrated in} \Cref{fig:CAT}, is an unplugged activity designed by \cite{piatti_2022} using the CT-cube to assess the development of algorithmic skills in compulsory schools.
\l{It involves a student and a researcher in a classroom setting, seated at a table and separated by a removable screen. The student has to communicate an algorithm to colour a white cross array to match a reference schema.
Components and characteristics of this CTP have been analysed using the graphical template in} \Cref{fig:CAT-features}, {while the profile, outlining the relationship between its characteristics and competencies, is illustrated in} \cref{tab:CAT-mapping}.

\subsection{Components}
\begin{itemize}[noitemsep,nolistsep]
\item \textit{Problem solver}: the student who has to communicate an algorithm corresponding to the sequence of instructions to reproduce the colouring of the reference schema.
The artefactual environment comprises cognitive tools such as {support and reference schemas, which are} available to the problem solver to reason about the task.
Additionally, the problem solver can interact with the system to communicate the algorithm. This can be achieved using a natural language such as the voice (symbolic) or gestures (embodied) on the {colouring schema, an} empty cross array {used from the agent}. 
Moreover, by removing the screen that separates the problem solver from the agent, he can have visual feedback (embodied) of the cross array being coloured.
\item \textit{Agent}: the researcher, executor of the problem solver's instructions, responsible for filling the colouring schema according to the problem solver's algorithm. The agent's actions are not resettable.
\item \textit{Environment}: the cross array to be coloured, whose state is described by the colour of each dot (white, yellow, blue, green, or red). 
\item \textit{Task}: find the algorithm. The system's state is defined by the colouring cross status, initially white and, at the end the same as the reference schema. 
The algorithm is the set of agent instructions to achieve this transformation.
\end{itemize}

\subsection{Characteristics}

\begin{itemize}[noitemsep,nolistsep]
\item \textit{Tool functionalities}: voice and gestures provide various functionalities associated with algorithmic concepts, suitable to design the algorithm, including 
(i) \textit{variables} can represent different colours of the cross array dots; 
(ii) \textit{operators} are used to change the colour of the dots performing actions such as colouring a dot, a row, a square and so on;
(iii) \textit{sequences} determine the order in which the actions should be executed to achieve the desired outcome; 
(iv) \textit{repetitions} allow for repeating specific sequences of operations, such as colouring the first column in red and repeating it every two columns;
(v) \textit{functions} consist of operations that perform a specific task and can be applied to different inputs, for example, creating a pattern of alternating red and yellow dots in a square and applying it to different positions of the cross array;
(vi) \textit{parallelism} involves executing multiple actions simultaneously and can be associated with using symmetries to describe the pattern.
\item \textit{System resettability}: the system is not resettable since it is impossible to reverse the agent's actions.
\item \textit{System observability}: the system is partially observable since the cross array being coloured by default is not seen until the end of the task unless the problem solver demands otherwise.
\item \textit{Task cardinality}: the task has a one-to-one mapping, with given one initial and one final state, and an algorithm to be found.
\item \textit{Task explicitness}: all elements are given explicitly.
\item \textit{Task constraints}: the algorithm is unconstrained.    
\item \textit{Algorithm representation}: the algorithm is represented through voice commands or gestures. It is considered manifest, because {it is externalised}, but not written since it is not stored in a permanent format.
\end{itemize}
\clearpage

 \clearpage

\subsection{Competencies}
 \paragraph{Enabling features for competencies development}
\begin{itemize}[noitemsep,nolistsep]
\item \textit{Problem setting}: all competencies can be activated thanks to the presence of variables, sequences, repetitions and functions in the tool functionalities.
The presence of many tool functionalities, the non-resettability of the system and the algorithm representation positively affect and boost problem setting skills.
The system observability supports data collection and pattern recognition. The one-to-one cardinality, in addition, stimulates decomposition. 
The explicit and unconstrained definition of the task elements also promotes pattern recognition, decomposition and abstraction. 

\item \textit{Algorithm}: all competencies associated with the algorithmic concepts enabled by the tool functionalities, meaning variables, operators, sequences, repetitions and functions, can be activated and promote one another.
The form of representation of the algorithm, the system observability, and the explicit and unconstrained definition of the task elements further enhance these. The one-to-one cardinality helps to enhance some of these skills as well.

\item \textit{Assessment}: since the system is not resettable, no assessment skills can be developed.
\end{itemize}

\paragraph{Inhibiting features for competencies development}
\begin{itemize}[noitemsep,nolistsep]
\item \textit{Conditionals and events}: non-activable as these functionalities are unavailable in the platform. 
A way to make conditionals available in the tool functionalities would be allowing the problem solver to change a dot colour, for example by communicating instructions such as: ``if the dot is red, then colour it yellow''. By doing this, the problem solver engages with the concept of conditionals and can develop their algorithmic skills.
The completion of each row in the cross array can be considered an event. The problem solver can specify that they want to fill the cross line by line, and once a line is complete, the researcher will move on to the next line. This allows the problem solver to list only the sequence of colours without repeating the instructions for where to go. The change in the environment (completing a row) triggers the researcher to move to the next row.
Using conditionals and events can greatly enhance the complexity of the solutions that can be generated and help develop advanced CT skills.

\item \textit{Assessment skills}: the inability to reset the system impairs the development of the student's skills. 
One possible solution to this issue is enabling the student to reset the colouring scheme using a voice command. This would return the schema to its initial blank state, allowing the student to start the task from the beginning and practice their assessment skills.
To develop system state verification, it is also essential to not reveal the initial or final states. 
Moreover, constraints should be imposed on the algorithm to develop constraint validation skills, for example, limiting the use of specific operators or the number of times they can be used, allowing the problem solver to develop the ability to think about the constraints and limitations in their algorithms.
\end{itemize}



\section{{Graph Paper Programming}}\label{sec:GPP}
\begin{figure*}[!h]
  \centering
  \includegraphics[width=\textwidth]{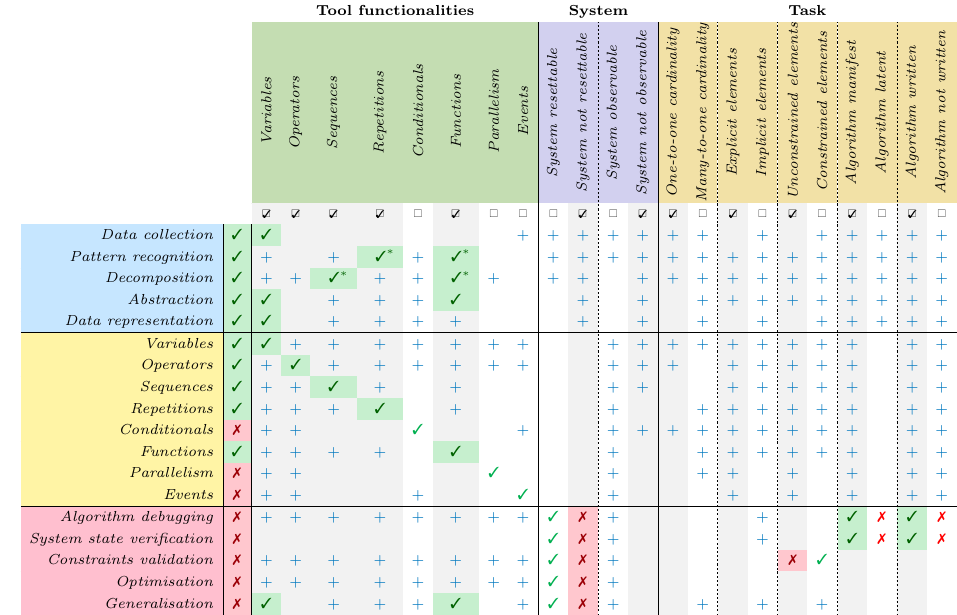}
    \caption{\textbf{Graph Paper Programming (part 1) profile.}}
\label{tab:GPPfa-mapping}
\end{figure*}

\begin{figure*}[!h]
		\centering
            \includegraphics[width=\textwidth]{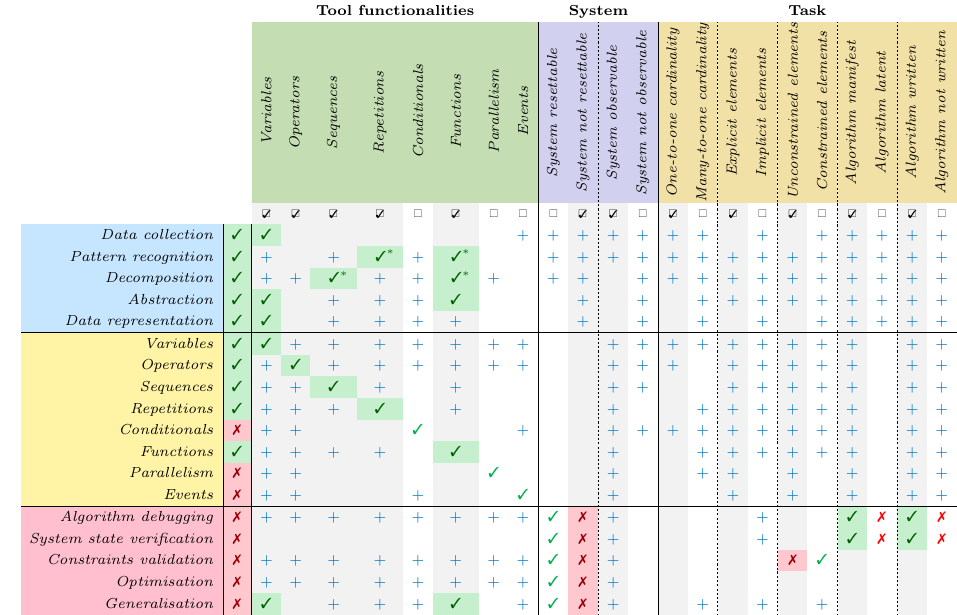}
		\caption{\textbf{Graph Paper Programming (part 2) profile.}}
		\label{tab:GPPffs-mapping}
\end{figure*}

Graph Paper Programming is an unplugged activity from \cite{org2015web}, a non-profit organisation that aims to provide students with the opportunity to learn computer science as part of their education, offering various activities designed to increase diversity in computer science and reach students at their skill level and in ways that inspire them to continue learning. 
This CTP has two variants. 
In the first, the student is given a $4\times4$ grid of white and black squares and asked to write explicit instructions for another classmate to reproduce the image without letting the other person see the original drawing. 
In the second, the same student follows the instructions they previously wrote to reproduce the image. 
By dividing the activity into these two parts, we can gain a deeper understanding of the cognitive processes and skills involved in each task and the potential for this activity to support the development of CT abilities.
{Components and characteristics of this CTP have been analysed using the graphical templates in} \Cref{fig:GPPfa,fig:GPPffs}, {while the profiles, outlining the relationship between its characteristics and competencies, are illustrated in} \cref{tab:GPPfa-mapping,tab:GPPffs-mapping}.

\subsection{Components (part 1)}
\begin{figure}[hb]
\centering
\includegraphics[width=.95\columnwidth]{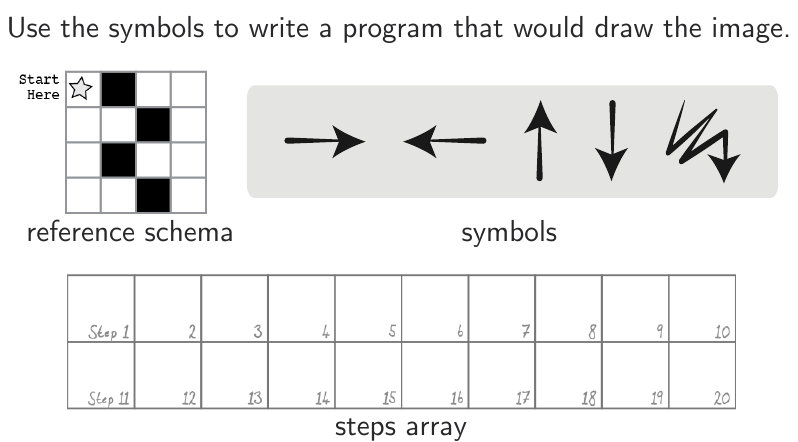}
\caption[GPP1]{\textbf{The first part of the Graph Paper Programming activity, adapted from \cite{org2015web}.} 
The task requires the problem solver to instruct the agent to reproduce the reference schema with instructions written on a steps array using a predefined set of arrow symbols.}
\label{fig:GPPfa}
\end{figure}

The first part of the activity is illustrated in \Cref{fig:GPPfa}.
\begin{itemize}[noitemsep,nolistsep]
\item \textit{Problem solver}: the student who writes the set of instructions for the agent to follow.
{The artefactual environment comprises cognitive tools such as the reference schema (embodied) and the support and the set of arrow symbols (embodied) available to the problem solver to reason about the task.}
Additionally, the problem solver can interact with the system to communicate the algorithm, writing the arrow symbols in the steps array (symbolic). These can be considered as a programming language and its programming platform. 
\item \textit{Agent}: the other student who executes the problem solver's instructions by filling the colouring schema accordingly. Its actions are not resettable.
\item \textit{Environment}: the schema to be coloured, described by the colour of each square (white or black). 
\item \textit{Task}: find the algorithm. The system's state is defined by the colouring schema status, initially white and, at the end, the same as the reference schema. 
The algorithm is the set of agent instructions to achieve this transformation.
\end{itemize}

\begin{figure*}[!t]
  \centering
  \includegraphics[width=\textwidth]{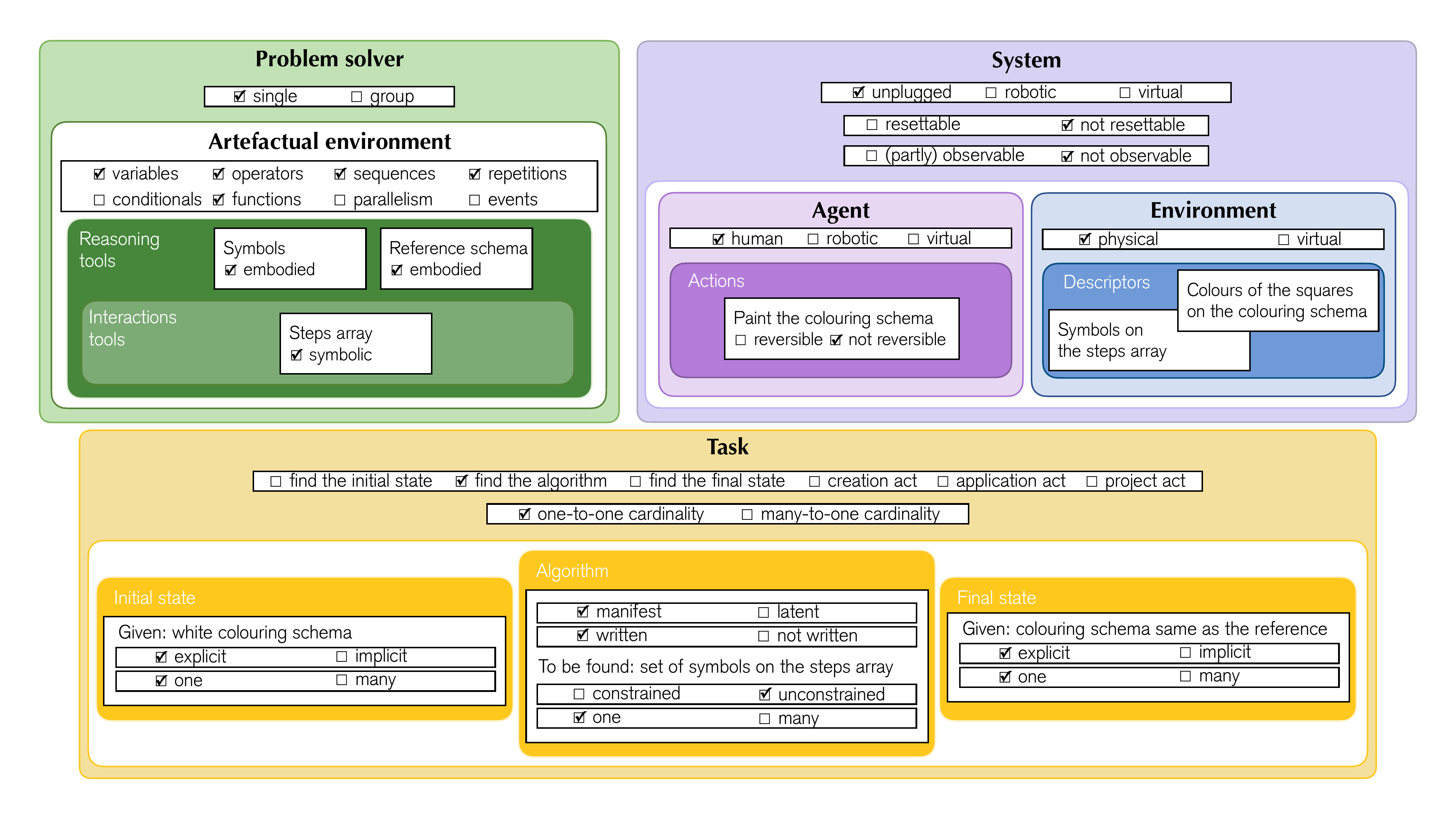}
    \caption{\textbf{Graph Paper Programming (part 1) components and characteristics.}}
\label{fig:GPPfa-features}
\end{figure*}

\FloatBarrier
\subsection{Components (part 2)}
\begin{figure}[!htb]
\centering
\includegraphics[width=.95\columnwidth]{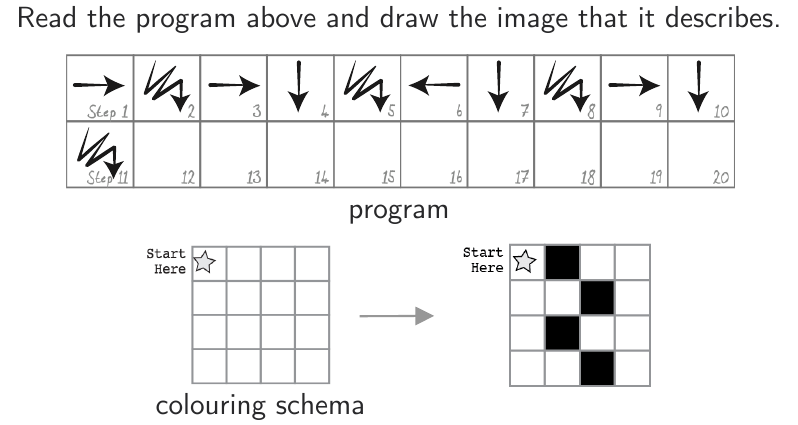}
\caption[GPP2]{\textbf{The second part of the Graph Paper Programming activity, adapted from \cite{org2015web}.} The task requires the problem solver to fill the empty colouring schema following the program provided. The figure illustrates the expected final state.}
\label{fig:GPPffs}
\end{figure}

The second part of the activity is illustrated in \Cref{fig:GPPffs}.
\begin{itemize}[noitemsep,nolistsep]
\item \textit{Problem solver and Agent}: they overlap and correspond to the student who follows the instructions to recreate the image. 
The only action the agent can perform is to paint the colouring schema without the possibility of undoing it.
The artefactual environment comprises cognitive tools such as the colouring schema (embodied) available to the problem solver to reason about the task. 
As before, the arrow symbols on the steps array (symbolic) are also used to interact with the system. Moreover, as the agent and the problem solver overlap, visual feedback (embodied) is always given.
\item \textit{Environment}: the schema to be coloured, described by the colour of each square (white or black). 
\item \textit{Task}: find the final state. The system's state is defined by the colouring schema status, which is initially white. However, the final state is not given and has to be found using the provided algorithm.    
\end{itemize}

\subsection{Characteristics}
\begin{itemize}[noitemsep,nolistsep]
\item \textit{Tool functionalities}: in both parts of the activity, the tools provide various functionalities associated with algorithmic concepts suitable to design the algorithm, including 
(i) \textit{variables} can represent different colours of the schema squares; 
(ii) \textit{operators} correspond to the arrow symbols used to instruct the agent to move from one square to another, determining whether a square is coloured black or white;
(iii) \textit{sequences} determine the order in which the actions should be executed to achieve the desired outcome; 
(iv) \textit{repetitions} allow for repeating specific sequences of operations;
(v) \textit{functions} can be represented by a group of instructions that perform a specific task, such as colouring a particular shape on the grid, that can be used multiple times.
\item \textit{System resettability}: the system is not resettable since it is impossible to reverse the agent's actions.
\item \textit{System observability}: in the first part of the activity, the system is not observable, as there is no visual feedback about the agent's actions; in the second part of the activity, {the visual feedback consented thanks to the problem solver and agent overlapping, makes the system totally observable.}
\item \textit{Task cardinality}: both tasks have a one-to-one mapping, with one initial state, final state and algorithm.
\item \textit{Task explicitness}: the task elements are explicit.
\item \textit{Task constraints}: the final state is unconstrained.    
\item \textit{Algorithm representation}: the algorithm is manifest and written, represented through the arrow symbols written in the steps array.
\end{itemize}
\begin{figure*}[!t]
\centering
\includegraphics[width=\textwidth]{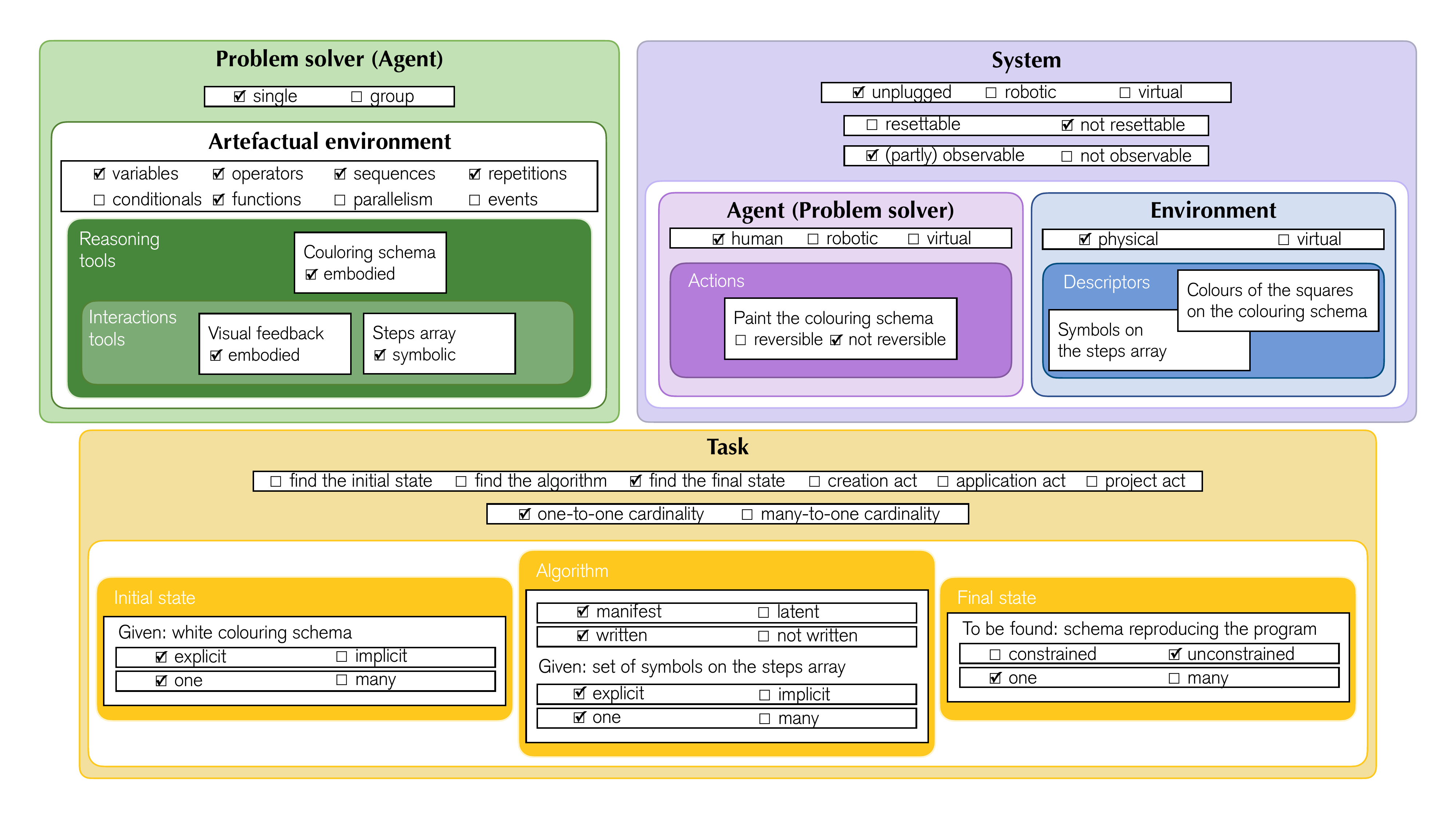}
\caption{\textbf{Graph Paper Programming (part 2) components and characteristics.}}
\label{fig:GPPffs-features}
\end{figure*}

\subsection{Competencies}
\paragraph{Enabling features for competencies development}
\begin{itemize}[noitemsep,nolistsep]
\item \textit{Problem setting}: all competencies can be activated thanks to the presence of variables, sequences, repetitions and functions in the tool functionalities.
The presence of many tool functionalities, the non-resettability of the system and the algorithm representation positively affect and boost problem setting skills.
In the first part of the activity, the inability to observe the systems further supports the development of all these competencies. In contrast, in the second part, the system observability sustains only data collection and pattern recognition. 
The one-to-one cardinality, in addition, stimulates data collection and pattern recognition but also decomposition. 
The explicit and unconstrained definition of the task elements also promotes pattern recognition, decomposition and abstraction. 

\item \textit{Algorithm}: {the competencies associated with algorithmic concepts, including variables, operators, sequences, repetitions, and functions, can be developed using the tool functionalities.} 
The algorithm representation, the system observability, and the explicit and unconstrained definition of the task elements further enhance all these skills. 

\item \textit{Assessment}: since the system is not resettable, no assessment skills can be developed.
\end{itemize}

\paragraph{Inhibiting features for competencies development}
\begin{itemize}[noitemsep,nolistsep]
\item \textit{Conditionals, parallelism and events}: non-activable as these functionalities are unavailable in the platform. 
Activating conditionals would be allowed by providing a new arrow symbol that determines the colour of a square based on certain conditions, such as the colour of the square above.
Parallelism can be enabled by operating a new arrow symbol instructing the agent to colour two squares simultaneously.
An event that could be considered is that every time a cell is filled with black, the instructions in the steps array move to the next row rather than continuing from that specific cell.

\item \textit{Assessment skills}: the inability to reset the system impairs the development of the student's skills. 
In the first part of the activity, where the problem solver is writing the set of instructions for the agent, to solve this issue the system can be reset by simply starting over with a new blank graph array and writing a new set of instructions, or simply offering the possibility to use pencil and eraser. 
In the second part of the activity, where the problem solver and the agent are the same people, the system can be reset by either using a new colouring schema or erasing the previously produced image and starting over. 
\end{itemize}

\section{{Triangular Peg Solitaire}}\label{sec:TPS}
\begin{figure*}[!t]
		\centering
		\includegraphics[width=\textwidth]{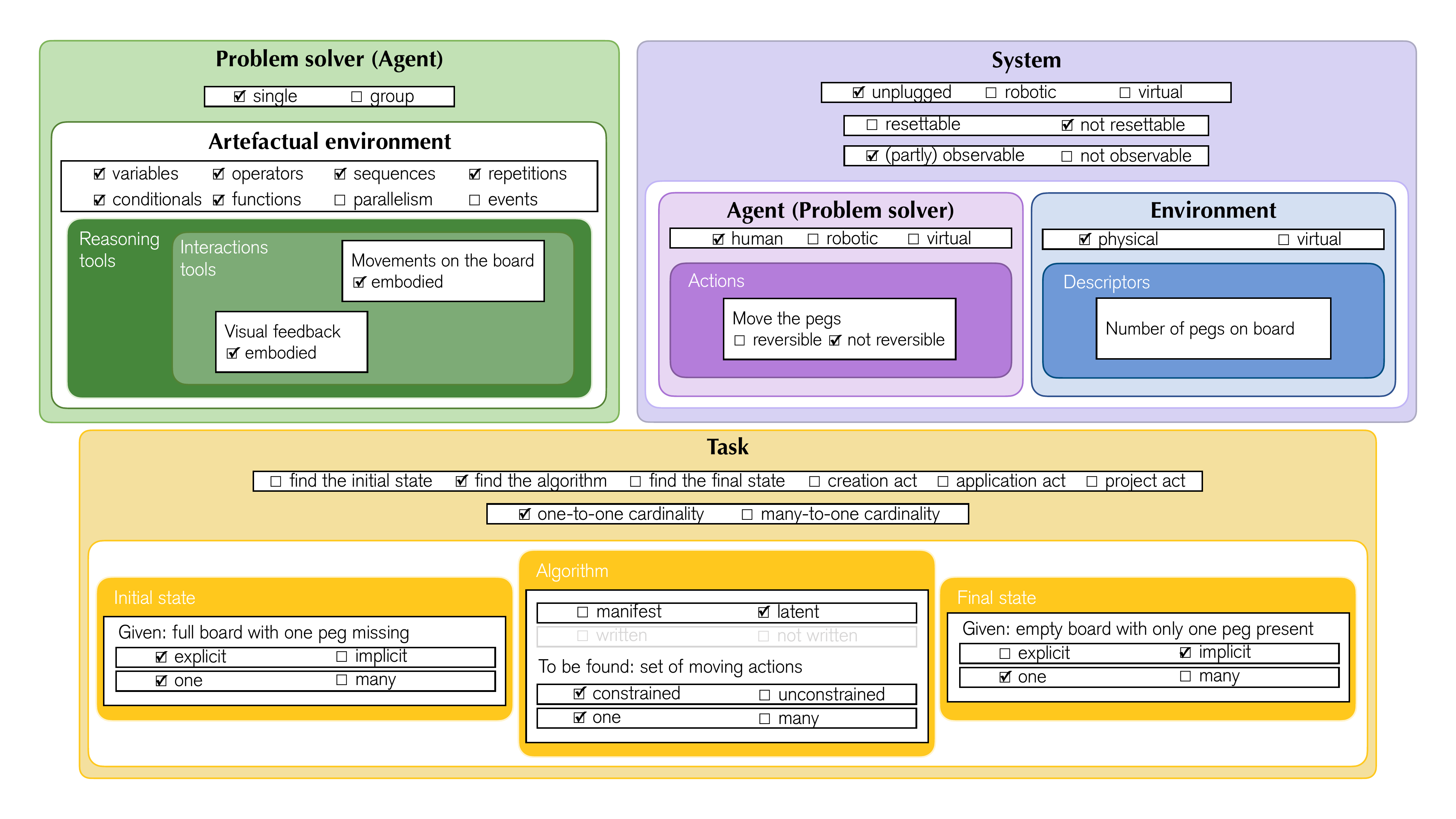}
		\caption{\textbf{Triangular Peg Solitaire (board variant) components and characteristics.}}
		\label{fig:TPSboard-features}
	\end{figure*}

\begin{figure*}[!h]
		\centering
		\includegraphics[width=\textwidth]{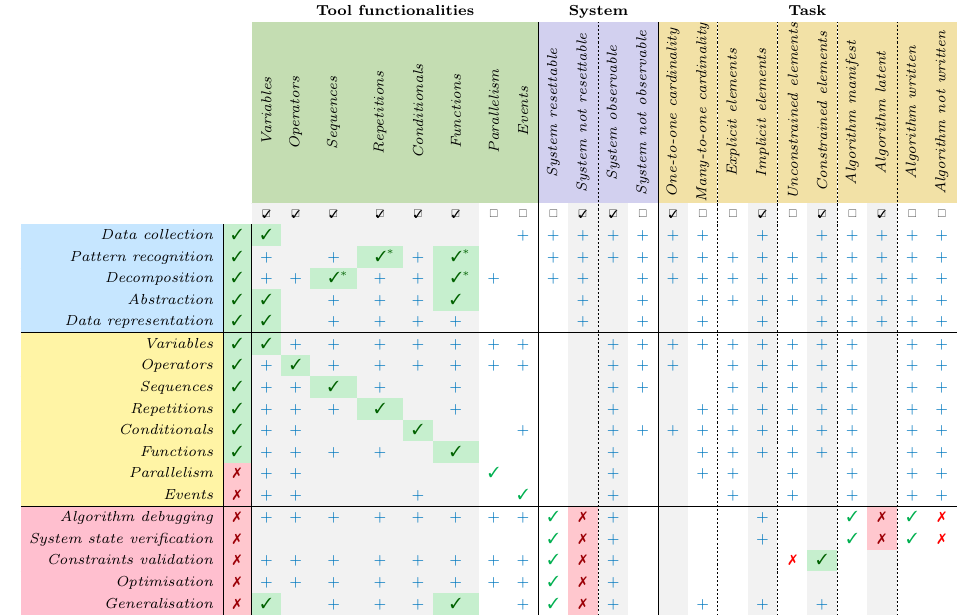}
		\caption{\textbf{Triangular Peg Solitaire (board variant) profile.}}
		\label{tab:TPSboard-mapping}
	\end{figure*}

\begin{figure*}[!t]
    \centering
    \includegraphics[width=\textwidth]{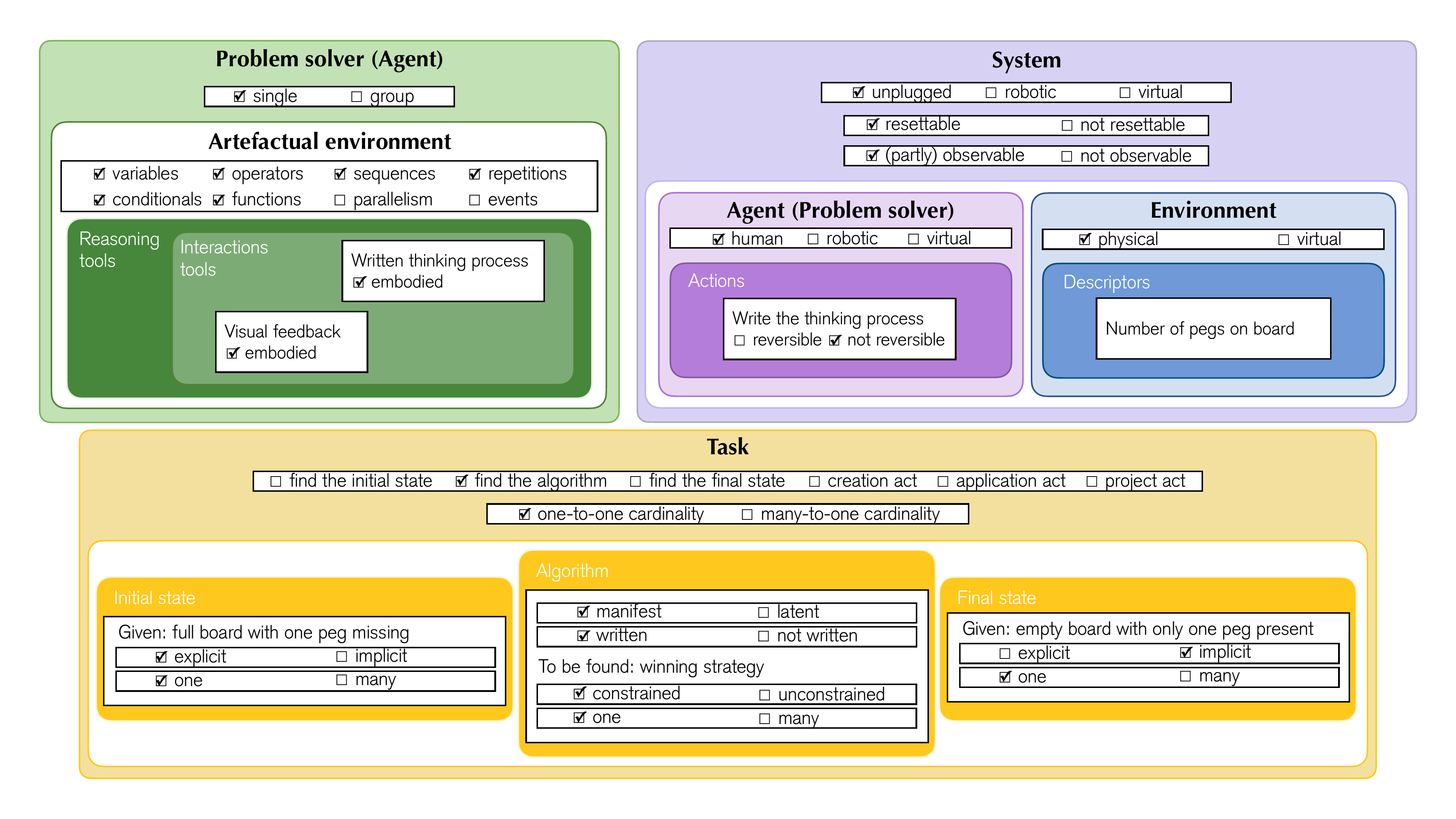}
    \caption{\textbf{Triangular Peg Solitaire (paper \& pencil variant) components and characteristics.}}
    \label{fig:TPSpp-features}
\end{figure*}

\begin{figure*}[!h]
		\centering
		\includegraphics[width=\textwidth]{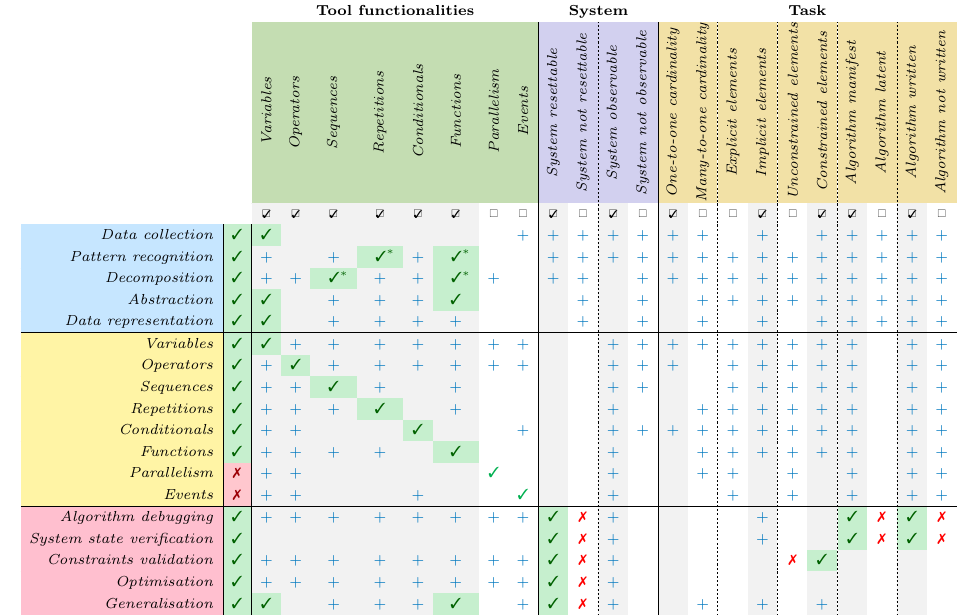}
		\caption{\textbf{Triangular Peg Solitaire (paper \& pencil variant) profile.}}
		\label{tab:TPSpp-mapping}
	\end{figure*}

Triangular Peg Solitaire is a strategy game that can be played in two variants: the classic, on a triangular board with 15 holes and pegs, and the paper and pencil modality.
The board is initially filled with pegs, except for one hole, which is left empty (see the top of \Cref{fig:TPSboard}).
This game is often used to teach problem-solving, logic and strategy skills, requiring the player to determine the most efficient sequence of moves to remove all the pegs on the board except one. Research has shown that the second activity variant can effectively promote problem-solving skills even in older students \citep{pegsolitaire}. 
The game is played by following the rules, which dictate that a peg can only be moved by jumping over a neighbouring peg on the diagonal or horizontal lines (see the bottom of \Cref{fig:TPSboard}).
By analysing the two versions of this game, we aim to understand their impact on promoting problem-solving and critical thinking skills, evaluating advantages and limitations and providing insights into how they can be used for CT development.
{Components and characteristics of this CTP have been analysed using the graphical templates in} \Cref{fig:TPSboard-features,fig:TPSpp-features}, {while the profiles, outlining the relationship between its characteristics and competencies, are illustrated in} \cref{tab:TPSboard-mapping,tab:TPSpp-mapping}.

\subsection{Components (board variant)}
\begin{figure}[!hbt]
\centering
\includegraphics[width=.95\columnwidth]{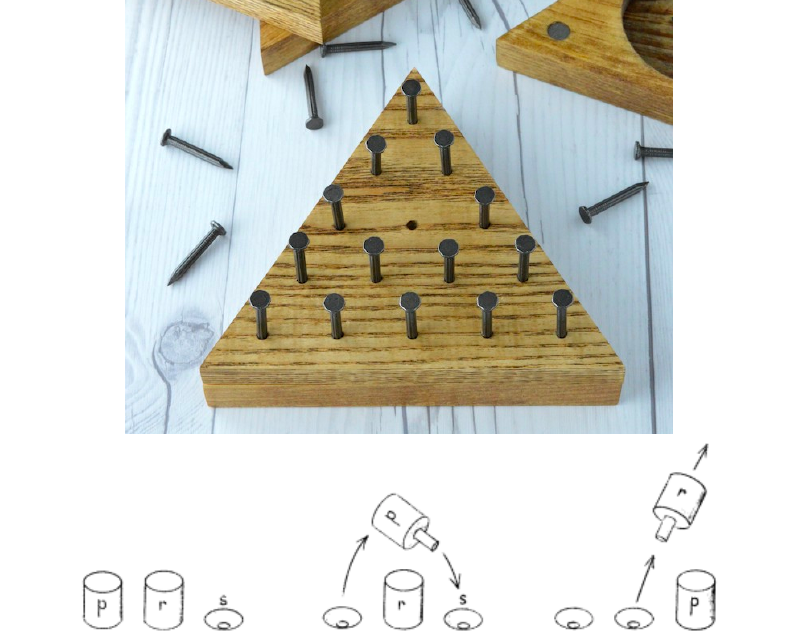}
\caption[TPS1]{\textbf{The Triangular Peg Solitaire.  }
The game is played on a board containing 15 spots, with 14 pegs placed on it at the start of the game (top). 
The task requires the problem solver to strategically move one peg at a time to eliminate all other pegs on the board until only one remains (bottom), jumping a peg over a neighbouring peg on the diagonal or horizontal lines, with the constraint that there must be a free landing spot for the jumping peg (adapted from \cite{berlekamp2004winning}).}
\label{fig:TPSboard}
\end{figure}

The first variant of the activity, illustrated in \Cref{fig:TPSboard}, requires solving the game on a physical board.
\begin{itemize}[noitemsep,nolistsep]
\item \textit{Problem solver and Agent}: they overlap and correspond to the player who must determine the most efficient sequence of moves to remove all the pegs on the board.
The only action the agent can perform is to move the pegs on the board without the possibility of undoing it.
The artefactual environment comprises tools for reasoning and interacting with the system. Being the agent and the problem solver overlapped, visual feedback (embodied) of the state of the board is always given. Moreover, the problem solver can physically interact with the system by moving the pegs on the board (embodied). 
\item \textit{Environment}: the wooden board, described by the number of pegs on it. 
\item \textit{Task}: find the algorithm. The system's initial state is the board with 14 pegs. The final state is the board with one peg. The algorithm to be found specifies the sequence of moves to remove all the pegs. 
\end{itemize}

\subsection{Components (paper \& pencil variant)}
\begin{figure}[!htb]
\centering
\includegraphics[width=.95\columnwidth]{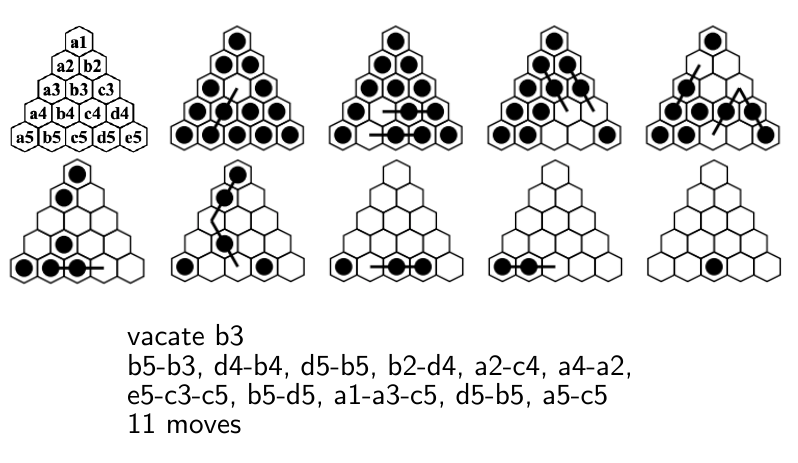}
\caption[TPS2]{\textbf{A Triangular Peg Solitaire solution adapted from \cite{bell_unlimited, bell2008solving} and \cite{barbero2018analysing}.} 
The task requires the problem solver to solve the game using paper and pencil by meticulously documenting their entire thought process. The solution can be presented in multiple ways, such as graphically using a Cartesian notation (top) or by numbering the boxes progressively and expressing the movements used (bottom).}
\label{fig:TPSpp}
\end{figure}

The first variant of the activity, illustrated in \Cref{fig:TPSpp}, requires solving the game with paper and pencil by documenting the thought process and devising a winning strategy.
\begin{itemize}[noitemsep,nolistsep]
\item \textit{Problem solver and Agent}: they overlap and correspond to the player who must determine the most efficient sequence of moves to remove all the pegs on the board.
The only action the agent can perform is to write the thinking process and strategy to remove the pegs on the board without the possibility of undoing it.
The artefactual environment comprises tools for reasoning and interacting with the system. Being the agent and the problem solver overlapped, visual feedback (embodied) of the state is always given. Moreover, the problem solver can physically interact with the system by writing the thinking process (symbolic).
\item \textit{Environment}: the board, drawn in the thinking process, described by the number of pegs on it. 
\item \textit{Task}: find the algorithm. The system's initial state is the board with 14 pegs. The final state is the board with one peg. The algorithm to be found specifies the sequence of moves to remove all the pegs. 
\end{itemize}

\subsection{Characteristics}
\begin{itemize}[noitemsep,nolistsep]
\item \textit{Tool functionalities}: in both variants of the activity, the tools provide various functionalities associated with algorithmic concepts suitable to design the algorithm, including 
(i) \textit{variables} can represent the state of the board and, in particular, the number of pegs on it; 
(ii) \textit{operators} correspond to the moves made by the player to change the state of the board by removing or moving pegs from one hole to another;
(iii) \textit{sequences} determine the order of moves made by the player; 
(iv) \textit{repetitions} allow for repeating certain moves or sequences of moves;
(v) \textit{conditionals} refer to the possible decisions that the plays may need to make, such as where to jump in one direction or another;
(vi) \textit{functions} can be represented by a group of instructions that perform a specific task, for example, a function to delete pegs in a row, which can be applied to several rows.
\item \textit{System resettability}: in the board variant, the system is not resettable, meaning that once the player has made a move, it cannot be undone. However, the system is resettable in the paper and pencil version, even though the player's actions are not reversible. This is because the informal setting, in which the player documents their thought process, allows for experimentation and exploration without fear of judgement or negative consequences. In other words, the player can freely make mistakes, express uncertainty, and experiment with different strategies without permanently impacting the game.
\item \textit{System observability}: in both variants, the system is observable because the agent, overlapping with the problem solver, can see the state of the board anytime.
\item \textit{Task cardinality}: both activity variants have a one-to-one mapping, with one initial state, one final state and one algorithm.
\item \textit{Task explicitness}: the initial state of the system is given explicitly, while the final one is given implicitly since the task instruction does not specify which is the position of the last remaining peg.
\item \textit{Task constraints}: the algorithm is constrained by the game rules, which dictate that a peg can only be moved by jumping over a neighbouring peg on the diagonal or horizontal lines and that there must be a free landing spot for the jumping peg.    
\item \textit{Algorithm representation}: in the board variant of the game, the algorithm is latent since it is performed physically through the player's moving the pegs. It is not permanently recorded and cannot be revisited.
On the other hand, in the paper and pencil modality, the player writes down the algorithm, and it becomes a permanent record that can be reviewed and used as a reference. This allows the player to experiment freely and change their approach without having to start over each time. The written representation of the algorithm in the paper and pencil modality provides a clear and concrete way to represent the player's thought process and strategy.
\end{itemize}

\subsection{Competencies}

\paragraph{Enabling features for competencies development}
\begin{itemize}[noitemsep,nolistsep]
\item \textit{Problem setting}: all competencies can be activated in both variants of the game thanks to the presence of variables, sequences, repetitions and functions in the tool functionalities.
The presence of many tool functionalities, the non-resettability of the system (in the board version of the game), the implicit and constrained definition of the task elements and the algorithm representation positively affect and boost problem setting skills.
The system's observability sustains data collection and pattern recognition skills, while the one-to-one cardinality and the resettability of the system (in the paper and pencil variant of the game) also stimulate decomposition. 

\item \textit{Algorithm}: the competencies associated with algorithmic concepts, including variables, operators, sequences, repetitions, conditionals and functions, can be developed. 
The system observability, the implicit and constrained definition of the task elements, and the manifest written representation of the algorithm (in the paper and pencil variant of the game)  can further enhance these skills.

\item \textit{Assessment}: {in the board variant of the activity, no assessment skills can be developed due to the lack of a resettable system. In contrast, the paper-and-pencil version enables the activation of all assessment skills, thanks to the system's resettable nature and a manifest, written representation of the algorithm. Additionally, the inclusion of constraints fosters constraint validation, while variables and functions support the development of generalisation skills.
}
\end{itemize}

\paragraph{Inhibiting features for competencies development}
\begin{itemize}[noitemsep,nolistsep]
\item \textit{Parallelism and events}: non-activable as these functionalities are unavailable in the platform. 
It is possible to develop parallelism and events in the game by incorporating a variant where multiple pegs can be moved simultaneously or by introducing multiple players who can make moves simultaneously. 
A way to incorporate events would be to use technology, such as a computer program, or an app to play the game, creating programmed events that the player's actions could trigger. For example, if the player removes a certain peg, the computer could trigger an event that changes the board's appearance or displays a message.
However, it is important to note that these changes would result in a different game with a different set of objectives and challenges and might not necessarily have the same educational benefits as the original Triangular Peg Solitaire.

\item \textit{Assessment skills}: the inability to reset the system impairs the development of the student's skills in the first variant of the activity. In the paper and pencil version, only system state verification is non-activable because both the initial and final states are provided. The task can be adjusted by modifying the game such that only the initial state is provided and the final state is unknown. 
For example, letting the player determine the specific peg position for the last peg
This would make the task more challenging and require the player to develop their skills in system state verification. 
\end{itemize}

\section{{Computational Thinking test (CTt)}}\label{sec:CTt}
\begin{figure}[!htb]
\centering
\includegraphics[width=.95\columnwidth]{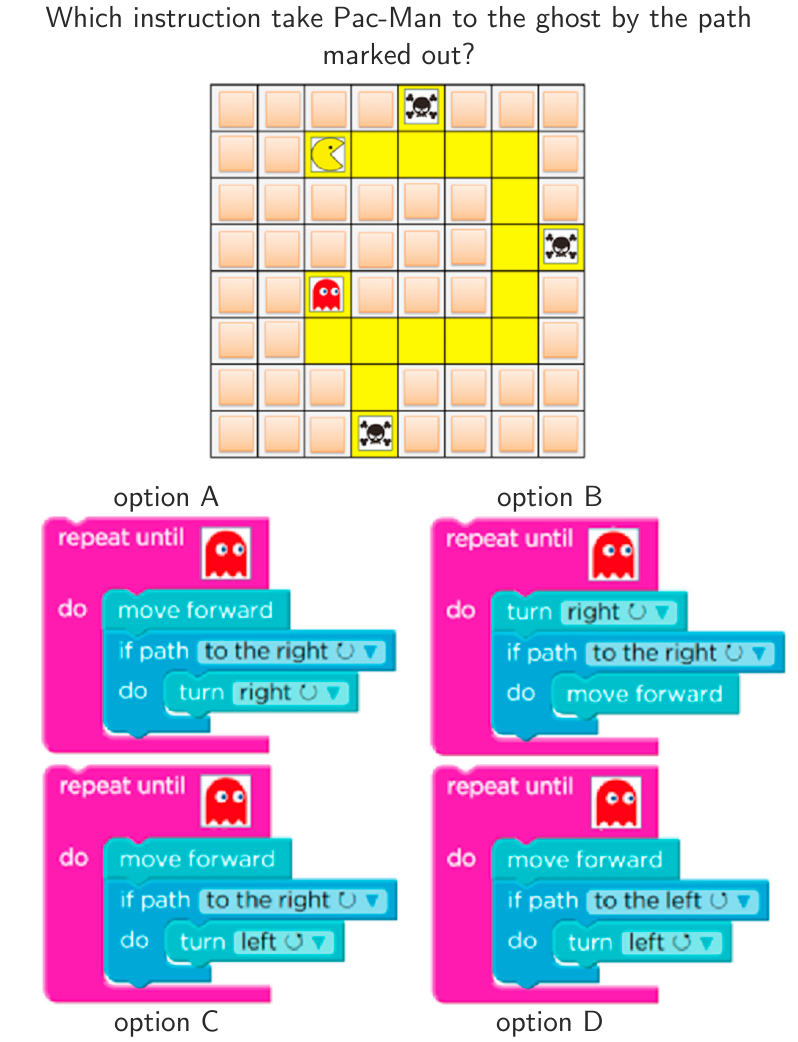}
\caption[CTt14]{\textbf{Item 14  of the CTt adapted from \cite{roman2017cognitive}.} The task requires the problem solver to select the correct set of instructions to make the agent cross a predefined path to reach a desired position.
}
\label{fig:CTt14}
\end{figure}

\begin{figure*}[!t]
		\centering
		\includegraphics[width=\textwidth]{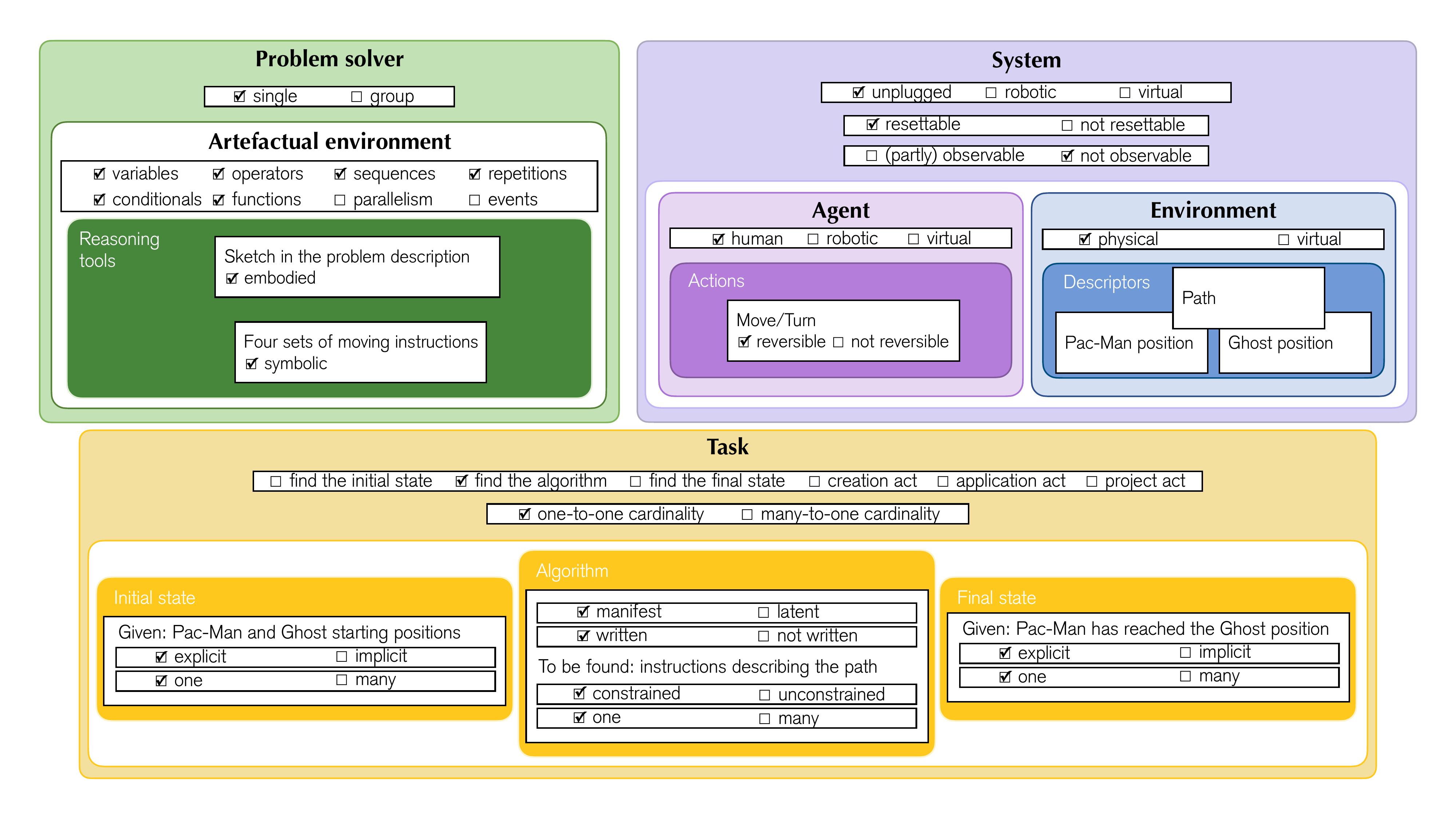}
		\caption{\textbf{CTt (item 14) components and characteristics.}}
		\label{fig:CTt14-features}
	\end{figure*}

\begin{figure*}[!h]
		\centering
            \includegraphics[width=\textwidth]{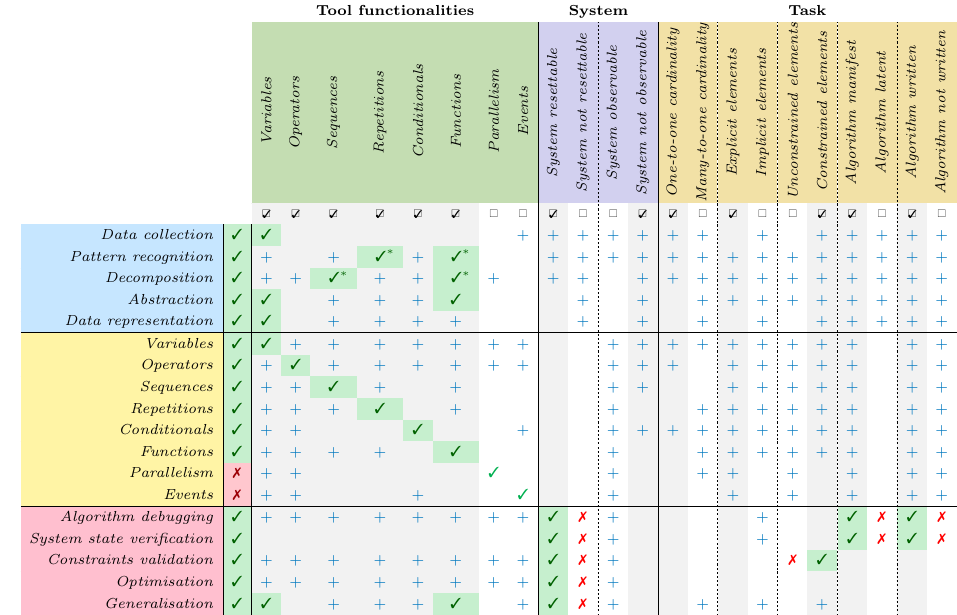}
		\caption{CTt profile.}
		\label{tab:CTt-mapping}
\end{figure*}

The Computational Thinking test (CTt) is an assessment tool designed to evaluate the CT skills of students between the ages of 12 and 13 \citep{roman2015computational,roman2017cognitive,roman2017complementary,roman2018detected}. 
It aligns with the work of researchers such as \cite{brennan2012new} and \cite{kaleliouglu2015new}, who have identified key computational concepts related to algorithmic skills. Additionally, the CTt is designed to align with the standard interfaces used by organisations universally recognised in this context, such as Code.org, which utilises visual blocks to teach coding. 
The test consists of 28 multiple-choice questions. 
{To illustrate this analysis,} \Cref{fig:CTt14} {portrays ``item 14'' of the test, designed to evaluate the student's ability to organise a set of commands in a logical and orderly manner in a script that does not involve computational nesting concepts but only two specific computational concepts: repetitions and conditionals.
Components and characteristics of this CTP have been analysed using the graphical template in} \Cref{fig:CTt14-features}, {while the profile, outlining the relationship between its characteristics and competencies, is illustrated in} \cref{tab:CTt-mapping}.

\subsection{Components (item 14)}

\begin{itemize}[noitemsep,nolistsep]
\item \textit{Problem solver}: the student taking the test, given with four sets of code scripts from which he must select the appropriate set of moving instructions.
The artefactual environment comprises only cognitive tools, including the sketch of the maze in the problem description (embodied) and the four sets of instruction in the form of visual blocks (symbolic); thus, it is impossible to interact with the system.
\item \textit{Agent}: Pac-Man, a representation of an abstract entity that can move in the maze to reach the Ghost following the predefined pattern marked out. 
Its actions are reversible since they must be corrected.
\item \textit{Environment}: the maze, described by the Pac-Man and the Ghost positions, and the path to be followed.
\item \textit{Task}: find the algorithm. The system's initial state corresponds to the Pac-Man and the Ghost in their starting position, while in the final state, Pac-Man is in front of the Ghost and has crossed the predefined path. The algorithm is not given since four sets of codes are provided, and the correct one has to be found to reach the desired outcome.
\end{itemize}


\subsection{Characteristics}
\begin{itemize}[noitemsep,nolistsep]
\item \textit{Tool functionalities}: the tools provide various functionalities associated with the visual blocks interface, including 
(i) \textit{variables}; 
(ii) \textit{operators} correspond to the agent actions contained in the blocks in turquoise;
(iii) \textit{sequences}; 
(iv) \textit{repetitions} represented by the loop in the pink blocks;
(v) \textit{conditionals} described by the if statements in the blue blocks (only in the second activity);
(vi) \textit{functions}.
\item \textit{System resettability}: even if the problem solver cannot interact with the system, the system is resettable since the algorithm has to be correct or selected from a set.
\item \textit{System observability}: the system is not observable because the agent in question, Pac-Man, is an imaginary entity whose actions (i.e., moving) are not physically visible. The problem solver must rely on the instructions provided to understand the actions taken by the agent and cannot observe their actual outcome. 
\item \textit{Task cardinality}: the CTP has a one-to-one mapping, with one initial state, final state and algorithm.
\item \textit{Task explicitness}: all elements are given explicitly.
\item \textit{Task constraints}: the algorithm is constrained since the computational concepts addressed are already determined and limited to the specific notions presented.
\item \textit{Algorithm representation}: the algorithm is manifest and written in both activities. 
\end{itemize}


\subsection{Competencies}
\paragraph{Enabling features for competencies development}
\begin{itemize}[noitemsep,nolistsep]
\item \textit{Problem setting}: all competencies can be activated, thanks to the presence of variables, sequences, repetitions and functions in the tool functionalities.
The presence of many tool functionalities, the system's non-observability, the algorithm's constrained definition, and its written representation promote the development of all problem setting skills. The system resettability, the one-to-one cardinality and the explicit representation of elements support other competencies.

\item \textit{Algorithm}: the competencies associated with algorithmic concepts provided by the tool functionalities can be developed. 
The non-observability of the system, the explicit and constrained definition of the task elements, and the manifest written representation of the algorithm can further enhance these skills.

\item \textit{Assessment}: {all competencies can be developed}, due to the resettability of the system and the written algorithm; since the system can be reset, also the constraints on the algorithm can be checked and corrected, allowing for the development of constraint validation; similarly optimisation can be activated since the resetting capability is sufficient; generalisation can be developed thanks to the system's resettability and the presence of variables and functions. 
The tool functionalities available further encourage the development of these competencies.
\end{itemize}

\paragraph{Inhibiting features for competencies development}
\begin{itemize}[noitemsep,nolistsep]
\item \textit{Conditionals}: non-activable in the first activity since this functionality is not present in the visual blocks provided to the students.
\item \textit{Parallelism and events}: non-activable in both activities, as before, because they are not available in the visual blocks provided to the students. To activate these skills, the visual blocks must include the tools for creating parallelism and events in the algorithm.
\end{itemize}


\section{{Thymio Lawnmower Mission}}\label{sec:tlm}
\begin{figure*}[!hbt]
\centering
\includegraphics[width=\textwidth]{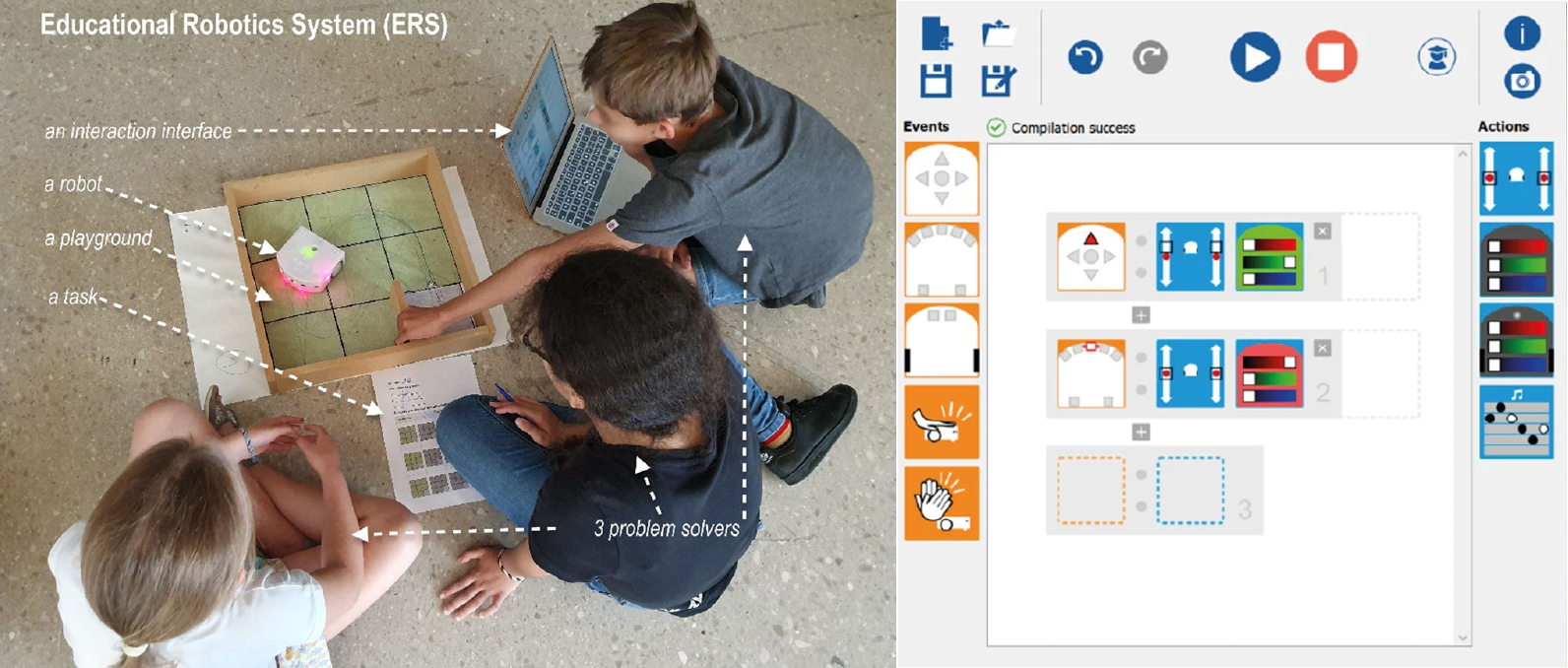}
\caption[TLM]{\textbf{The Thymio Lawnmower Mission adapted from \cite{Chevalier2020}.} A group of pupils must program the Thymio II robot to pass over all eight green lawn squares and avoid the fence (left). A special visual programming language platform, with graphical icons that are straightly interpretable, is used for this task (right).}
\label{fig:Thymio lawnmower}
\end{figure*}
\begin{figure*}[!t]
		\centering
		\includegraphics[width=\textwidth]{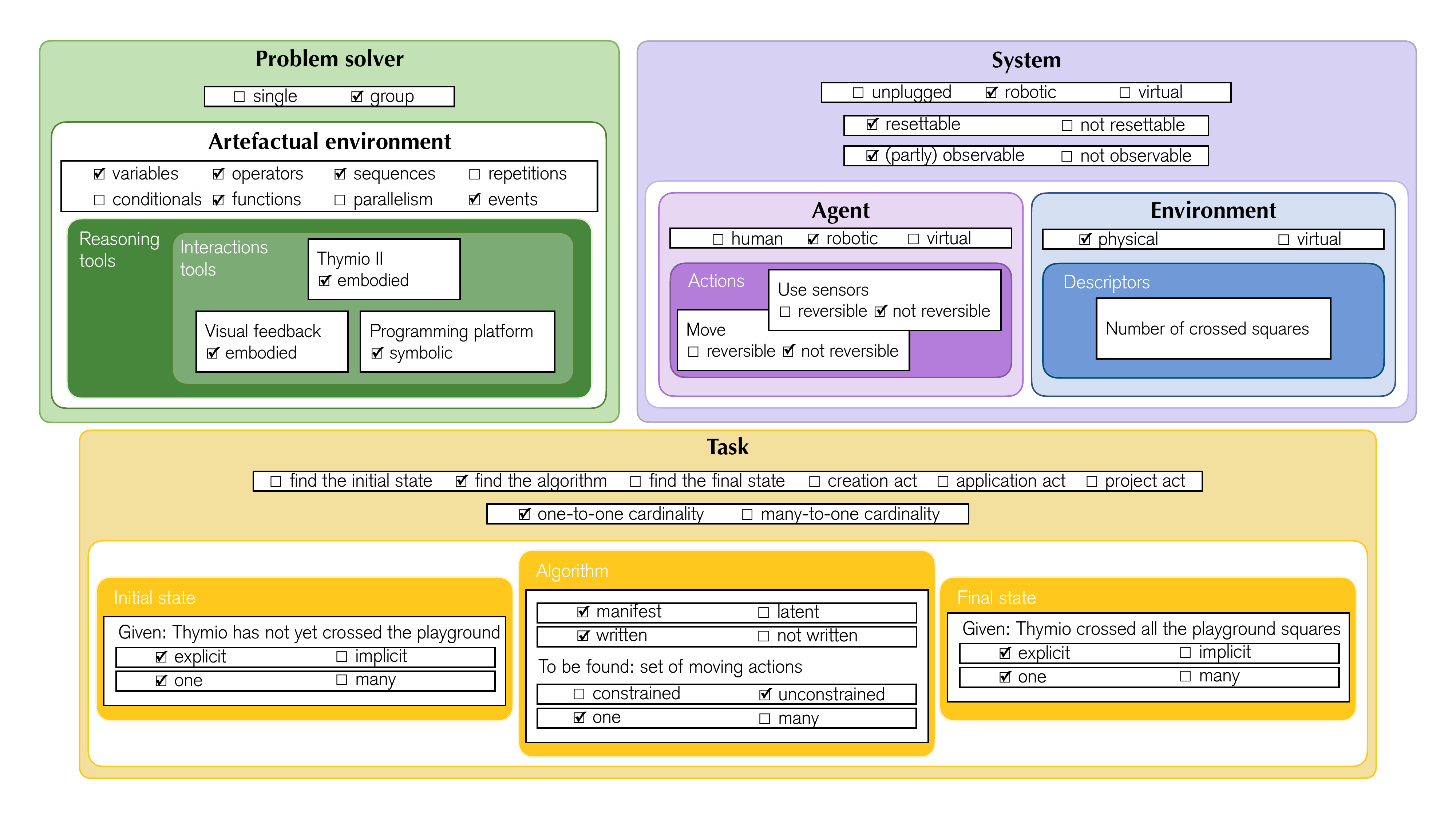}
		\caption{\textbf{Thymio Lawnmower Mission (control group) components and characteristics.}}
		\label{fig:TLMcontrol-features}
	\end{figure*}
\begin{figure*}[!h]
		\centering
            \includegraphics[width=\textwidth]{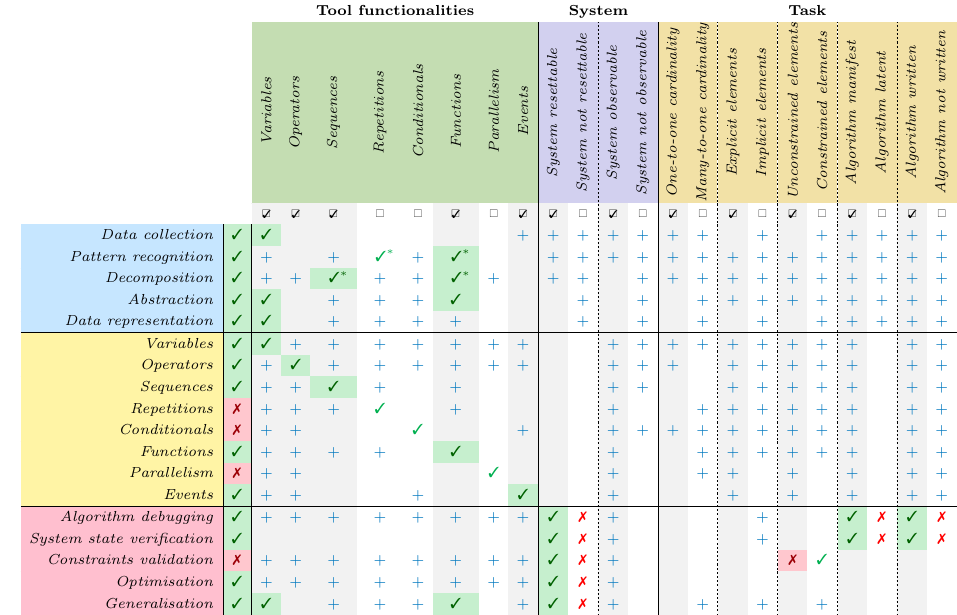}
		\caption{\textbf{Thymio Lawnmower Mission (control group) profile.}}
		\label{tab:TLMcontrol-mapping}
\end{figure*}
\begin{figure*}[!t]
		\centering
		\includegraphics[width=\textwidth]{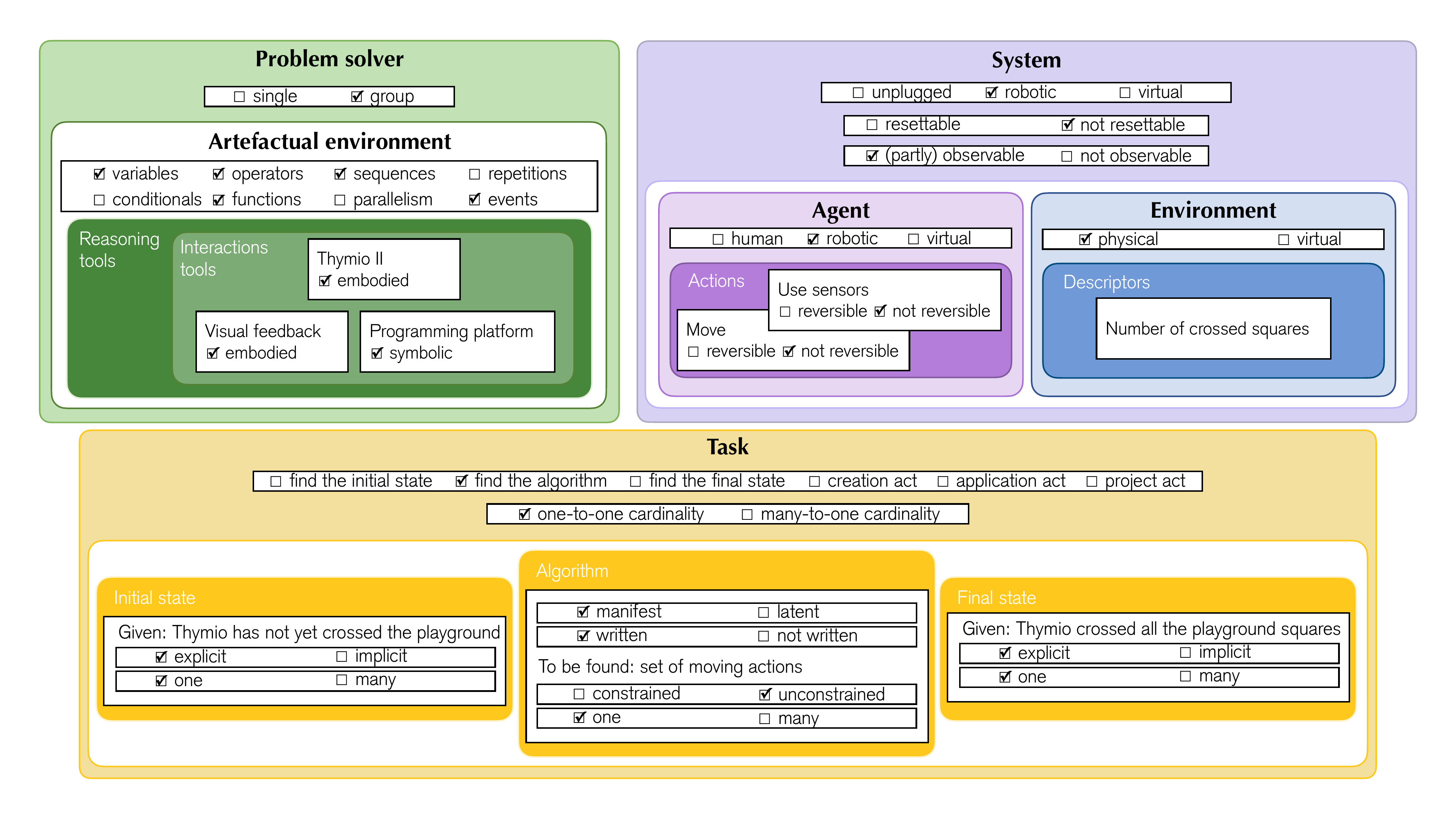}
		\caption{\textbf{Thymio Lawnmower Mission (test group) components and characteristics.}}
		\label{fig:TLMtest-features}
	\end{figure*}
\begin{figure*}[!h]
		\centering
            \includegraphics[width=\textwidth]{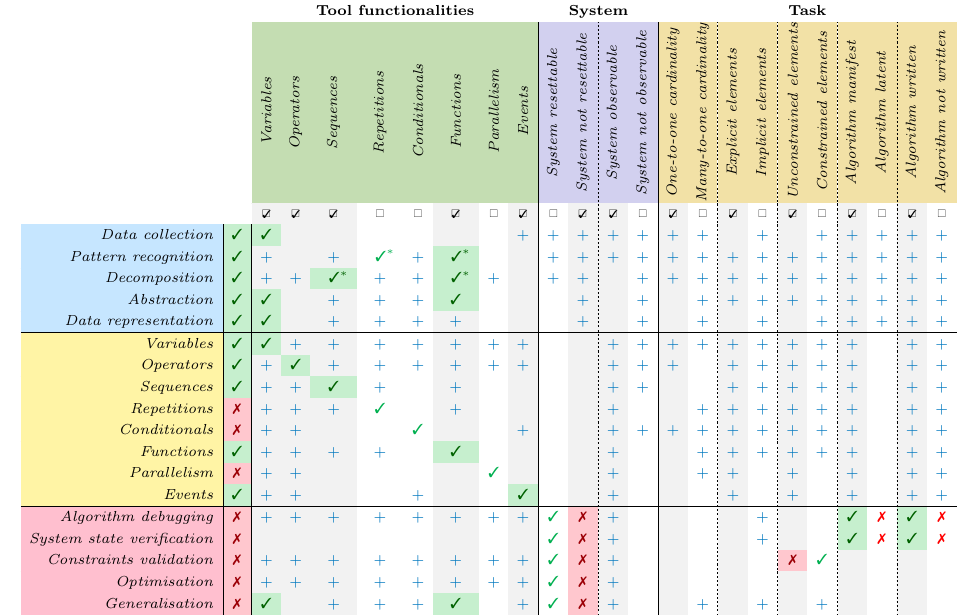}
		\caption{\textbf{Thymio Lawnmower Mission (test group)profile.}}
		\label{tab:TLMtest-mapping}
\end{figure*}

Thymio Lawnmower Mission is an educational robotics activity designed by \cite{Chevalier2020} to promote the development of students' CT skills through the Thymio II robot. 
The Thymio II, for instance, is a widely used educational robot equipped with various sensors, including proximity sensors, an accelerometer, a remote control receiver, motors, a speaker, and LEDs, distributed throughout its body \citep{riedo2013thymio,shin2014visual}.
{In this CTP, illustrated in} \Cref{fig:Thymio lawnmower}, {the Thymio II robot must systematically traverse all green lawn squares, much like a lawnmower would mow a lawn.
With this activity, the authors aimed to address the issue of students spending excessive time programming and not enough time problem-solving, referred to as the trial-and-error loop, by conducting an instructional intervention on two groups of primary school students. 
The control group was allowed to complete the task without any constraints. In contrast, the test group was subjected to an instructional intervention that blocked the programming interface at certain times to overcome the trial-and-error loop. 
Components and characteristics of this CTP have been analysed using the graphical templates in} \Cref{fig:TLMtest-features,fig:TLMcontrol-features}, {while the profiles, outlining the relationship between its characteristics and competencies, are illustrated in} \cref{tab:TLMcontrol-mapping,tab:TLMtest-mapping}.

\subsection{Components}
\begin{itemize}[noitemsep,nolistsep]
\item \textit{Problem solver}: the group of students performing the task who must program the agent's lawnmower behaviour. 
The artefactual environment comprises tools designed for reasoning and interaction, including a graphical programming environment called VPL (symbolic), which allows for the creation of sensor-action relationships to determine the robot's behaviour, the agent (embodied) and the visual feedback (embodied).
\item \textit{Agent}: the Thymio II, whose actions, {which are irreversible}, consist of moving around the playground by changing velocity and orientation and using sensors to detect obstacles. 
\item \textit{Environment}: the playground, i.e., a lawn area surrounded by a fence, divided into eight green squares and one grey square, the garage. Its state is defined by the grass condition or the number of squares passed over by the agent. 
\item \textit{Task}: find the algorithm. The initial state is the lawn with tall grass, meaning the robot is not passed over any of the squares that compose it.
The final state is the same lawn with the grass mowed, meaning the robot passes over all green squares.
The algorithm is the set of moving actions to reach the system's final state from the initial.
\end{itemize}

\subsection{Characteristics}
\begin{itemize}[noitemsep,nolistsep]
\item \textit{Tool functionalities}: the system provides a comprehensive set of tools for the problem solver to create and control the behaviour of the agent to solve the task, including (i) \textit{variables} such as the values of sensors or the state of the robot; (ii) \textit{operators} represent the basic actions that the agent can perform; (iii) \textit{sequences} are not represented by a specific block in the VPL interface, but can be created by arranging blocks in a specific order; 
(iv) \textit{functions} are not represented by blocks in the VPL but refer to the possibility of conceptually grouping blocks of code associated with a particular behaviour to produce outputs given inputs; 
(v) \textit{events} are directly represented in the graphical interface by sensor-action relationships, allowing the robot to perform actions in response to stimuli, such as detecting an obstacle.
\item \textit{System resettability}: in the control group, the problem solvers can reset the system directly by physically moving the Thymio II agent in the environment and restarting the task by repositioning it in the garage. On the other hand, those in the test group do not have this option as they cannot directly interact with the agent and cannot modify the algorithm since it has to be first written and then executed.
\item \textit{System observability}: the platform provides real-time visual feedback, making the system observable.
\item \textit{Task cardinality}: the task has a one-to-one mapping, with an initial and final state and an algorithm.
\item \textit{Task explicitness}: the elements of the task are given explicitly, as the student is provided with clear instructions on what the outcome should look like.
\item \textit{Task constraints}: the algorithm is unconstrained.
\item \textit{Algorithm representation}: the algorithm is written in the workspace and expressed by the set of blocks and their connections. 
\end{itemize}

\subsection{Competencies}

\paragraph{Enabling features for competencies development}
\begin{itemize}[noitemsep,nolistsep]
\item \textit{Problem setting}: all competencies can be activated thanks to the presence of variables, sequences and functions in the tool functionalities.
The presence of many tool functionalities positively affects and boosts problem setting skills.
The manifest and written representation of the algorithm can further encourage the development of these skills.
In the control group, being the system resettable, data collection, pattern recognition and decomposition are promoted, while in the test group, abstraction and data representation are also encouraged.
The one-to-one cardinality of data elements also facilitates data collection, pattern recognition and decomposition. 
The system observability also supports data collection and pattern recognition. 

\item \textit{Algorithm}: all competencies associated with the algorithmic concepts enabled by the tool functionalities, meaning variables, operators, sequences, functions and events, can be activated and promote one another.
These are further enhanced by the manifest and written algorithm representation, system observability, and the explicit and unconstrained definition of the task elements. The one-to-one cardinality helps to develop variables and operators further.

\item \textit{Assessment}: regarding the control group, {a part form constraint validation, all assessment competencies can be developed, like algorithm debugging and system state verification, since the system is resettable and the algorithm is manifest and written;} optimisation can be developed thanks to the resettability of the system; generalisation can be activated through the system's resettability and the presence of variables and functions. 
The tool functionalities available and the system observability help develop these skills as well. 
On the other hand, the system is non-resettable in the test group, and no assessment skills can be developed.
\end{itemize}

\paragraph{Inhibiting features for competencies development}
\begin{itemize}[noitemsep,nolistsep]
\item \textit{Repetitions, conditionals and parallelism}: non-activable in both control and test groups, as these functionalities are unavailable in the VPL programming platform. Therefore, to develop these skills, it is necessary to change the programming language to a textual programming language such as ASEBA.
\item \textit{Algorithm debugging}: non-activable in the test group since the system is not resettable.
\item \textit{System state verification}: non-activable {in the test group} because the system is not resettable. 
\item \textit{Constraint validation}: non-activable, in both control and test groups, due to the absence of constraints on the algorithm. 
\item \textit{Optimisation}: non-activable in the test group because the system is not resettable.
\item \textit{Generalisation}: non-activable in the test group due to the non-resettability of the system.
\end{itemize}



\section{{Remote Rescue with Thymio II (R2T2)}}\label{sec:r2t2}
\begin{figure*}[!htb]
\centering
\includegraphics[width=\textwidth]{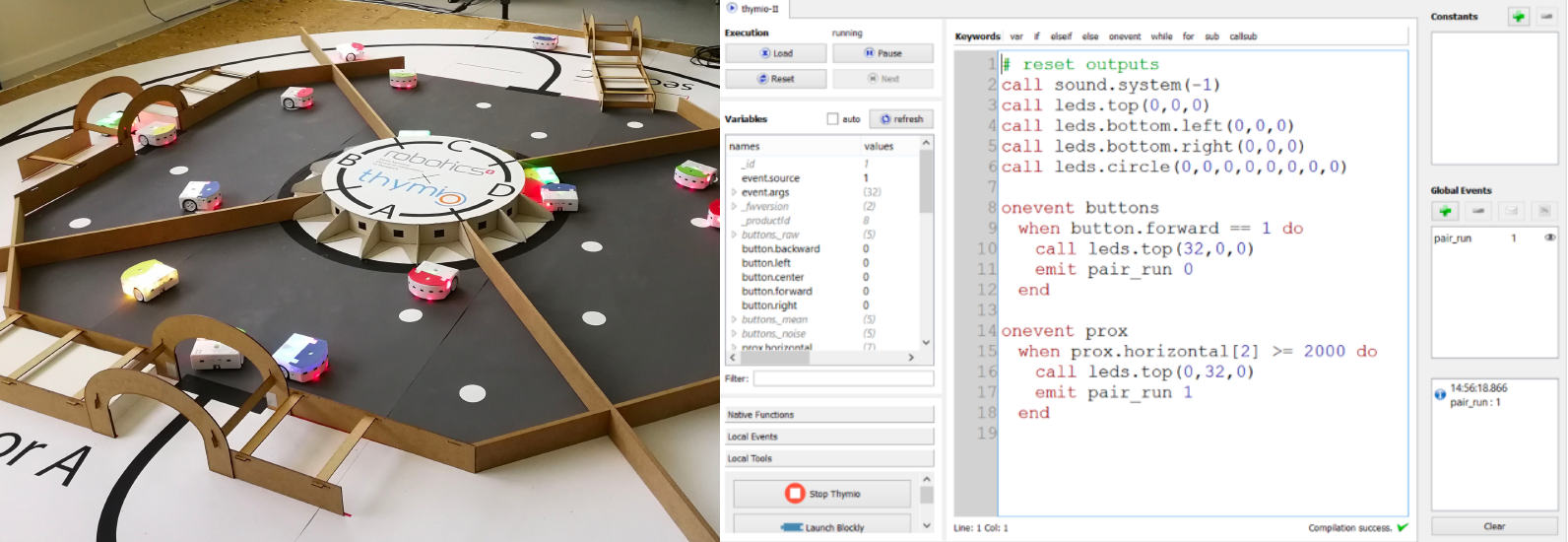}
\caption[R2T2 mission]{\textbf{The Remote Rescue with Thymio II (R2T2) mission on Mars adapted from \cite{mondada2016r2t2}.} 
Sixteen worldwide teams of pupils collaborate with 16 Thymio to restart the main generator of a simulated damaged power Mars station (left) in five phases using a visual programming language or textual programming language programming platforms (right).}
\label{fig:r2t2}
\end{figure*}
\begin{figure*}[!h]
		\centering
		\includegraphics[width=\textwidth]{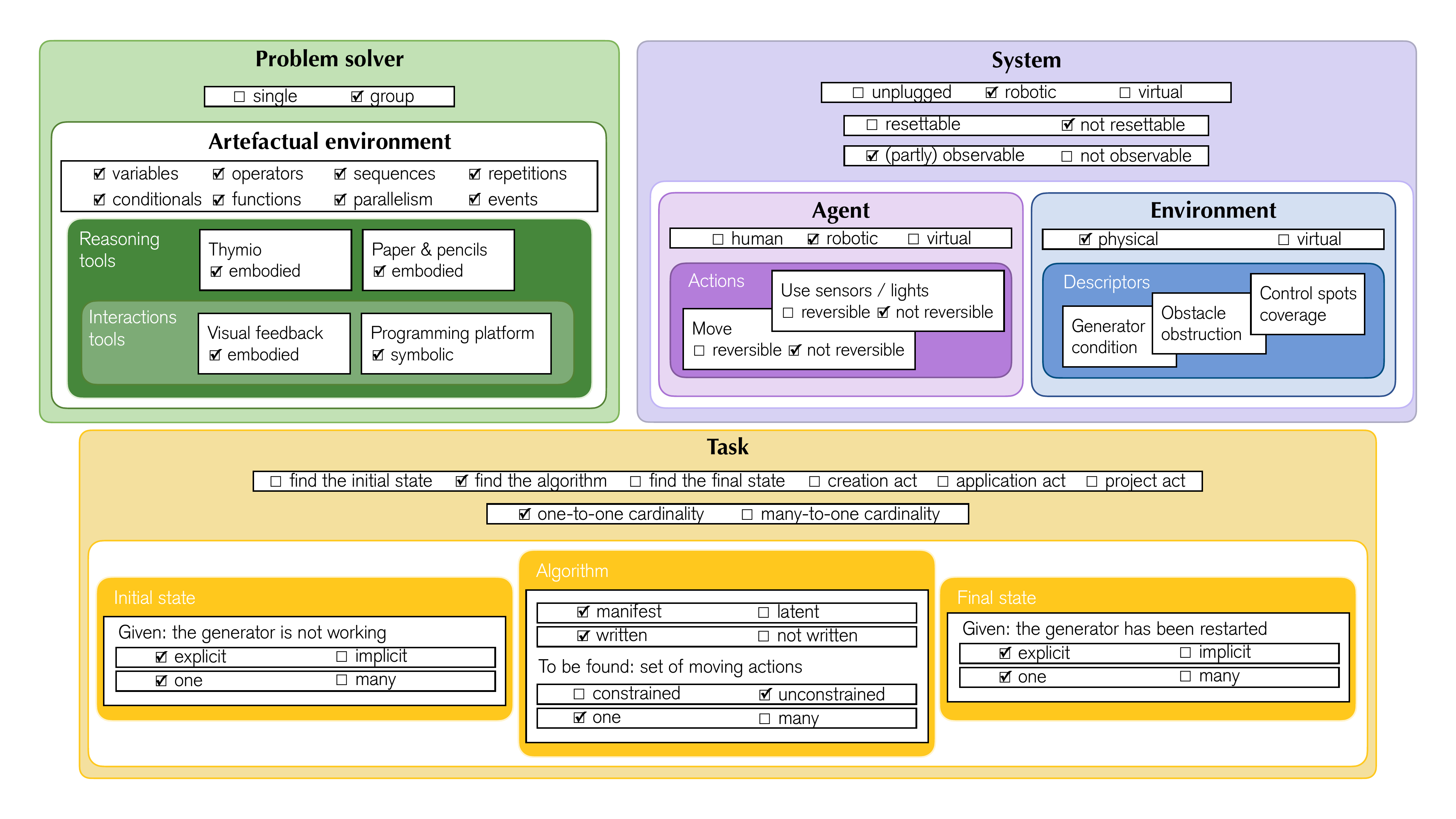}
		\caption{\textbf{R2T2 mission components and characteristics.}}
		\label{fig:R2T2-features}
	\end{figure*}
\begin{figure*}[!h]
		\centering
		\includegraphics[width=\textwidth]{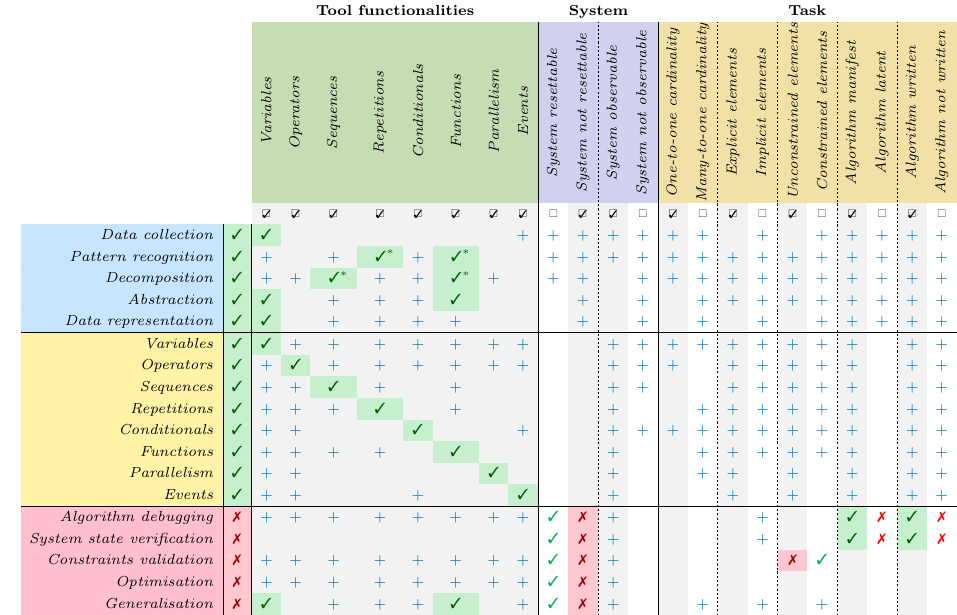}
		\caption{\textbf{R2T2 mission profile.}}
		\label{tab:R2T2-mapping}
	\end{figure*}

Remote Rescue with Thymio II (R2T2) is another collaborative educational robotics activity, presented by \cite{mondada2016r2t2} to promote STEM education in schools and encourage students towards careers in these fields. 
{This CTP activity, illustrated in} \Cref{fig:r2t2}, {is a rescue operation on a Mars station, whose goal is to assess the damage of the power plant and restart the main generator by remotely controlling 16 Thymio II robot.
The activity is divided into five phases, each with a specific objective. 
In the first phase, the robots must enter the station and push away an obstacle blocking the main door. 
In the second phase, the robots must stand on control spots to activate access to the generator. 
In the third phase, the robots must look into the generator through a small window. 
In the fourth phase, the robots must turn on a light when detecting the generator rotor using proximity sensors and off when it is no longer visible, thus estimating the generator speed. 
In the final phase, the generator is restarted, and the mission is  completed.}
{Components and characteristics of this CTP have been analysed using the graphical template in} \Cref{fig:R2T2-features}, {while the profile, outlining the relationship between its characteristics and competencies, is illustrated in} \cref{tab:R2T2-mapping}.


\subsection{Components}
\begin{itemize}[noitemsep,nolistsep]
\item \textit{Problem solver}: the group of students performing the task who must program the agents' behaviours to restart the main generator. 
The artefactual environment comprises tools designed for reasoning, such as paper and pencils and another robot that is physically accessible. Tools available also to interact with the system include the programming platform (symbolic) with the two available programming environments, VPL and ASEBA, a textual programming language \citep{magnenat2011aseba}, and the five webcams, installed around the playground, provide a delayed continuous visual feedback (embodied) through YouTube video streams.
\item \textit{Agent}: the 16 remote-controlled Thymio II, which can move around the playground accelerating and rotating, use proximity sensors and turn some lights on and off.
All actions are considered irreversible.
\item \textit{Environment}: the playground, i.e., the Mars station, characterised by different descriptors used in the different mission stages, such as the obstruction by the obstacle, covering of the control spots and finally, the restart of the generator.
\item \textit{Task}: find the algorithm. In the initial state, the generator is not working, while it has been restarted at the end.
The algorithm is the set of moving actions to reach the system's final state from the initial.
\end{itemize}


\subsection{Characteristics}
\begin{itemize}[noitemsep,nolistsep]
\item \textit{Tool functionalities}: the system provides a comprehensive set of tools for the problem solver to create and control the agent's behaviour to solve the task, depending on the programming environment used. VPL offers the possibility to use \textit{variables}, \textit{operators}, \textit{sequences}, \textit{functions} and \textit{events}. Additionally, ASEBA offers control flows such as \textit{repetitions} and \textit{conditionals}. Furthermore, \textit{parallelism} is possible since it refers to the ability to run multiple processes simultaneously, in this case, the Thymio II robots performing the rescue operation in parallel. The agents can execute their tasks concurrently without waiting for each other to complete them.
\item \textit{System resettability}: the system cannot be reset due to the irreversible nature of the actions carried out by the robots. Once the robots take an action, the change in the system's status is permanent and cannot be undone. Furthermore, the physical separation between the problem solvers and the system means no immediate way to reset the system.
\item \textit{System observability}: the delayed but continued system visual feedback makes it totally observable. 
\item \textit{Task cardinality}: the task has a one-to-one mapping, with an initial and final state and an algorithm.
\item \textit{Task explicitness}: all elements are given explicitly.
\item \textit{Task constraints}: the algorithm is unconstrained.
\item \textit{Algorithm representation}: the algorithm is manifest and written in the programming platform. 
\end{itemize}

 
\subsection{Competencies}

\paragraph{Enabling features for competencies development}
\begin{itemize}[noitemsep,nolistsep]
\item \textit{Problem setting}: all competencies can be activated thanks to the presence of variables, sequences and functions in the tool functionalities. The presence of all tool functionalities, the manifest written representation of the algorithm and the non-resettability of the system further encourage the development of these skills.
The system observability supports data collection and pattern recognition. The one-to-one cardinality, in addition to these skills, stimulates decomposition. 
The explicit and unconstrained definition of the task elements also promotes pattern recognition, decomposition and abstraction.

\item \textit{Algorithm}: competencies can be triggered since the tool's functionalities enable all algorithmic concepts.
The system observability, the explicit and unconstrained definition of the task elements, and the manifest written algorithm representation further enhance these. 

\item \textit{Assessment}: since the system is not resettable, there are no assessment skills activable.
\end{itemize}

\paragraph{Inhibiting features for competencies development}
The inability to reset the system restricts the activation of all assessment skills. 
The task can be adjusted by incorporating a mechanism for resetting the system to a previous state, enabling the problem solver to correct errors made during the implementation of the algorithm, explore different solutions, and learn from their mistakes. 
Additionally, it is necessary to omit the initial or final states to verify the system's state. Furthermore, to develop the constraint validation skill, constraints must be incorporated into the algorithm.

\begingroup
\let\clearpage\relax
\FloatBarrier
\mbox{ ~ }
\endgroup

\section{{Ozobot maze}}\label{sec:ozobot}
\begin{figure}[!h]
\centering
\includegraphics[width=\columnwidth]{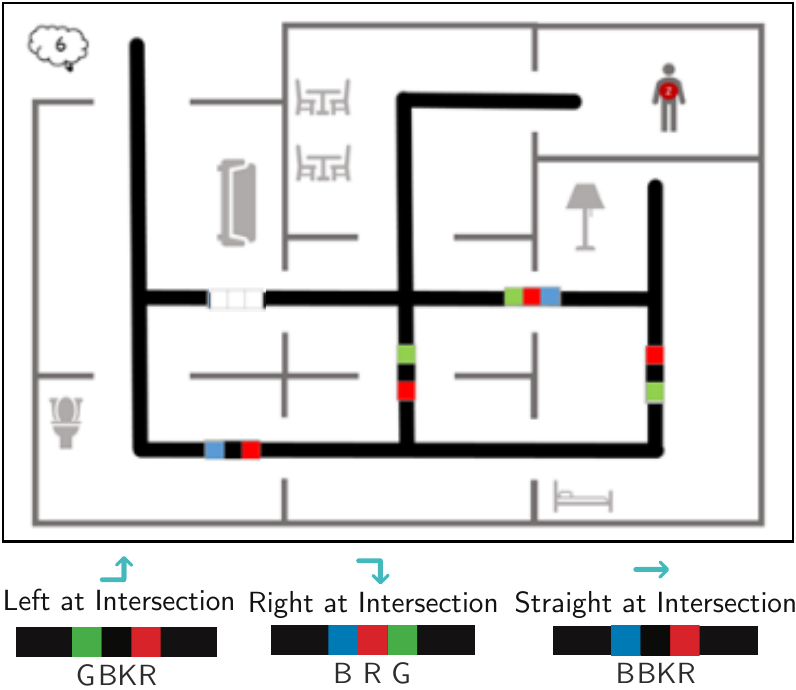}
\caption[Ozobot]{\textbf{The Ozobot maze adapted from \cite{Bryndova}.} The task requires the pupil to instruct the Ozobot to cross a maze avoiding obstacles and reaching the room where the red person is. Commands such as increasing the speed, changing direction and making some cool movements (spinning like a tornado) are given in Color Codes.}
\label{fig:Ozobot}

\end{figure}

\begin{figure*}[!t]
		\centering
		\includegraphics[width=\textwidth]{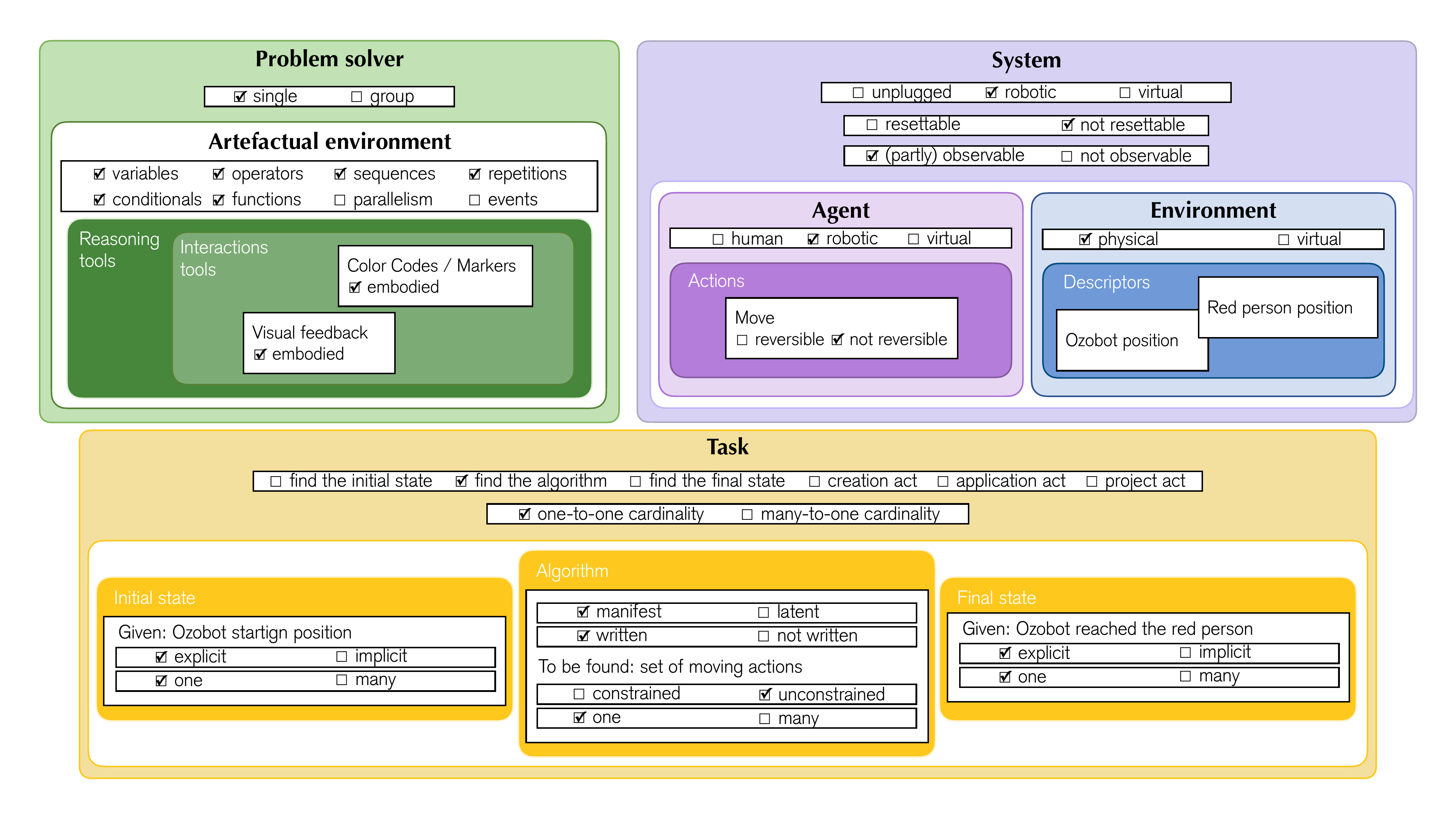}
		\caption{\textbf{Ozobot maze components and characteristics.}}
		\label{fig:ozobot-features}
	\end{figure*}

\begin{figure*}[!h]
		\centering
		\includegraphics[width=\textwidth]{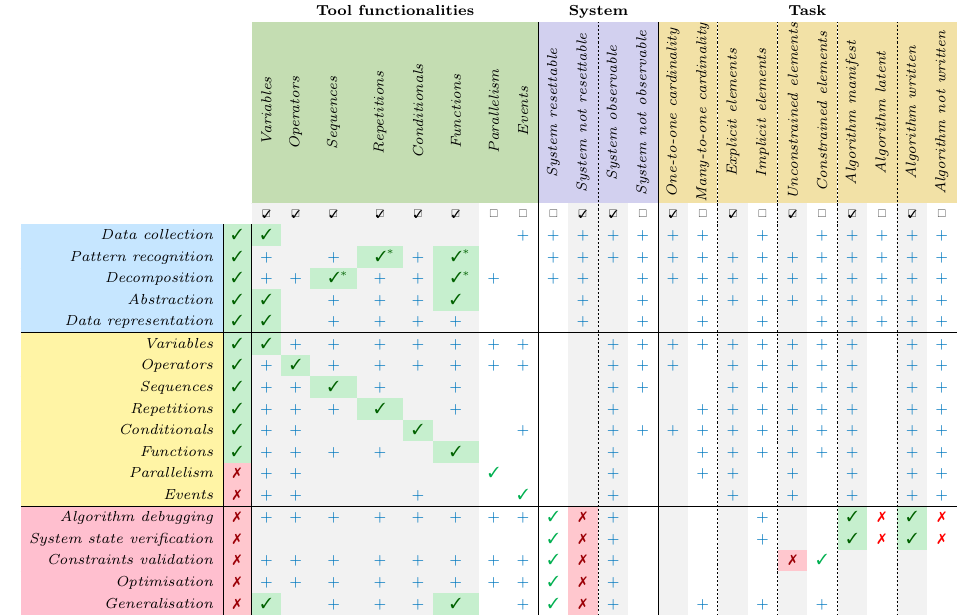}
		\caption{\textbf{Ozobot maze profile.}}
		\label{tab:ozobot-mapping}
	\end{figure*}

\begingroup
\let\clearpage\relax
\mbox{ ~ }
\endgroup
The Ozobot Maze activity is a screenless robotics task proposed by \cite{Bryndova} aimed at teaching primary school students in the Czech Republic CT skills. 
The educational robot used in this task is the Ozobot, a small programmable robot, used to introduce students to coding, equipped with sensors to follow black lines and read colour patterns called Color Codes to change speed, direction and movements.
{In this CTP activity, illustrated in} \Cref{fig:Ozobot}, {the robot should be guided through a maze to reach the room where the red person is.
Components and characteristics of this CTP have been analysed using the graphical template in} \Cref{fig:ozobot-features}, {while the profile, outlining the relationship between its characteristics and competencies, is illustrated in} \cref{tab:ozobot-mapping}.

\subsection{Components}

\begin{itemize}[noitemsep,nolistsep]
\item \textit{Problem solver}: the student who creates a suitable sequence of instructions using Color Codes to guide the Ozobot through the maze.
The artefactual environment comprises tools for reasoning and interacting with the system. Predefined stickers or markers to fill the empty Codes with colour sequences (embodied) are used to give the robot the correct instructions to achieve the goal. The visual feedback (embodied) lets the problem solver observe the agent and its movements in the playground.
\item \textit{Agent}: the Ozobot agent, which can move around in the playground by changing velocity and orientation. This action is not reversible.
\item \textit{Environment}: the playground, i.e., the house map, whose state is defined by the agent's position relative to the red person.
\item \textit{Task}: find the algorithm. The initial state is the empty maze with the Ozobot positioned near the starting point. The system's final state is the Ozobot reaching the end of the maze, in the room with the red person, and all the Color Codes being filled.
The algorithm is the set of agent instructions, shown by the Color Codes, to reach the system's final state from the initial.
\end{itemize}

\subsection{Characteristics}
\begin{itemize}[noitemsep,nolistsep]
\item \textit{Tool functionalities}: the system provides a comprehensive set of tools for the problem solver to create and control the behaviour of the agent to solve the task, including 
(i) \textit{variables} can be used to store values such as the position of the robot in the maze; 
(ii) \textit{operators} are basic actions that the agent can perform, represented by Direction Codes such as moving straight, turning left or right;
(iii) \textit{sequences} represent the set of instructions used to control the behaviour of the robot in a step-by-step manner and that the Ozobot must follow to complete the task; 
(iv) \textit{repetitions} are a way of repeating the same instructions multiple times and refer to the possibility of repeating the same Color Code multiple times, for example, if the agent encounters the same type of intersection repeatedly in the path and the same Color Code is used to specify the direction the agent should take;
(v) \textit{conditionals} are used to make decisions based on certain conditions, for example, understanding what to do at an intersection;
(vi) \textit{functions} can be reflected in the different types of Color Codes that can be reused in different situations and map inputs to outputs, such as mapping a specific type of intersection to a specific direction.
\item \textit{System resettability}: the system is not resettable since it is impossible to change the Color Codes once they have been filled in.
\item \textit{System observability}: the real-time visual feedback makes the system observable.
\item \textit{Task cardinality}: the task has a one-to-one mapping.
\item \textit{Task explicitness}: the elements of the task are given explicitly, as the student is provided with clear instructions on what the outcome should look like.
\item \textit{Task constraints}: the algorithm is unconstrained.
\item \textit{Algorithm representation}: the algorithm is manifest and written, expressed by the set of Color Codes. 
\end{itemize}

\subsection{Competencies}

\paragraph{Enabling features for competencies development}
\begin{itemize}[noitemsep,nolistsep]
\item \textit{Problem setting}: all competencies can be activated thanks to tool functionalities such as variables, sequences and functions.
The manifest and written algorithm representation, as well as the non-resettability of the system, can further encourage the development of these skills.
The system observability supports data collection and pattern recognition. The one-to-one cardinality, in addition, stimulates decomposition. 
The explicit and unconstrained definition of the task elements promotes pattern recognition, decomposition and abstraction. 

\item \textit{Algorithm}: all competencies associated with the algorithmic concepts enabled by the tool functionalities, meaning variables, operators, sequences, repetitions, conditionals and functions, can be activated and promote one another.
The manifest and written representation of the algorithm, the system observability, and the explicit and unconstrained definition of the task elements further enhance these. The one-to-one cardinality helps to enhance some of these skills as well.

\item \textit{Assessment}: since the system is not resettable, there are no assessment skills activable.
\end{itemize}

\paragraph{Inhibiting features for competencies development}
\begin{itemize}[noitemsep,nolistsep]
\item \textit{Parallelism and events}: non-activable since the related features are missing in the tool functionalities. To develop these skills, it is possible to switch to a different interaction tool, such as OzoBlockly, a visual programming language designed to code Ozobots Evo and includes these functionalities.

\item \textit{Assessment skills}: The non-resettable feature of the system hinders the development of assessment abilities. In this sense, by allowing the problem solver to reset the system to a previous state, for example, by letting them change the Color Codes stickers and move the robot back to the starting position, the activity can be improved. This way, the problem solver can correct any mistakes made during the implementation of the algorithm, experiment with different solutions, and learn from their mistakes.
To develop system state verification, it is also essential to not reveal the initial or final states. 
Moreover, constraints should be imposed on the algorithm to develop constraint validation skills.
\end{itemize}

\section{{Mini-golf challenge with micro:bit}}\label{sec:minigolf}
\begin{figure}[!htb]
\centering
\includegraphics[width=\columnwidth]{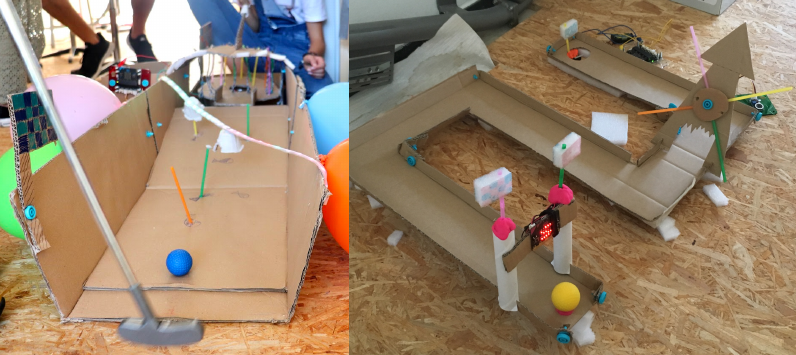}
\caption[Mini-golf]{\textbf{The Mini-golf challenge challenge adapted from \cite{assaf2021imake}.}
The task requires a group of pupils to define the behaviour of the mini-golf lane movable obstacles, sounds, and lights by programming the BBC micro:bit.}
\label{fig:minigolf}
\end{figure}

\begin{figure*}[!t]
		\centering
		\includegraphics[width=\textwidth]{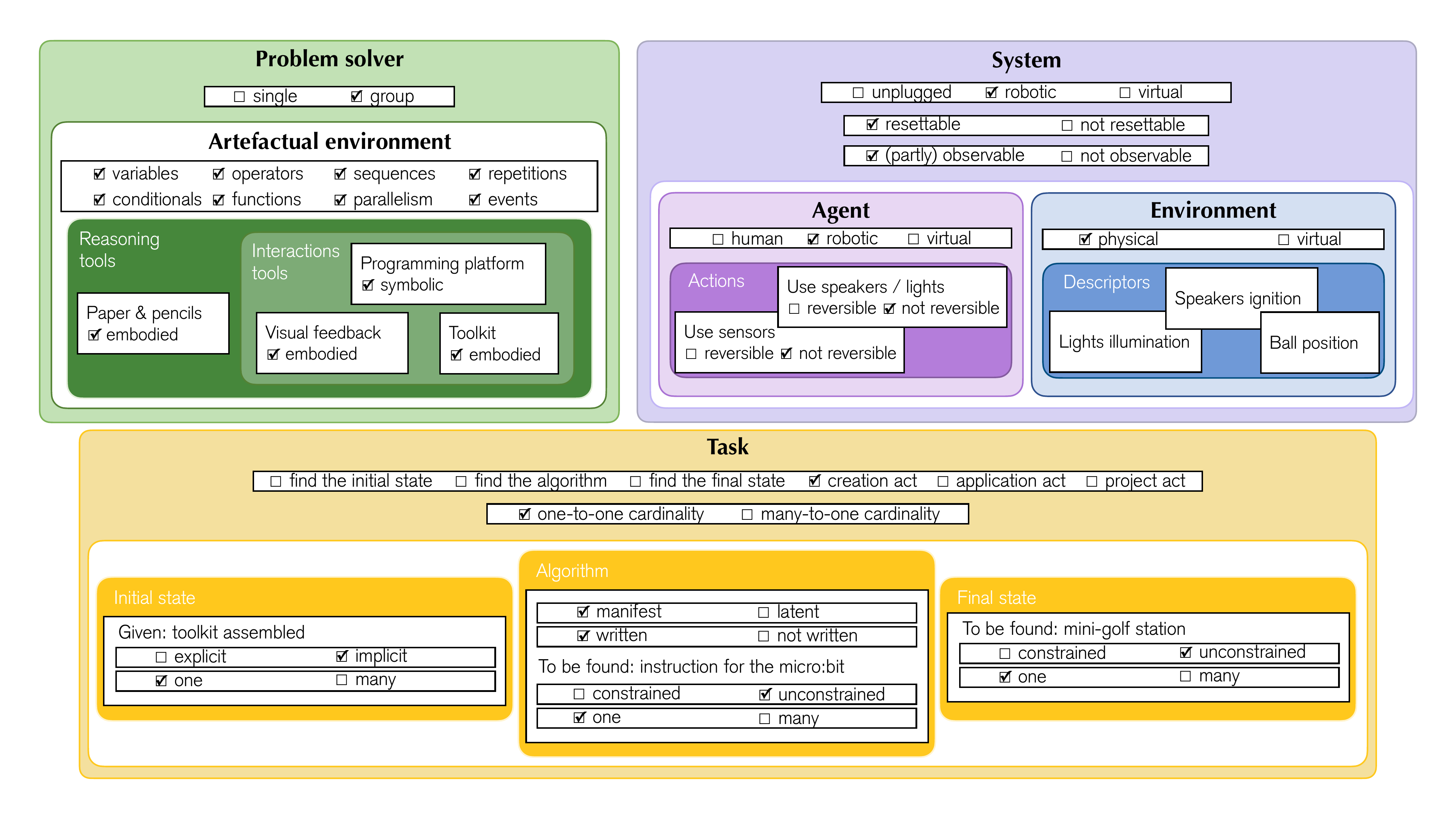}
		\caption{\textbf{Mini-golf challenge components and characteristics.}}
		\label{fig:minigolf-features}
	\end{figure*}

\begin{figure*}[!h]
		\centering
		\includegraphics[width=\textwidth]{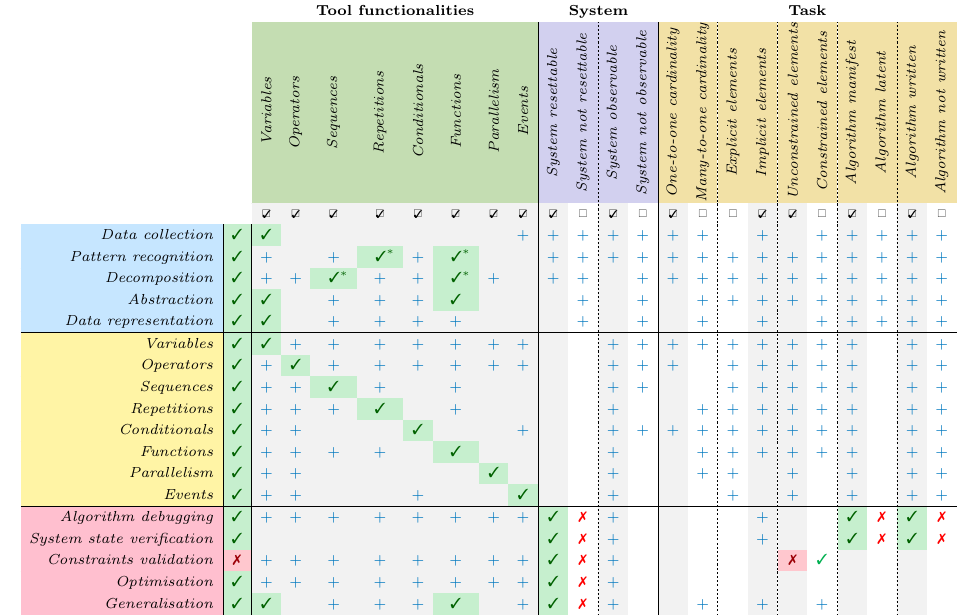}
		\caption{\textbf{Mini-golf challenge profile.}}
		\label{tab:minigolf-mapping}
	\end{figure*}

In robotics, physical computing activities involve using microcontrollers, sensors, and other electronic components to build and program interactive systems. 
To enhance the learning experience, various off-the-shelf robotic kits have been developed that allow students to construct robots easily and control them through a graphical user interface.
In these activities, students are often engaged in an initial phase of actively constructing the system using recycled materials, electronic circuits, and robot programming.
These activities evaluate the students' understanding of algorithmic concepts, problem-solving skills, knowledge of physics and engineering, creativity, and ability to work collaboratively.
{The Mini-golf challenge, illustrated in} \Cref{fig:minigolf}, {is a CTP with the objective is to program a mini-golf lane's moving and interactive elements.
Components and characteristics of this CTP have been analysed using the graphical template in} \Cref{fig:minigolf-features}, {while the profile, outlining the relationship between its characteristics and competencies, is illustrated in} \cref{tab:minigolf-mapping}.

\begin{figure}[!htb]
\centering
\includegraphics[width=\columnwidth]{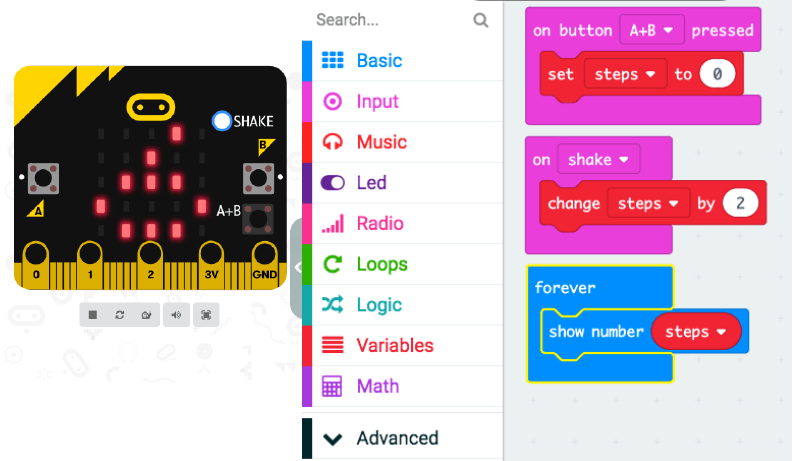}
\caption[Micro:bit]{\textbf{The BBC micro:bit (left) and its block programming interface (right).}}
\label{fig:microbit}
\end{figure}
One such physical computing activity is the Mini-golf challenge, proposed by \cite{assaf2021imake}.
In this activity, students are tasked with programming a mini-golf lane's moving and interactive elements using the BBC micro:bit \citep{ball2016microsoft,microbit2016web}.
The micro:bit, depicted in \Cref{fig:microbit}, is a pocket-sized computer that can be programmed using Microsoft's MakeCode editor \citep{makecode2016web}, which provides a user-friendly interface with colour-coded blocks similar to Scratch and the ability to switch to JavaScript to view the text-based code.

\subsection{Components}
\begin{itemize}[noitemsep,nolistsep]
\item \textit{Problem solver}: the group of students who must program the micro:bit.
The artefactual environment disposed of paper and pencil (embodied), a cognitive tool to support the thinking phase. Other tools are provided to interact with the system, including the toolkit (embodied), whose components can be assembled and disassembled at will, the visual programming language offered by the MakeCode editor (symbolic), and the visual feedback (embodied).
\item \textit{Agent}: the micro:bit, which can use sensors, control the movement and actions of the mini-golf elements, and turn LEDs and speakers on and off. All actions are considered not reversible.
\item \textit{Environment}: the assembled toolkit, which consists of various components, including a ball, speakers, and lights. Its state is described by the state of its elements, including the ball position, lights illumination and speakers ignition.
\item \textit{Task}: creation act. The initial state is given by the toolkit assembled. 
The system's final state and the students' algorithm are open-ended, which defines the behaviour of mini-golf station elements.
\end{itemize}

\subsection{Characteristics}
\begin{itemize}[noitemsep,nolistsep]
\item \textit{Tool functionalities}: the MakeCode editor allows using all the tool functionalities we defined.
\item \textit{System resettability}: the system can be directly reset by physically moving the ball back to its starting position, and resetting the state of the lights and speakers, for example, by turning them off. Additionally, the MakeCode editor includes a convenient button to streamline the agent's reset process.
\item \textit{System observability}: the real-time visual feedback makes the system observable.
\item \textit{Task cardinality}: the task has a one-to-one mapping.
\item \textit{Task explicitness}: the initial state of the toolkit, including the ball's position, the state of the lights, and the speakers, is not specified.
\item \textit{Task constraints}: no constraints are imposed on the two elements to be found.    
\item \textit{Algorithm representation}: the algorithm is manifest and written in the programming platform. 
\end{itemize}


\subsection{Competencies}
\paragraph{Enabling features for competencies development}
\begin{itemize}[noitemsep,nolistsep]
\item \textit{Problem setting}: all competencies can be activated thanks to the presence of variables, sequences and functions in the tool functionalities.
The availability of numerous tool functionalities, the written algorithm representation and the non-resettability of the system encourage the development of problem setting skills.
The implicit description of the task elements creates an environment of uncertainty, allowing for multiple interpretations and solutions, stimulating problem setting skills.
The system's observability allows the development of data collection and pattern recognition, while the one-to-one cardinality also encourages decomposition. 
The unconstrained definition of task elements fosters pattern recognition, decomposition, and abstraction.

\item \textit{Algorithm}: all competencies can be activated since the tool functionalities enable all algorithmic concepts.
The system observability, the implicit and unconstrained definition of the task elements, and the algorithm's manifest written representation further enhance these. The one-to-one cardinality helps to enhance some algorithmic skills as well.

\item \textit{Assessment}: algorithm debugging and system state verification can be developed by the direct resettability of the system and the written representation of the algorithm; optimisation can be activated as the resettability of the system alone is sufficient; generalisation is enabled through the system's resettability and the presence of variables and functions. 
Tool functionalities, system observability and the implicit definition of the task elements further support their development.    
\end{itemize}

\paragraph{Inhibiting features for competencies development}
The constraint validation skill cannot be developed as the algorithm and final state to be found are unconstrained. 
To encourage the development of this competence, the activity can be adapted by introducing defined constraints, such as a maximum number of moves for the ball to reach the hole or the activation of certain elements of the mini-golf station in a specific order. These constraints will require students to evaluate the feasibility of their solutions within the specified limitations and assess their adherence to the established criteria.

\section{{Classic Maze}}\label{sec:classicmaze}

\begin{figure*}[!t]
    \centering
    \includegraphics[width=.8\textwidth]{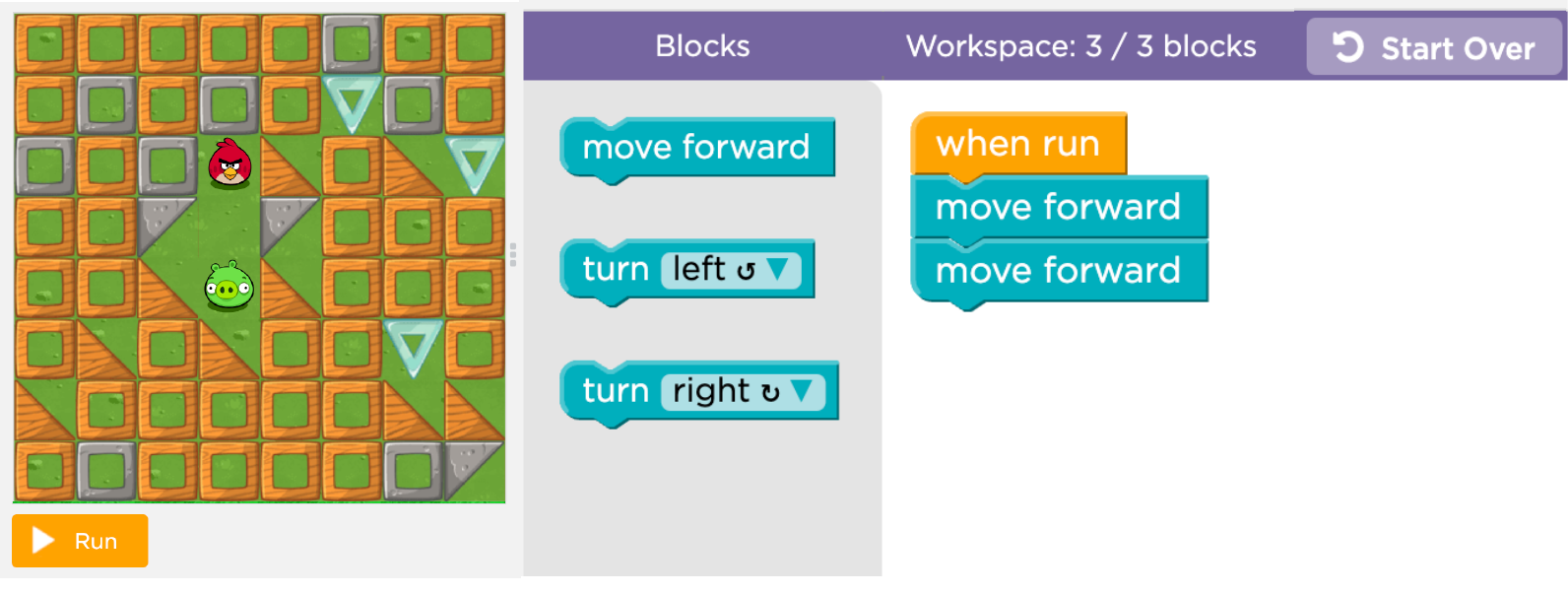}
    \caption[AB]{\textbf{The Angry Bird hitting the Green Pig maze adapted from \cite{AB}.} The problem solver must write a program to get the Angry Bird through the maze to hit the Green Pig (left) by selecting the instruction blocks (middle) and assembling them in the workspace (right).}
    \label{fig:AB}
\end{figure*}
\begin{figure*}[!t]
    \centering
    \includegraphics[width=\textwidth]{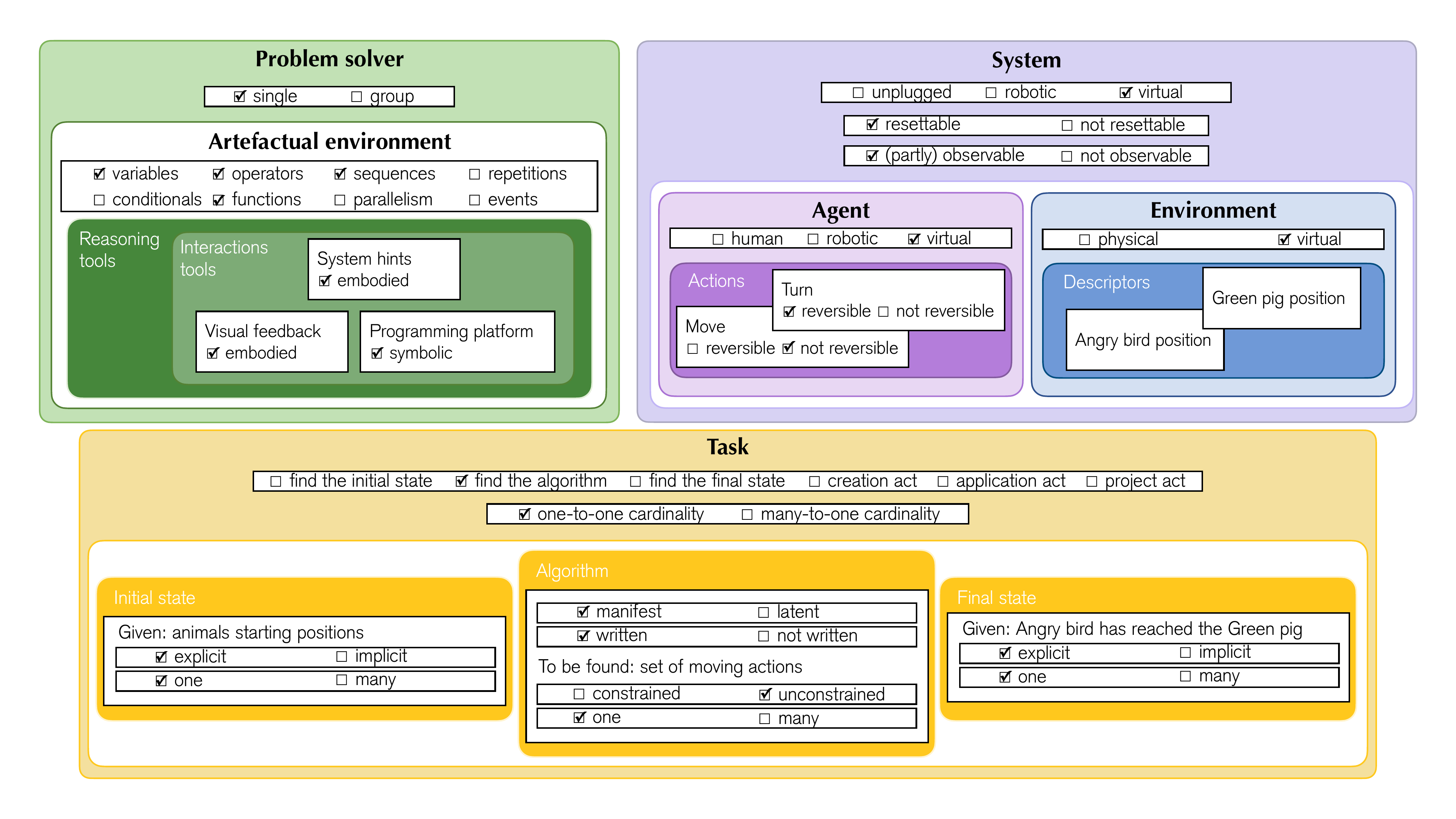}
    \caption{\textbf{Angry Bird maze components and characteristics.}}
    \label{fig:ABHGP-features}
\end{figure*}

\begin{figure*}[!t]
    \centering
    \includegraphics[width=.8\textwidth]{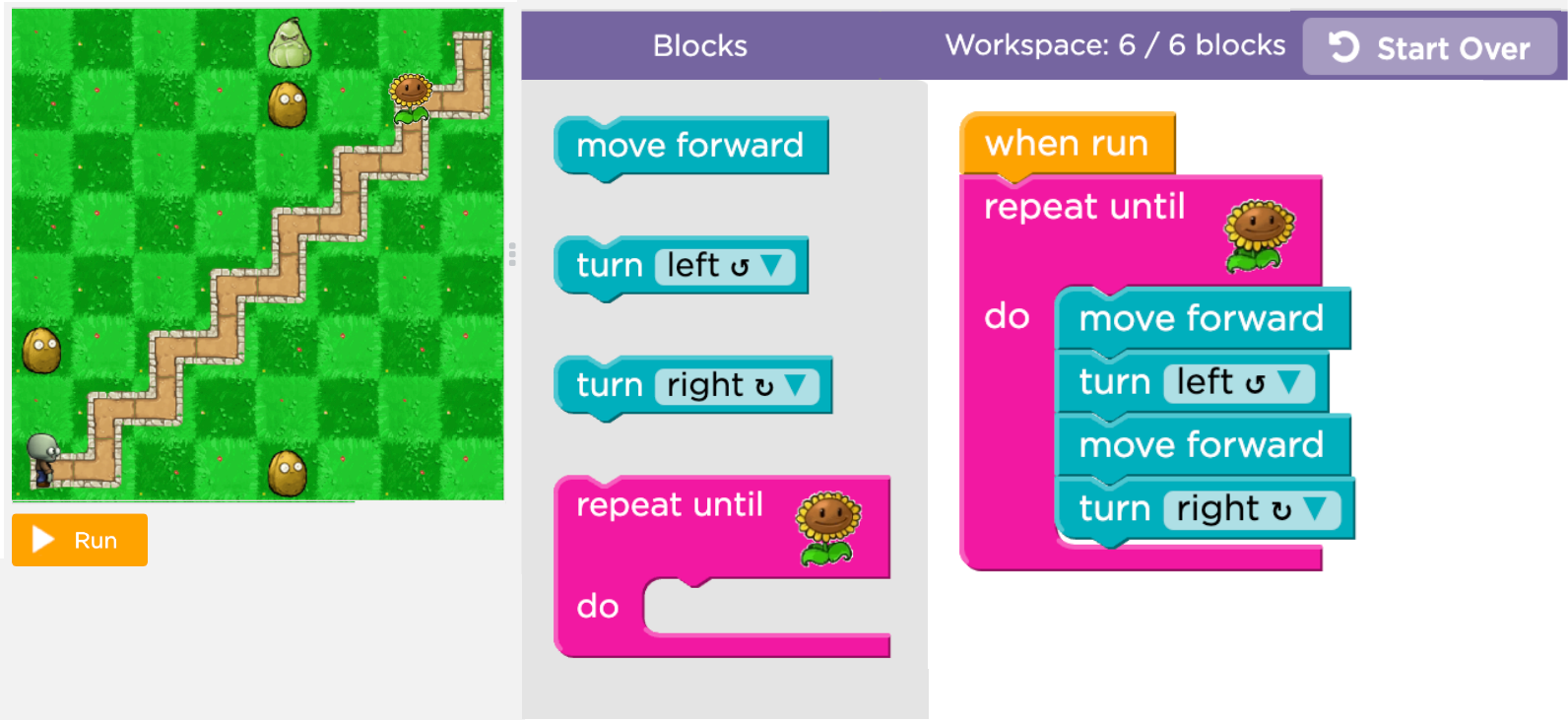}
    \caption[PZ]{\textbf{The Plants vs Zombies maze adapted from \cite{PZ}.} The problem solver must write a program to get the Zombie through a maze to eat the plant (left) by selecting the instruction blocks (middle) to be assembled in the workspace (right).}
    \label{fig:PZ}
\end{figure*}

\begin{figure*}[!h]
    \centering
    \includegraphics[width=\textwidth]{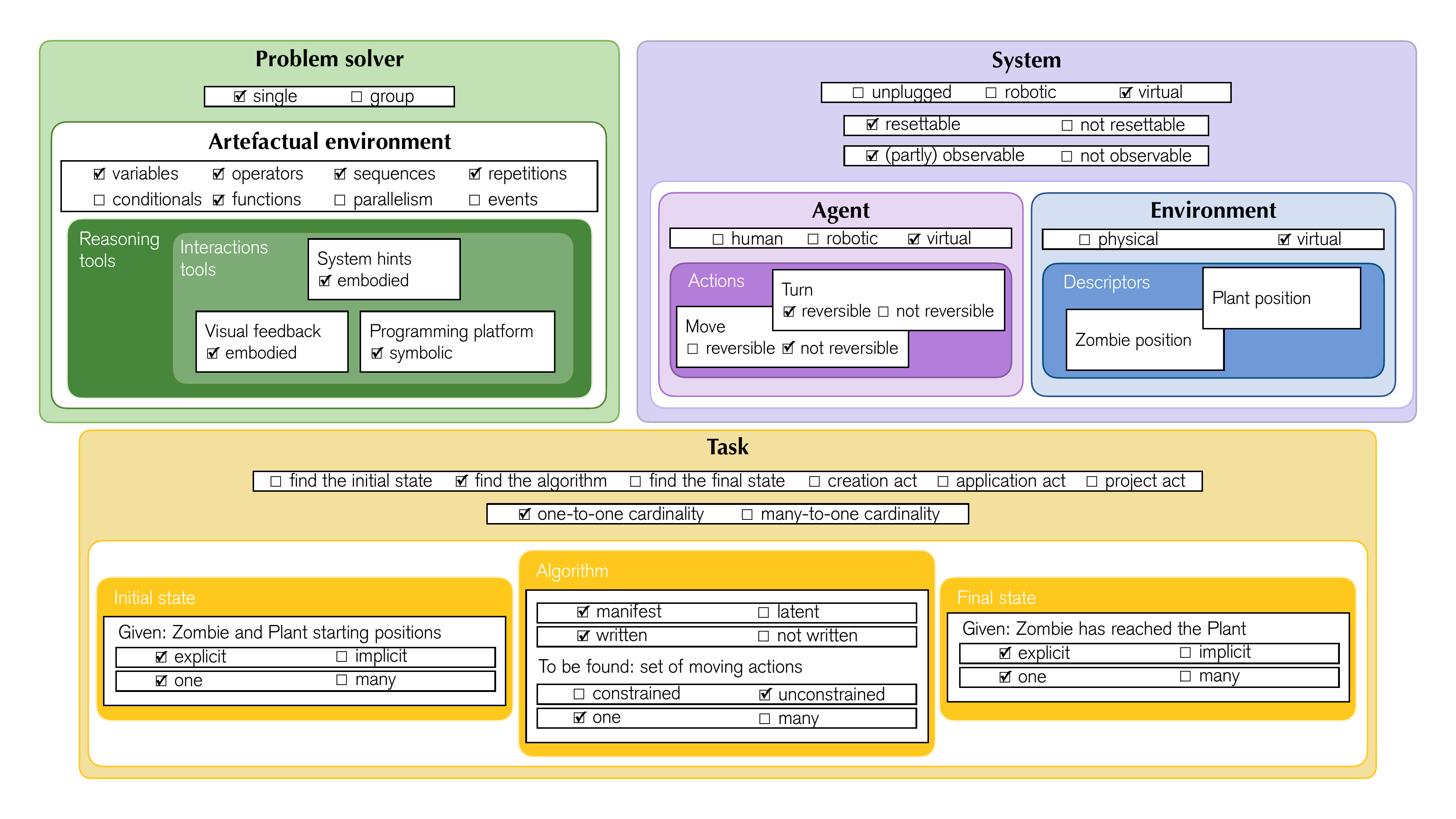}
    \caption{\textbf{Plants vs Zombies maze components and characteristics.}}
    \label{fig:PZ-features}
\end{figure*}

\begin{figure*}[!h]
    \centering
    \includegraphics[width=\textwidth]{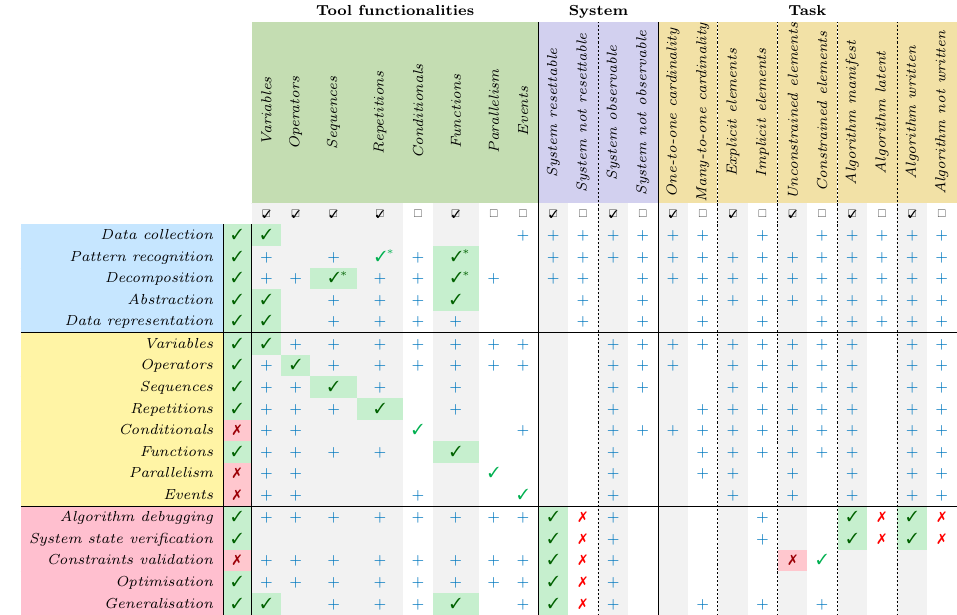}
    \caption{\textbf{Classic maze profile.}}
    \label{tab:classicmaze-mapping}
\end{figure*}

{Code.org, in addition to providing unplugged activities for CT development, as the Graph Paper Programming in} \Cref{sec:GPP}, {it delivers a platform with many coding activities for children based on Blockly, a Google framework for block-based programming} \citep{lovett2017coding}. 
The Classic Maze is part of the Hour of Code offered by Code.org, a worldwide effort to broaden participation in the computer science field \citep{classicmaze}.
Participants must use block-based programming to guide different characters through a maze in this activity. The creatures include ones from popular franchises such as Angry Birds, Plants vs Zombies, and Scrat from Ice Age. In this way, they learn the foundations of computer science and algorithmic concepts by successfully guiding the characters through the maze.
{The activities proposed in the Classic Maze exhibit a progressive increase in difficulty.
For this analysis, we explored the Angry Bird hitting the Green Pig maze, illustrated in} \Cref{fig:AB}, {where the character should be guided through a maze to reach and hit a Green Pig} \cite{roman2018detected}, {and the Plants vs Zombies maze, illustrated in} \Cref{fig:PZ} {a similar task differing for a different character involved, a Zombie, which must traverse the maze to reach the plant.}
{Components and characteristics of this CTP have been analysed using the graphical templates in} \Cref{fig:ABHGP-features,fig:PZ-features}, {while the profile, outlining the relationship between its characteristics and competencies, is illustrated in} \cref{tab:classicmaze-mapping}.

\subsection{Components (Angry Bird maze)}
\begin{itemize}[noitemsep,nolistsep]
\item \textit{Problem solver}: the student performing the task who must program the agent's behaviour. The artefactual environment comprises tools designed for reasoning and interacting with the system simultaneously, including the programming platform composed of the virtual scenario (embodied artefact), the blocks and the workspace (symbolic artefacts). The system also furnishes various hints to users (embodied artefact), including video tutorials, guidance on using the platform, command recommendations, suggestions on the number of blocks required to solve the task, and feedback on the problem solver's progress towards a solution.
\item \textit{Agent}: the Angry Bird, programmed to navigate a maze and hit the other character. The agent's actions comprise moving forward, turning left, and right. Moving forward is considered a non-reversible action, whereas turning is reversible, as a turn right can easily undo a turn left and vice versa. 
\item \textit{Environment}: the virtual scenario where the two creatures are located, described by their positions. 
\item \textit{Task}: find the algorithm. The initial state corresponds to the animals' initial positions, while in the final, the characters are in the same position. The algorithm is the set of moving actions to reach the system's final state from the initial.
\end{itemize}

 
\subsection{Characteristics}
\begin{itemize}[noitemsep,nolistsep]
\item \textit{Tool functionalities}: {for both mazes,} the programming platform enables problem solvers to reason about the task at hand and program the movements of the red bird by employing a set of predefined blocks, each representing a specific action the agent is authorised to perform: (i) \textit{variables}, while not explicitly delivered in the blocks of the programming platform, can be inferred from the visual feedback provided by the system, allowing the problem solver to store values such as the position of the characters; (ii) \textit{operators} represent the basic actions that the agent can perform, such as moving or turning and are represented by distinct blocks in the visual programming language depicted in cyan by the platform; (iii) \textit{sequences} are a series of blocks to be executed in a specific order and are implicitly conveyed by the collection of blocks; (iv) \textit{functions}, which are self-contained blocks of code that perform a specific task and can be executed multiple times with different inputs (e.g., different initial positions of characters), are a concept of relative complexity. It is not certain that the problem solver will recognise them as such rather than just blocks.
{For the Plants vs Zombies maze, the programming platform also includes the possibility of using \textit{repetitions}, represented in pink, adding another layer of difficulty.}
\item \textit{System resettability}: the platform provides a direct means of resetting the task through the ``start over'' button, even if some agent actions are irreversible. This allows the problem solver to start over and try a different approach if necessary.
\item \textit{System observability}: the system provides real-time visual feedback through animations and graphical representations of the system state and its changes, making it observable. The problem solver can monitor the effects of the agent's actions on the system, allowing him to have a complete understanding of the system's current state and to make informed decisions in their problem-solving process.
\item \textit{Task cardinality}: the task has a one-to-one mapping, with only one starting position for the animal elements, one final position for the Angry Bird to be placed, and only one algorithm to be found.
\item \textit{Task explicitness}: the elements of the task are given explicitly through the depiction of the scenario that clearly shows the animals' positions.
\item \textit{Task constraints}: no constraints exist on the elements to be found. All the blocks provided are available without limitations. 
\item \textit{Algorithm representation}: the algorithm is written in the workspace and expressed by the set of blocks and their connections. 
\end{itemize}

\subsection{Competencies}
\paragraph{Enabling features for competencies development}
\begin{itemize}[noitemsep,nolistsep]
\item \textit{Problem setting}: all competencies can be activated thanks to the presence of variables, sequences, and functions in the tool functionalities.
The manifest and written algorithm representation can further encourage the development of these skills.
The resettability of the system and the one-to-one cardinality of data elements facilitate data collection, pattern recognition and decomposition. 
Observing the system also supports data collection and pattern recognition. 
Using variables boosts pattern recognition and decomposition, while explicit and unconstrained elements, in addition to these two skills, encourage abstraction.
Sequences positively affect pattern recognition, abstraction and data representation, which is also facilitated by functions. Operators also encourage decomposition.

\item \textit{Algorithm}: all competencies associated with the algorithmic concepts enabled by the tool functionalities, meaning variables, operators, sequences, and functions, can be activated.
Features such as variables, operators, sequences, and functions help activate the algorithmic skills in a task.
These are further enhanced by the manifest and written algorithm representation, system observability, and the explicit and unconstrained definition of the task elements. The one-to-one cardinality enhances variables and operators.
{In the Plants vs Zombies maze, being the repetition functionality available in the tool, the related competence can be developed.}

\item \textit{Assessment}: {all competencies can be developed, i.e., algorithm debugging and system state verification} can be activated due to the resettability of the system and the manifest and written representation of the algorithm; optimisation can be activated as it only requires the resettability of the system; generalisation can be activated through the system's resettability and the presence of variables and functions. 
Features such as variables, operators, sequences, functions and system observability help develop algorithm debugging and optimisation, while sequences can also foster generalisation. 
\end{itemize}

\paragraph{Inhibiting features for competencies development}
This paragraph explores the impact of missing features on skill activation, focusing on how adjusting the task can enable the development of certain competencies.
\begin{itemize}[noitemsep,nolistsep]
\item \textit{Repetitions, conditionals, parallelism, and events}:non-activable due to the absence of specific tool functionalities. These competencies cannot be triggered until these functionalities are added to the tool.
\item \textit{Constraint validation}: non-activable because the algorithm to be found is unconstrained. Some constraints can be imposed on the blocks used in the task's programming platform to activate this skill, such as limiting the agent's ability to turn right. 
\end{itemize}

\section{{Store the Marbles}}\label{sec:RLB}
\begin{figure*}[!htb]
\centering
\includegraphics[width=.95\textwidth]{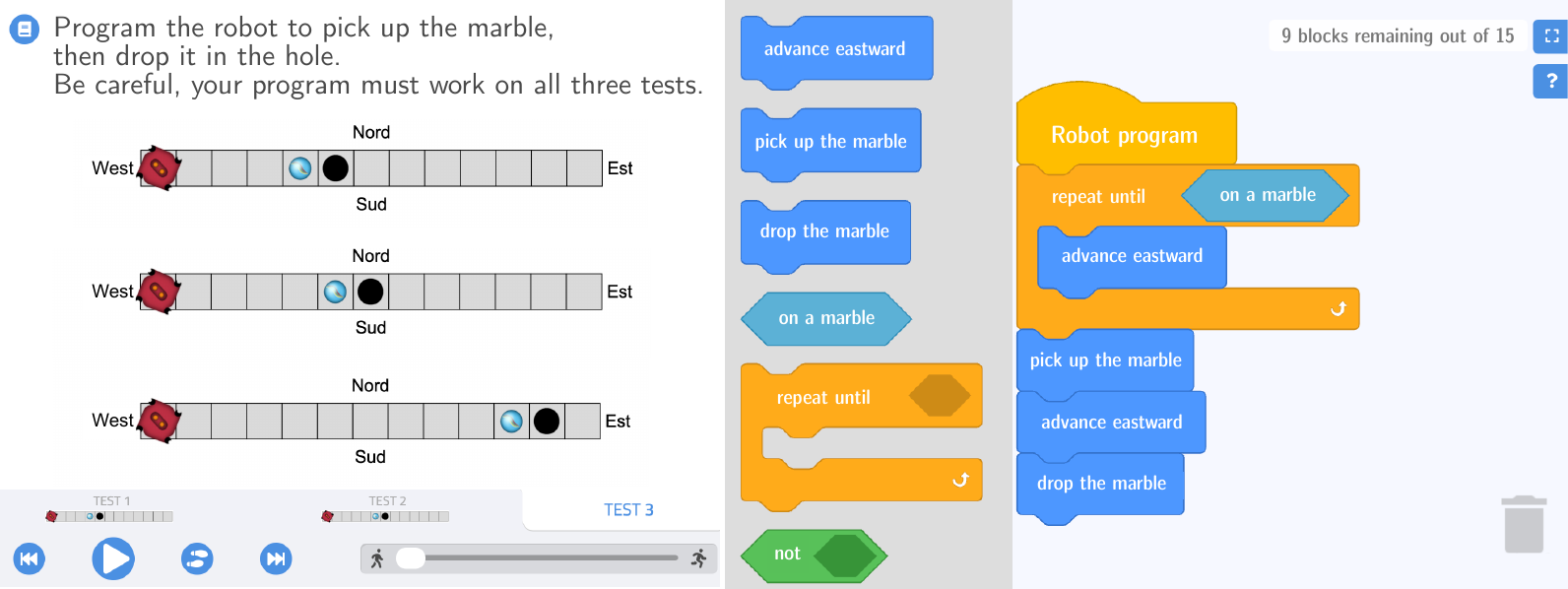}
\caption[store the marbles ]{\textbf{The Store the Marbles activity adapted from \cite{RLB}. }The task requires the problem solver to program the robot to produce an algorithm valid for different situations using a visual programming language. }
\label{fig:rlb}
\end{figure*}

\begin{figure*}[!t]
		\centering
		\includegraphics[width=\textwidth]{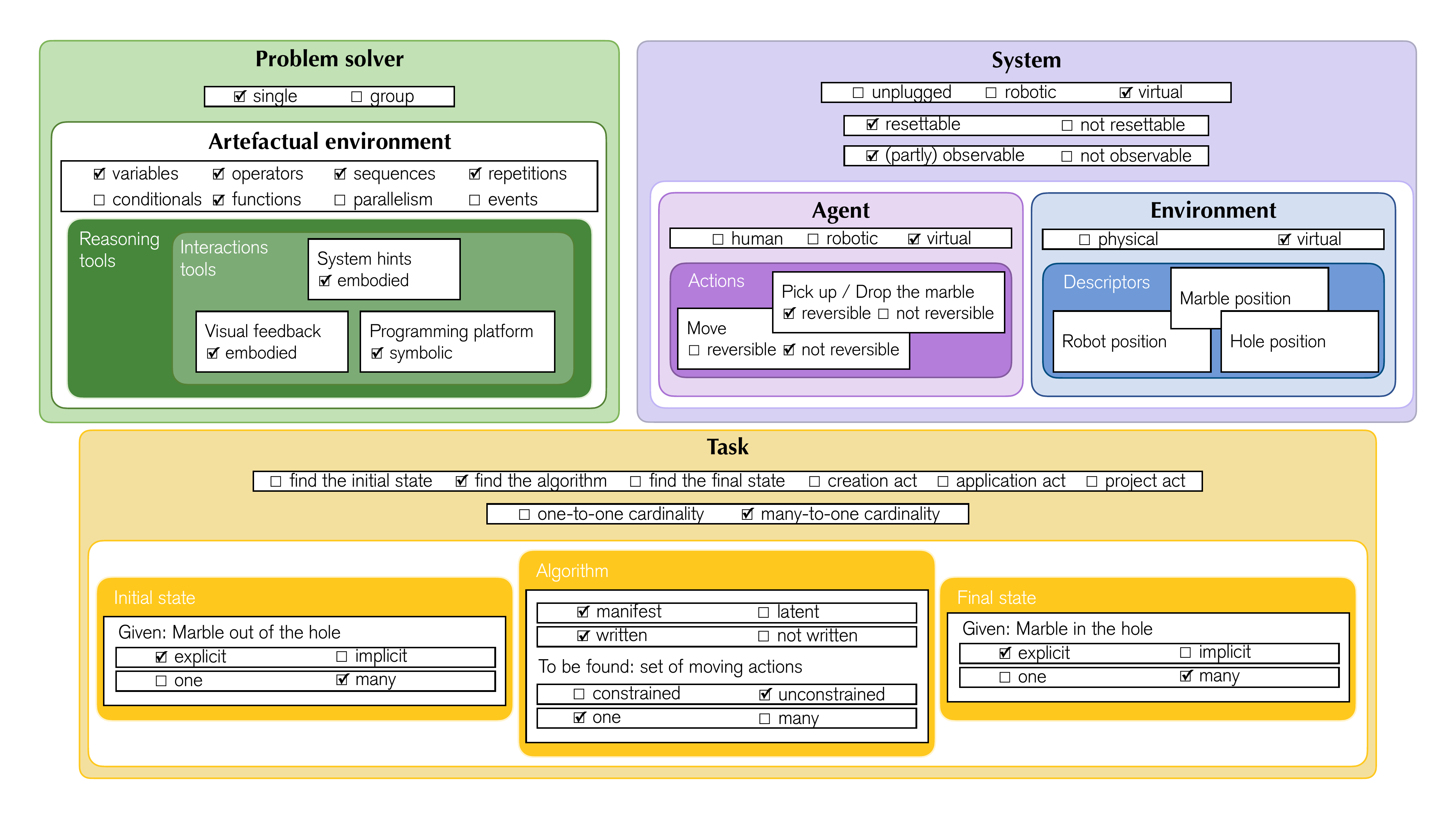}
		\caption{\textbf{Store the Marbles components and characteristics.}}
		\label{fig:RLB-features}
	\end{figure*}

\begin{figure*}[!h]
		\centering
		\includegraphics[width=\textwidth]{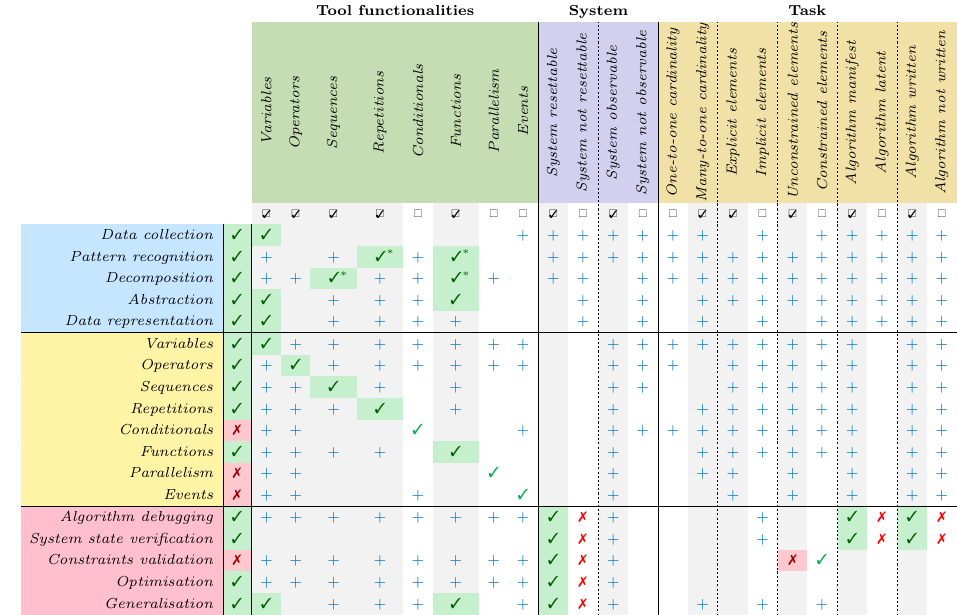}
		\caption{\textbf{Store the Marbles profile.}}
		\label{tab:RLB-mapping}
	\end{figure*}

Store the Marbles is a virtual programming activity, presented by Algorea, an online resource designated by France-IOI, for learning the basics of programming \citep{RLB, FIOI}. 
{This CTP, illustrated in} \Cref{fig:rlb}, {is part of a series of progressive difficulty courses and exercises available on the France-IOI website and is designed to teach students problem-solving skills and programming concepts using a visual block-based programming language. 
In the activity, the robot should pick up the marble on his path and drop it into a hole. 
Components and characteristics of this CTP have been analysed using the graphical template in} \Cref{fig:RLB-features}, {while the profile, outlining the relationship between its characteristics and competencies, is illustrated in} \cref{tab:RLB-mapping}.

\subsection{Components}
\begin{itemize}[noitemsep,nolistsep]
\item \textit{Problem solver}: the student who must program the agent's behaviour. 
The artefactual environment comprises tools designed for reasoning and interacting with the system simultaneously, including the programming platform composed of the virtual scenario (embodied artefact), the blocks and the workspace (symbolic artefacts). The system also furnishes suggestions on the number of blocks required to solve the task (embodied artefact), but it does not provide additional hints beyond this information. 
\item \textit{Agent}: the virtual robot, programmed to move on the path to collect the marbles and drop them in the hole. 
The agent's actions comprise moving eastward, picking up a marble, and dropping a marble. The simple movement is considered irreversible, whereas picking up and dropping a marble are reversible actions, as they are the reverse of each other.
The agent's actions comprise moving forward, turning left, and right. Moving forward is considered a non-reversible action, whereas turning is reversible, as a turn right can easily undo a turn left and vice versa. 
\item \textit{Environment}: the virtual scenario where the robot and the marble are located, described by their respective positions. 
\item \textit{Task}: find the algorithm. The initial state corresponds to the robots' initial positions and the maze distribution of marbles. The final state corresponds to the robot positioned on the hole and the marbles inside it.
The algorithm is the set of moving actions to reach the system's final state from the initial.
\end{itemize}


\subsection{Characteristics}
\begin{itemize}[noitemsep,nolistsep]
\item \textit{Tool functionalities}: the system provides a comprehensive set of tools for the problem solver to implement the algorithm to solve the task, including 
(i) \textit{variables} allows the problem solver to store values such as the position of the robot, marble, and hole; 
(ii) \textit{operators} represent the basic actions that the agent can perform, such as moving or turning and are represented by the blue blocks of the visual programming language; 
(iii) \textit{sequences} are obtained by concatenating a series of blocks to be executed in a specific order, allowing the problem solver to plan and execute a series of actions to reach the goal; 
(iv) \textit{repetitions} are offered by a specific orange block and allow the problem solver to repeat a sequence of actions until a specific condition is met; 
(v) \textit{functions} are self-contained blocks of code that perform a specific task and can be executed multiple times, providing a flexible and reusable component in the algorithm design. The platform places importance on the design of the algorithm as a function that can adapt to changing inputs, such as the different starting conditions of the task.
\item \textit{System resettability}: the platform allows for resetting the task through a button, enabling the problem solver to start over and try a different approach.
\item \textit{System observability}: the platform provides real-time visual feedback, making the system observable.
\item \textit{Task cardinality}: the task has a many-to-one cardinality, with many initial and final states given and only one algorithm to be found. The challenge in this activity is for the program to work with multiple initial states, as the problem solver is given three different configurations of marble and hole positions. The solution must successfully sort the marbles into the correct locations. 
\item \textit{Task explicitness}: the task elements are given explicitly through the depiction of the scenario that clearly shows the positions of the robot, marble, and hole.
\item \textit{Task constraints}: the algorithm is unconstrained.
\item \textit{Algorithm representation}: the algorithm is manifest and written, expressed through a set of blocks.
\end{itemize}

\subsection{Competencies}
\paragraph{Enabling features for competencies development}
\begin{itemize}[noitemsep,nolistsep]
\item \textit{Problem setting}: all competencies can be activated thanks to the presence of variables, sequences, repetitions and functions in the tool functionalities.
The manifest and written representation of the algorithm and the many-to-one cardinality can further encourage the development of these skills.
The resettability of the system facilitates data collection, pattern recognition and decomposition. 
The system observability supports data collection and pattern recognition.
Using variables boosts pattern recognition and decomposition, while explicit and unconstrained elements in addition to these two skills encourage abstraction.
The use of sequences positively affects pattern recognition, abstraction and data representation, which is also facilitated by functions. Operators also encourage decomposition, while repetitions promote decomposition abstraction and data representation.
\item \textit{Algorithm}: all competencies associated with the algorithmic concepts enabled by the tool functionalities, i.e., variables, operators, sequences, repetitions and functions, can be activated.
Features such as variables, operators, sequences, repetitions and functions help activate the algorithmic skills in a task.
These are further enhanced by the manifest and written representation of the algorithm, system observability, the explicit and unconstrained definition of the task elements, and the many-to-one cardinality.
\item \textit{Assessment}: {all assessment competencies can be developed, like algorithm debugging and system state verification} can be activated due to the resettability of the system and the manifest and written representation of the algorithm; optimisation can be developed as it only requires the resettability of the system; generalisation can be activated through the system's resettability and the presence of variables and functions. 
Features such as variables, operators and functions help develop algorithm debugging and optimisation. 
Sequences, repetitions and system observability foster algorithm debugging, optimisation, and generalisation. 
The many-to-one cardinality influences the development of generalisation as well. 
\end{itemize}

\paragraph{Inhibiting features for competencies development}
The absence of certain features can limit the activation of certain skills.
\begin{itemize}[noitemsep,nolistsep]
\item \textit{Conditionals, parallelism, and events}: non-activable as the platform does not supply them. 
\item \textit{Constraint validation}: non-activable due to the algorithm's absence of constraints. The task is adjustable by imposing constraints on it.
\end{itemize}

\section{{Zoombinis Allergic Cliffs puzzle}}\label{sec:zoombinis}
\begin{figure}[!htb]
\centering
\includegraphics[width=\columnwidth]{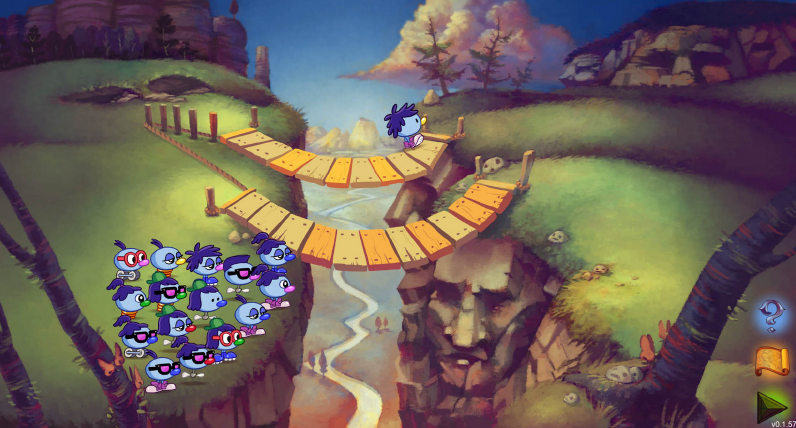}
\caption[Allergic Cliffs Puzzle]{\textbf{The Zoombinis Allergic Cliffs puzzle from \cite{zoombinis2021web}.} The player must determine which characteristics allow the Zoombinis to cross the bridge without being sent back. The bottom cliff does not accept creatures with flat hair, while the top cliff rejects all others.}
\label{fig:Zoombinis}
\end{figure}

\begin{figure*}[!t]
		\centering
		\includegraphics[width=\textwidth]{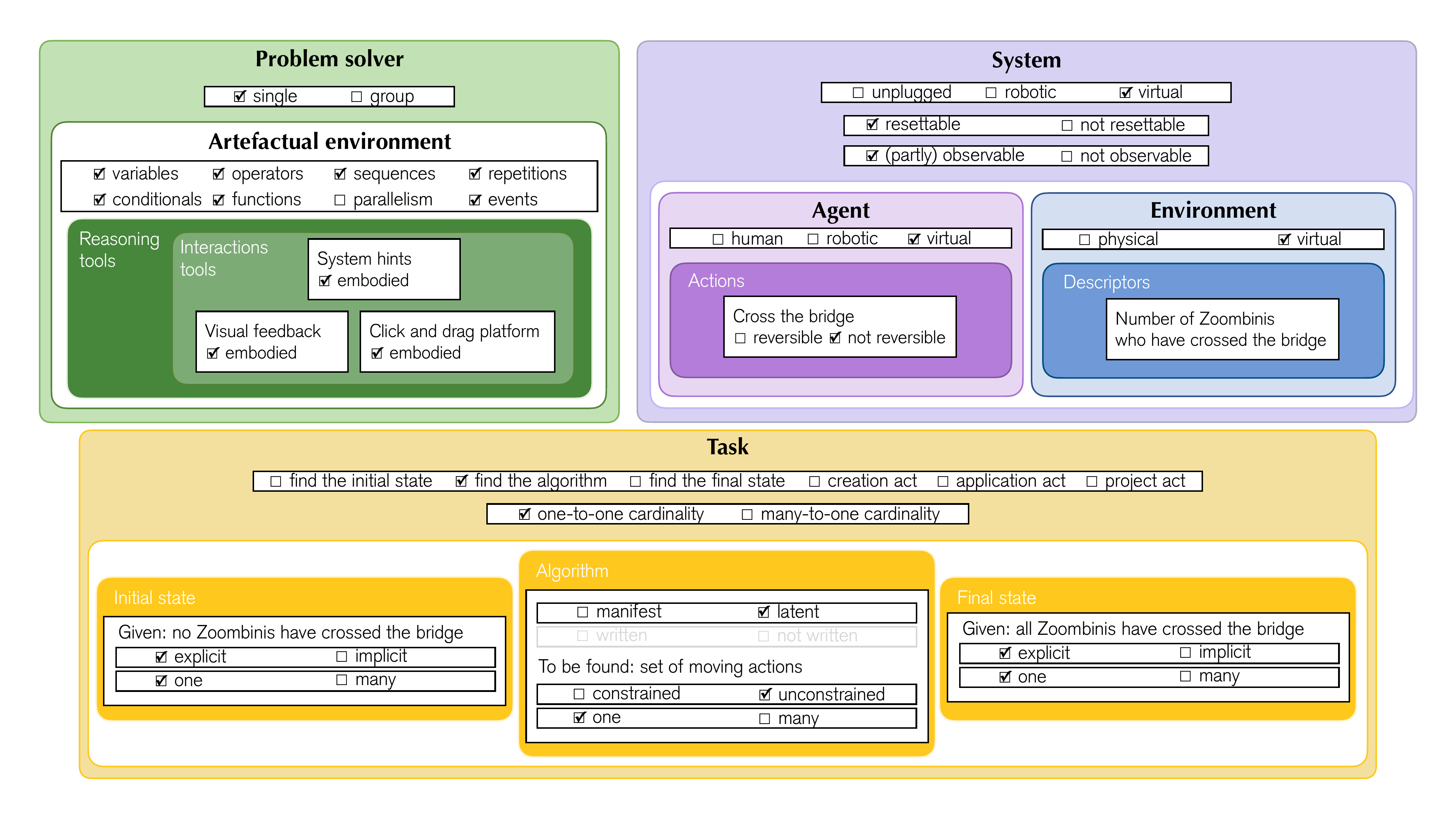}
		\caption{\textbf{Zoombinis Allergic Cliffs puzzle components and characteristics.}}
		\label{fig:zoombinis-features}
	\end{figure*}

\begin{figure*}[!h]
		\centering
		\includegraphics[width=\textwidth]{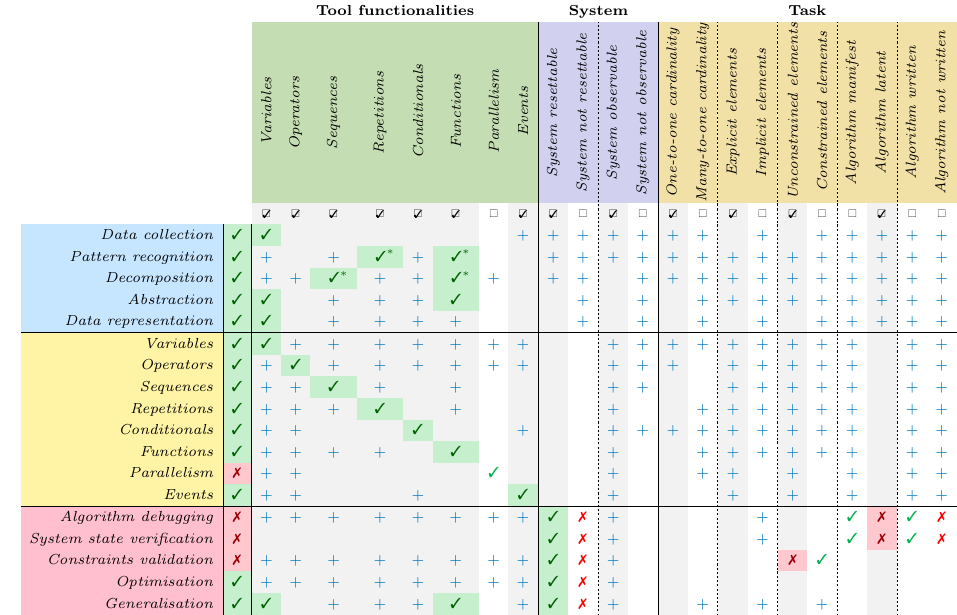}
		\caption{\textbf{Zoombinis Allergic Cliffs puzzle profile.}}
		\label{tab:zoombinis-mapping}
	\end{figure*}

The Zoombinis activity is a popular CT learning game where the player must guide little blue creatures with distinct personalities and appearances through different puzzles to escape imprisonment \citep{zoombinis2021web, zoombinis2021}.
{In the Allergic Cliffs Puzzle, illustrated in} \Cref{fig:Zoombinis}, {the task is to find a procedure that allows all Zoombinis to cross the bridge.
Each cliff accepts different attributes of the creatures, and the player must find the correct combination of Zoombini attributes that will allow each creature to cross the bridge.
Components and characteristics of this CTP have been analysed using the graphical template in} \Cref{fig:zoombinis-features}, {while the profile, outlining the relationship between its characteristics and competencies, is illustrated in} \cref{tab:zoombinis-mapping}.

\subsection{Components}
\begin{itemize}[noitemsep,nolistsep]
\item \textit{Problem solver}: the player who must find the correct combination of Zoombini attributes that will allow each creature to cross the bridge.
The artefactual environment comprises tools for reasoning and interacting with the system simultaneously, including the click-and-drag platform (embodied artefact) and the visual feedback of the virtual scenario (embodied artefact). 
The system provides feedback on whether the Zoombini's attributes match the requirements of the bridge (embodied artefact), allowing the player to adjust their strategy accordingly.
\item \textit{Agent}: the Zoombinis character, which can be dragged to the entrance of a bridge and the only action is allowed to do is to cross it or be sent back in case of failure. This action cannot be reset.
\item \textit{Environment}: the virtual scenario described by the Zoombinis' positions and the number of Zoombinis that have crossed a bridge. 
\item \textit{Task}: find the algorithm. The system's state depends on the number of creatures that crossed the bridge, initially none and at the end all. The algorithm is the set of moving actions to reach the final state.
The problem solver can try different combinations of attributes, testing one attribute's value by holding one attribute constant and testing the others, or even proceed by changing values and attributes until all Zoombinis have crossed the bridges.
\end{itemize}

 
\subsection{Characteristics}
\begin{itemize}[noitemsep,nolistsep]
\item \textit{Tool functionalities}: (i) \textit{variables} are used to refer to the attributes of the Zoombinis, such as hair colour or eye shape, which can have different values and must be combined correctly to cross the bridge; 
(ii) \textit{operators} can be used to combine the conditions for the Zoombinis to cross the bridge, for example, the player could use a logical AND operator to require that the creature have two specific attributes to cross the bridge; 
(iii) \textit{sequences} can be the order in which the Zoombinis are moved to cross the bridge; 
(iv) \textit{repetitions} are used to try multiple combinations of Zoombini attributes until the problem solver finds the correct solution;
(v) \textit{conditionals} are reflected in the requirement of the bridges to accept certain attributes of the Zoombinis, as the player must determine if the Zoombini's attributes match the requirements of the bridge;
(vi) \textit{functions} are the combination of Zoombini attributes that allows all creatures to cross the bridges, as it maps inputs (the Zoombini attributes) to outputs (the success or failure of the Zoombinis crossing the bridges);
(vii) \textit{events} correspond to the Zoombini being sent back as it occurs due to a wrong player's combination of Zoombini attributes.
\item \textit{System resettability}: the platform provides a direct means of resetting the game.
\item \textit{System observability}: the system provides real-time visual feedback through animations and graphical representations of the system state and its changes, making the system observable. 
\item \textit{Task cardinality}: the task presented in the scenario is a one-to-one mapping.
\item \textit{Task explicitness}: the system's state is explicit and is represented by the number of Zoombinis that have crossed the bridge.
\item \textit{Task constraints}: no constraints exist on the elements to be found. All the blocks provided are available without limitations.    
\item \textit{Algorithm representation}: the algorithm is latent because the problem solver only performs actions such as dragging the characters across the bridge, but the underlying logic and steps to reach the final goal of having all Zoombinis cross the bridge are not explicitly defined by the player.
\end{itemize}

\subsection{Competencies}

\paragraph{Enabling features for competencies development}
\begin{itemize}[noitemsep,nolistsep]
\item \textit{Problem setting}: all competencies can be activated thanks to the presence of variables, sequences, and functions in the tool functionalities.
The latent algorithm can further encourage the development of these skills.
The resettability of the system and the one-to-one cardinality facilitate data collection, pattern recognition and decomposition. 
Observing the system also supports data collection and pattern recognition. 
Explicit and unconstrained elements boost pattern recognition, decomposition and abstraction.
Moreover, all tool functionalities have a positive impact on the problem setting learning process, helping the problem solver to understand the problem and find a solution more effectively.
\item \textit{Algorithm}: all competencies associated with the algorithmic concepts enabled by the tool functionalities can be activated and foster the development also of the other algorithmic concepts related skills.
The system observability and the explicit and unconstrained definition of the task elements further enhance these. 
The one-to-one cardinality helps to enhance variables, operators and conditionals further. 
\item \textit{Assessment}: optimisation can be activated by the resettability of the system, while generalisation can be developed through the system's resettability and the presence of variables and functions. The tool's functionalities and the system's observability aid in developing these skills.
\end{itemize}

\paragraph{Inhibiting features for competencies development}
\begin{itemize}[noitemsep,nolistsep]
\item \textit{Parallelism}: non-activable due to the absence of the specific tool functionalities. The design of the activity could be altered to include multiple Zoombinis acting simultaneously, allowing for the simultaneous execution of multiple actions.
\item \textit{Algorithm debugging}: non-activable because the algorithm is implicit and in the form of dragging and dropping the characters to cross the bridge. Without the ability to debug, the problem solver must rely on their understanding of the rules and ability to identify any issues through trial and error.
To integrate debugging, the algorithm needs to be made explicit, either by writing down the steps as text or in code form. Then, the user can test and debug the algorithm by checking if each step is performed correctly and if the final result matches expectations, thereby improving their debugging skills.
\item \textit{System state verification}: non-activable because both the initial and final state are provided. The task can be adjusted to activate this skill by making one of these states to be found. 
\item \textit{Constraint validation}: non-activable since the algorithm to be found is unconstrained. Constraints can be imposed to activate this skill, such as the order in which the Zoombinis can cross the bridge. 
\end{itemize}
\end{document}